\documentclass[review,1p]{elsarticle}

\usepackage{amssymb}
\usepackage{amsmath}
\usepackage{amsbsy}
\usepackage{multirow,multicol}
\usepackage{epsfig}
\usepackage[utf8]{inputenc}
\usepackage{multirow}
\usepackage{todonotes}

\usepackage[draft]{fixme}
\fxusetheme{color}
\FXRegisterAuthor{ak}{apek}{\color{blue}AK}
\FXRegisterAuthor{db}{adb}{\color{red}DB}

\begin{document}


\begin{frontmatter}
\title{A Stabilised Nodal Spectral Element Method for \\ Fully Nonlinear Water Waves }

\author[apek]{A.~P.~Engsig-Karup}
\ead{apek@dtu.dk, Office phone: +45 45 25 30 73}
\address[apek]{Department of Applied Mathematics and Computer Science \\ Center for Energy Resources Engineering (CERE) \\ Technical University of Denmark, 2800 Kgs. Lyngby, Denmark. }

\author[cles]{C.~Eskilsson}
\address[cles]{Department of Shipping and Marine Technology \\ Chalmers University of Technology, SE-412 96 Gothenburg, Sweden.}

\author[apek,dabi]{D.~Bigoni}
\address[dabi]{Department of Aeronautics and Astronautics\\ Massachusetts Institute of Technology, 02139 Cambridge, MA, USA. }

\date{\today.}

\begin{abstract}
We present an arbitrary-order spectral element method for general-purpose simulation of non-overturning water waves, described by fully nonlinear potential theory. The method can be viewed as a high-order extension of the classical finite element method proposed by Cai et al (1998) \cite{CaiEtAl1998}, although the numerical implementation differs greatly. Features of the proposed spectral element method include: nodal Lagrange basis functions, a general quadrature-free approach and gradient recovery using global $L^2$ projections.
The quartic nonlinear terms present in the Zakharov form of the free surface conditions can cause severe aliasing problems and consequently numerical instability for marginally resolved or very steep waves. We show how the scheme can be stabilised through a combination of over-integration of the Galerkin projections and a mild spectral filtering on a per element basis. This effectively removes any aliasing driven instabilities while retaining the high-order accuracy of the numerical scheme. The additional computational cost of the over-integration is found insignificant compared to the cost of solving the Laplace problem. 
The model is applied to several benchmark cases in two dimensions. The results confirm the high order accuracy of the model (exponential convergence), and demonstrate the potential for accuracy and speedup. The results of numerical experiments are in excellent agreement with both analytical and experimental results for strongly nonlinear and irregular dispersive wave propagation. The benefit of using a high-order -- possibly adapted -- spatial discretisation for accurate water wave propagation over long times and distances is particularly attractive for marine hydrodynamics applications. 

\end{abstract}


\end{frontmatter}

\section{Introduction}

Robust and cost-efficient time-dependent simulation of the propagation and transformation of water waves in both shallow near-shore and deeper off-shore areas is a computationally challenging and longstanding scientific problem for ocean, coastal and naval engineering applications. For example, fully non-linear wave simulations have been subject to research for a long time, and have still not yet entered common coastal and ocean engineering practice. One remaining key challenge is to resolve accurately highly nonlinear and dispersive wave propagation in maritime areas while taking into account varying bathymetry, the geometry of complex structures and their nonlinear interaction with fixed and floating structures. Resolving this problem lead to improved opportunities for using simulations in realistic marine regions as well as enabling experiments in numerical wave tanks of increasing fidelity. Furthermore, it is attractive to design a flexible computational framework that entails use on modern commodity workstations as well as high-performance computing systems. These goals dictate stringent requirements on the design of engineering tools, and this suggests that a generic tool for wave propagation should be based on 
\begin{enumerate}[(i)]
\item a general modelling basis for broadly describing relevant {\em physics},
\item a generalised computational framework for the {\em numerics}, and
\item software design for efficient mapping to modern and emerging many-core {\em architectures}. 
\end{enumerate}
Our main objective is to meet each of these requirements via a spectral element based computational framework. Such a framework has already been used for several different applications within marine hydrodynamics such as the fully nonlinear potential flow equations (present work), Boussinesq and shallow water equations \cite{EskilssonEngsigKarup2014}. In this work we focus exclusively on the first two requirements and we pave the way for the fulfilment of the last one, which is being addressed in ongoing work. The use of modern many-core hardware is attractive for acceleration of the high-order spectral element framework and for enabling practical computations of realistic engineering problems \cite{Markall2013,Markidis01082015}. 

\subsection{Choice of modelling basis for description of physics}

During the last decades computationally efficient depth-integrated Boussinesq-type models have been widely adopted as essential tools for water wave modelling in the near-shore region; see e.g. \cite{MS99, Brocchini2013}. 
For shorter waves, such as the ones arising in offshore and naval engineering, Boussinesq-type models are not applicable due to the limited accuracy in terms of dispersive and nonlinear properties. 
For these cases we have to turn to Computational Fluid Dynamics (CFD) models based on the Navier-Stokes equations, or fully nonlinear potential flow (FNPF) models. 
The CFD models take viscous effects into account; effects that may be important for breaking waves, load computations, boundary layer effects, etc. 
Even though CFD is often prohibitively expensive in terms of computational resources when considering simulation of entire sea states \cite{EskilssonEtAl2015}, it is widely used to quantify  breaking wave loads and wave run-up on structures. CFD models are typically too dissipative as a result of the low-order accuracy imposed by computational limitations for large-scale wave simulations. 
In contrast, already today FNPF models can be used for long-time and large-scale wave simulations \cite{DucrozetEtAl2007,Glimberg2013}. 
FNPF solvers can be used for resolution of full sea states in large marine or coastal areas where nonlinear waves interact with fixed or floating structures.
The cons of the FNPF models are that they cannot account for non-overturning breaking waves and viscous effects. 
For these reasons it can be attractive to combine FNPF models (far-field) with CFD (near-field) in hybrid modelling approaches for wave structure interaction, cf. \cite{PaulsenEtAl2014}. 
This hybrid approach enables better simulation of strong nonlinear wave structure interactions in areas where the local wave climates can not be predicted accurately via a FNPF model.

\subsection{On the quest towards developing numerical strategies for real-world applications}

A review of existing conventional discretisation methods and applications reveals that historically the main emphasis has been on Finite Difference Methods (FDM), boundary element methods (BEM) and finite element methods (FEM) \cite{KimEtAll1999,HarrisEtAl2014}. 
These methods have been designed for the concept of Numerical Wave Tanks (NWT). 
The main computational bottleneck in all such numerical solvers is the solution of a large linear system. 
In FDM and FEM the discretisation procedure leads to sparse linear systems due to the local support of discrete operators, while in BEM it is only the domain boundary that needs representation. 
The discretisation procedure for BEM is based on a surface integral formulation together with Green's identities.
This leads to dense non-symmetric matrix operators that cannot be solved in a straightforward way with linear asymptotic scaling. 
There has been some recent progress in bringing the asymptotic cost (scaling rate) down for BEM \cite{HarrisEtAl2014} for both matrix-vector multiplications and storage requirements using both high-order basis functions and the Fast Multipole Method (FMM) \cite{GreengardRokhlin1987}. 
While this strategy can asymptotically achieve linear complexity $\mathcal{O}($n$)$ ($n$ number of computational nodes) in work effort for the spatial solver, it has a large constant in front of this asymptotic scaling term due to the need of solving a dense linear system of equations. This leaves BEM less efficient compared to volume-based discretisation methods such as FDM and FEM solvers as suggested in \cite{WuTaylor1995,ShaoEtAl2014}. 
We note, that BEM is particularly attractive as a near-field solver for cases where waves interact with complex geometries \cite{ZhouEtAl2015} and may be combined with a far-field solver such as FEM \cite{WuTaylor2003}.
The overall efficiency and scalability of BEM \cite{HarrisEtAl2014b} can be compared to  efficient and massively parallel free surface hydrodynamics solvers such as \cite{EngsigKarupEtAl2011,EGNL13,ShaoEtAl2014} which can achieve very high efficiency and scalability using multigrid-type methods \cite{LiFleming1997,EngsigKarup2014} for arbitrary sized discrete problems, in particular when the (possibly curvilinear multi-block) meshes are logically structured, e.g. as in \cite{Glimberg2013}. 

\subsection{State-of-the-art in finite element methods for fully nonlinear water waves}

Reviews on the state-of-the art of numerical models for freely propagating water waves are given in \cite{KimEtAll1999,Tanizawa2000,DiasBridges2006,SpinnekenEtAll2012,NimmalaEtAl2012}. Our scope in the present work is restricted to FNPF solvers and FEM.

The use of FEM for fully nonlinear water waves was pioneered by Wu \& Eatock Taylor (1994) \cite{WuTaylor1994}. 
Since then, the majority of solvers for fully nonlinear potential flow equations have been limited to second (low) order FEM schemes \cite{WuTaylor1995,GreavesBorthwickTaylor1997,MaWuTaylor2001,MaWuTaylor2001b} based on a Mixed Eulerian Lagrangian method \cite{LonguetHigginsCokelet1976} for updating the free surface variables. 
This approach requires expensive mesh update techniques and may suffer from stability problems for deformed meshes \cite{RobertsonSherwin1999}. 
However, it is particularly well suited for dealing with the interaction between waves and fixed or freely floating bodies \cite{WangWu2011}. To overcome this expensive re-meshing problem a less expensive  Quasi Arbitrary Lagrangian Eulerian finite element method (QALE-FEM) was developed \cite{MaYan2006} and a novel mesh update technique for ALE is described in \cite{BouffanaisDeville2006}. 

Very little work has been done based on purely Eulerian formulations, e.g. the classical approach where a $\sigma$-transformed formulation is used. 
Previous studies based on using the $\sigma$-transformation \cite{CaiEtAl1998,WesthuisAndonowati1998,ClaussSteinhagen:1999tr,TurnbullEtAl2003} have been limited to two spatial dimensions and flat bathymetries for wave-structure interaction applications. 
These are essentially limited to the inclusion of bottom-mounted possibly surface-piercing structures, despite that the method may be used as an efficient base solver for very large domains and from shallow to deep waters. 
More details about free surface solvers based on FEM can be found in \cite{Zienkiewicz2014,TurnbullEtAl2003,WangWu2011}. 

While there are very few studies on high-order finite element methods for water wave applications, it is well known that large efficiency gains for time-dependent wave problems can be achieved by the use of high-order accurate methods \cite{KO72}. 
One attractive class of methods are the Spectral Element Methods (SEM) \cite{EESHB05,DFM02,KS2005} which rely on the strong theoretical foundations of Spectral Methods \cite{CHQZ2006,HGG07}. 
The Spectral Element Method was first used by Patera (1984) \cite{PAT84} for fluid dynamics problems and has since then gained popularity. 
It combines the best properties of spectral methods and classical finite element methods, namely, high accuracy and flexibility in the spatial representation of domains. 
For smooth problems, the use of high-order discretisation is generally an efficient way to balance accuracy and cost, since a high-order discretisation allow for more coarse spatial representation compared to a low-order method. 
High-order methods may also reduce the robustness of explicit time-stepping methods due to global restriction on the stable time step sizes, although, in recent works \cite{EGNL13,EskilssonEngsigKarup2014} it is demonstrated that for certain wave models that require operator inversion in the time-stepping, high-order discretization combined with explicit time stepping methods need not reduce robustness - not even for unstructured mesh methods -- and thereby provide a strong basis for efficient tools. 
Furthermore, with the geometric flexibility provided by adaptive meshes and the support for cost-efficient simulation through $h$- and $p$-refinement strategies, the SEM is suitable for large-scale computing due to accurate temporal integration over long times via high-order discretisation that lead to small dispersive and dissipative errors. However, a key challenge for unstructured solvers such as FEM and SEM is achieving work effort that is linearly proportional to the number of computational processing nodes (weak scalability) for unstructured meshes used to represent general domains. 
This requires the design of efficient and scalable preconditioned iterative methods \cite{CaiEtAl1998} and of data structures that map efficiently to modern many-core architectures \cite{GoddekeEtAl2008}.

\subsection{Paper contributions}
The main objective of this work is the proper design and validation of the SEM framework for -- ultimately large-scale -- dispersive and nonlinear water wave propagation. 
This effort is crucial to enable engineering analysis of water waves in marine regions. 
This paper proposes for the first time a stabilised high-order spectral element method for solving the fully nonlinear potential flow equations. 
A rigorous assessment of the stabilised numerical model is carried out using numerical experiments and known benchmarks in two space dimensions. 

\section{Governing equations}

For the description of inviscid and irrotational fluid flows we introduce the set of fully nonlinear and dispersive free surface equations described by a two-variable potential flow formulation. This formulation can be derived from the Navier-Stokes equations, cf. \cite{EGNL13}. 
\begin{figure}[!htb]
\begin{center}
\includegraphics[height=3cm]{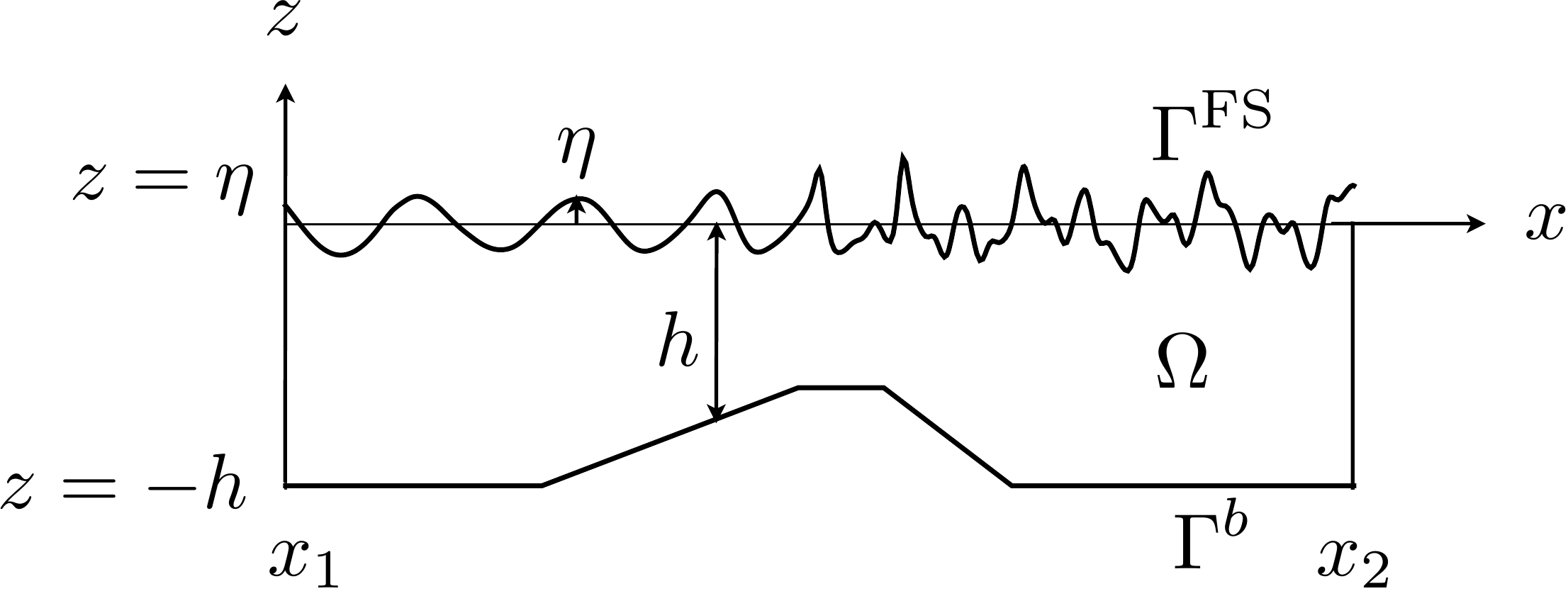}
\end{center}
\caption{Notations for physical domain ($\Omega$).}
\label{fig:domainorig}
\end{figure}
Let the fluid domain $\Omega\subset \mathbb{R}^d$ ($d=2$) be a bounded, connected domain with piecewise smooth boundary $\Gamma$ and introduce restrictions to the free surface $\Gamma^{\textrm{FS}}\subset \mathbb{R}^{d-1}$ and the bathymetry $\Gamma^{b}\subset \mathbb{R}^{d-1}$. Let $T:t\geq 0$ be the time domain. The mathematical problem is to find a scalar velocity potential function  $\phi(x,z,t):\Omega\times T \to \mathbb{R}$ and to determine the evolution of the free surface elevation $\eta(x,t):\Gamma^{\textrm{FS}} \times T \to \mathbb{R}$. The notations are illustrated in Figure \ref{fig:domainorig}.

The Eulerian description of the unsteady kinematic and dynamic free surface boundary conditions can be expressed in the Zakharov form \cite{ZAK1968}. Find $\eta,\tilde{\phi}$ such that
\begin{subequations}
\begin{align}
\partial_t\eta = -\partial_x\eta\partial_x\tilde{\phi}+\tilde{w}(1+\partial_x\eta\partial_x\eta) \quad \textrm{in} \quad \Gamma^{\textrm{FS}} \times T & \label{FSeta} \\
\partial_t \tilde{\phi} = -g\eta - \frac{1}{2}\left(\partial_x\tilde{\phi}\partial_x\tilde{\phi}-\tilde{w}^2(1+\partial_x\eta\partial_x\eta)\right) \quad \textrm{in} \quad \Gamma^{\textrm{FS}} \times T &
\label{FSphi}
\end{align}
\label{FSeqs}
\end{subequations}
We have introduced the '$\sim$' symbol to denote functionals defined only on the free surface plane. The vertical component of the velocity $\tilde{w}\equiv\partial_z\phi|_{z=\eta}$ is determined by first solving a Laplace problem 
\begin{subequations}
\begin{align}
\phi  =  \tilde{\phi}, \quad z = \eta \quad &\textrm{on} \quad \Gamma^{\textrm{\textrm{FS}}} &\\
\partial_{xx}\phi + \partial_{zz}\phi  =  0, \quad -h(x)<z<\eta \quad &\textrm{in}\quad \Omega \label{Laplace} \\
\partial_z \phi + \partial_{x}h\partial_{x}\phi = 0, \quad z=-h(x) \quad &\textrm{on} \quad \Gamma^b \label{KB}
\end{align}
\label{eq:laplaceproblem}
\end{subequations}
where $h(x):\Gamma^{\textrm{FS}}\mapsto \mathbb{R}$ describes the still water depth.
\begin{figure}[!htb]
\begin{center}
\includegraphics[height=3cm]{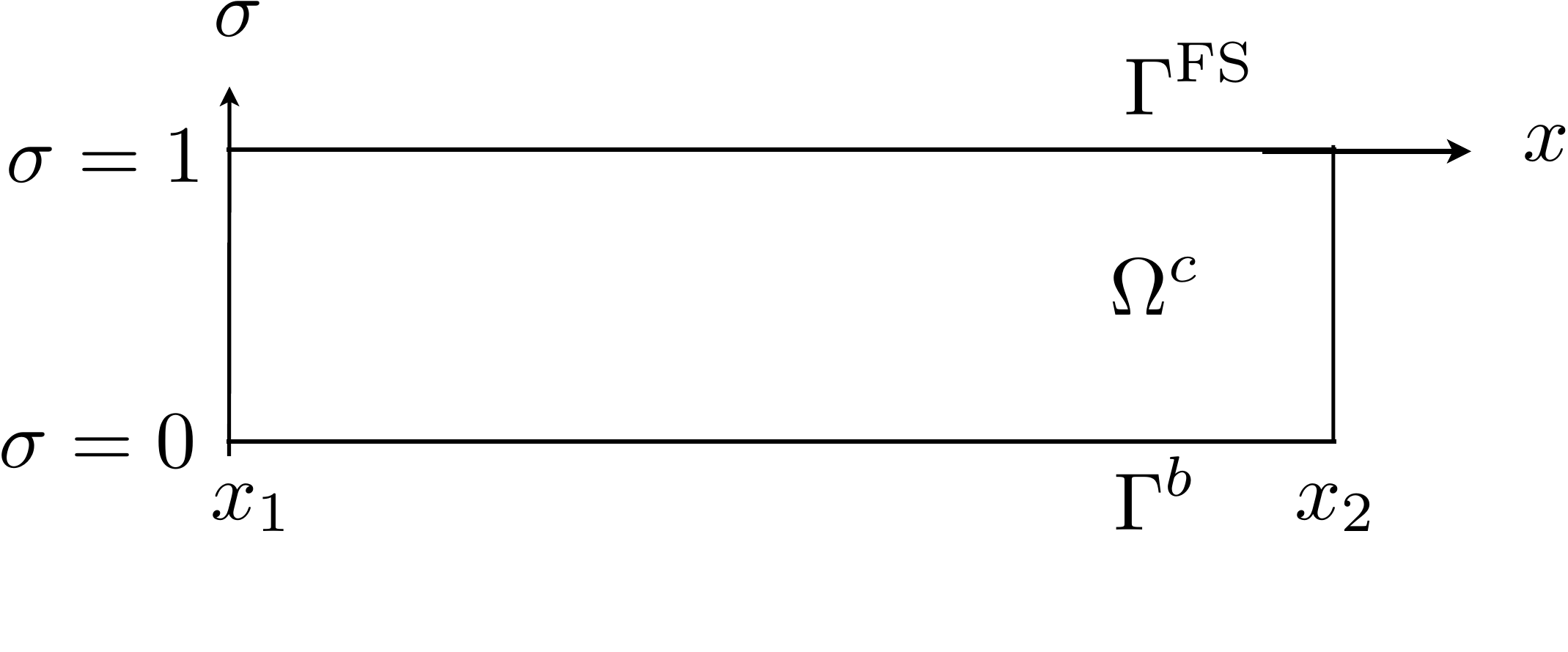}
\end{center}
\caption{Notations for computational domain ($\Omega^c$).}
\label{fig:domaintrans}
\end{figure}
We are interested in a set of governing equations that can be used as a basis for efficient simulations with support for representing structures accurately. A basis for efficient simulations is the classical $\sigma$-transformation of the vertical coordinate 
\begin{align}
\label{sigtrans}
\sigma \equiv (z+h(x))d(x,t)^{-1}, \quad 0\leq \sigma \leq 1,
\end{align}
where $d(x,t)=\eta(x,t) + h(x)$ is the height of the water column above the bottom. This transforms the fluid domain to a time-independent computational domain at the expense of time-varying coefficients. The notations are illustrated in Figure \ref{fig:domaintrans}.

The main drawback of using the $\sigma$-transformation is that it needs to be non-singular, and thus exclude the geometric modelling of breaking waves. For general wave-structure problems, this restriction can be removed by discretising and solving the Laplace problem \eqref{eq:laplaceproblem} directly. 
Following \cite{CaiEtAl1998}, we express the $\sigma$-transformed system in a form  where variable depth is accounted for. Let $\Omega^c\subset \mathbb{R}^d$ be the time-independent computational domain $\Omega^c=\{ (x,\sigma) | (x)\in\Gamma^{\textrm{FS}}, 0\leq \sigma \leq 1 \}$. The Jacobian of the map $\chi:\Omega\to\Omega^{c}$ is then 
\begin{subequations}
\begin{align}
\label{eq:jacobian}
\mathcal{J}(x,z,t) =  \left[  
\begin{array}{ccc} 
\partial_x x & \partial_z x \\
\partial_x \sigma & \partial_z \sigma
\end{array} 
\right]
=
 \left[  
\begin{array}{ccc} 
1 & 0 \\ 
\frac{\partial_x h}{d}-\frac{\sigma \partial_xd}{d}  & \frac{1}{d} 
\end{array} 
\right]
\end{align}
enabling the $\sigma$-transformed system to be expressed in the differential form
\begin{align}
\label{TransformedLaplace2}
\boldsymbol{\nabla}^c \cdot(K\boldsymbol{\nabla}^c\Phi) = 0 \quad \textrm{in} \quad \Omega^c
\end{align}
where $\boldsymbol{\nabla}^c=(\partial_x,\partial_\sigma)$ is introduced and the symmetric coefficient matrix is
\begin{align}
\label{eq:symTransformedLaplace2}
K(x,t) = \frac{1}{\textrm{det} \mathcal{J}}\mathcal{J} \mathcal{J}^T =  \left[  
\begin{array}{cc} 
d  & -\sigma \partial_x\eta\\ 
-\sigma \partial_x \eta  & \frac{1+(\sigma \partial_x\eta)^2}{d}
\end{array} 
\right].
\end{align}
\end{subequations}
The artificial scalar velocity function $\Phi(x,\sigma,t)=\phi(x,z,t)$ contains all information about the flow kinematics in the entire fluid volume. The velocity field can be determined from $\Phi$ using the relation ${\boldsymbol u} = (u,w) = (\partial_x+\partial_x\sigma\partial_\sigma, \partial_z \sigma \partial_\sigma) \Phi$.
All metric coefficients in \eqref{eq:jacobian} can be evaluated from the known two-dimensional free surface and bottom positions at given instants of time.
It is possible to discretise \eqref{TransformedLaplace2} after completing the differentiations, cf. \cite{EBL08,WesthuisAndonowati1998}, however, this increases the complexity of this formulation.

\subsection{Boundary conditions}

For the solution of the Laplace problem at every time step, the following free surface boundary condition is specified
\begin{align}
\label{eq:FSeqs}
\Phi=\tilde{\phi}, \quad \textrm{on} \quad \Gamma^{\textrm{\textrm{FS}}}
\end{align}
while at vertical boundaries, impermeable wall boundary conditions are assumed
\begin{align}
{\bf n}\cdot {\boldsymbol u} =0, \quad \textrm{on} \quad \Gamma\backslash (\Gamma^\textrm{\textrm{FS}}\cup\Gamma^b)
\end{align}
where ${\bf n}=(n_x,n_z)$ is an outward pointing unit normal vector at $\Gamma$. Due to symmetry at wall boundaries the domain boundary conditions at free surface variables imposed are
\begin{align}
\partial_n \eta = 0, \quad \partial_n \phi =0, \quad \textrm{on} \quad \Gamma \cap \Gamma^\textrm{\textrm{FS}}.
\end{align}

\subsection{Wave generation and absorption zones}

We employ a general-purpose embedded penalty forcing technique \cite{EGNL13} similar to the technique described in \cite{Cointe1989}. This technique can be used for both wave generation of regular and irregular waves as well as absorption in a numerical wave tank setup. It can be derived from the relaxation method described in \cite{LD83} to avoid the preprocessing step and turn it into an equivalent source term to the governing equations 
\begin{align}
\partial_t q = \mathcal{N}(q) + (1-\gamma(x))\tau^{-1}(q_a(x,t) - q(x,t)), \quad (x,t)\in \Gamma^{\textrm{FS}}\times [0,t]
\end{align}
where $\mathcal{N}(q)$ is the nonlinear function for the PDE, $q$ a free surface state variable ($\eta$,$\tilde{\phi}$ in Eq. \eqref{FSeqs}), the relaxation functions $\gamma(x)$ can be defined to avoid minimal reflections \cite{ENG06}, and the source function $q_a(x,t)$ defines the analytical representation of the wave input signal to be generated. We choose $\tau$ equal to time step size $\Delta t$.

\section{Numerical discretisation}

The governing partial differential equations are discretised in a generic computational framework based on the method of lines, where first a semi-discrete system of ordinary differential equations is formed by spatial discretisation using a nodal Spectral Element Method. We present the 2D formulation before presenting results of 2D numerical experiments.

\subsection{Weak Galerkin formulation and discretisation }

We form a partition of the domain $\Gamma_h^{\textrm{FS}}\subseteq\Gamma^{\textrm{FS}}$ to obtain a tessellation $\mathcal{T}_h^{\textrm{FS}}$ of $\Gamma^{\textrm{FS}}$ consisting of $N_{el}$ non-overlapping shape-regular elements $\mathcal{T}_k$ such that $\cup_{k=1}^{N_{el}}\mathcal{T}_k=\mathcal{T}_h$ with $k$ denoting the $k$'th element. We introduce the finite element approximation space of continuous, piece-wise polynomial functions of degree at most $P$, $V=\{ v_h\in C^0(\Gamma_h^{\textrm{FS}}); \forall k \in \{ 1,...,N_{el} \}, v_{h|\mathcal{T}_k}\in\mathbb{P}^P \}$. 

\subsubsection{Unsteady free surface equations}
The weak formulation of the free surface equations takes the following form. Find $f\in V$ where $f=\eta, \tilde{\phi}$ such that 
\begin{subequations}
\begin{align}
\int_{\mathcal{T}_h^{\textrm{FS}}}\partial_t \eta v(x) dx &= \int_{\mathcal{T}_h^{\textrm{FS}}}[ -\partial_x\eta\partial_x\tilde{\phi}+\tilde{w}(1+\partial_x\eta\partial_x\eta) ] v(x) dx, \\
\int_{\mathcal{T}_h^{\textrm{FS}}}\partial_t \tilde{\phi} v(x) dx &= \int_{\mathcal{T}_h^{\textrm{FS}}}\left[ -g\eta - \frac{1}{2}\left(\partial_x\tilde{\phi}\partial_x\tilde{\phi}-\tilde{w}^2(1+\partial_x\eta\partial_x\eta)\right) \right]v(x)dx,
\end{align}
\label{FSproblem}
\end{subequations}
for all $v\in V$.
We introduce the finite-dimensional approximations
\begin{align}
f_h = \sum_{i=1}^{N_{\textrm{FS}}} f_i(t) N_i(x),
\end{align}
where $\{N_i\}_{i=1}^{N_{\textrm{FS}}}\in V$ is the set of global finite element basis functions with cardinal property $N_i(x_j)=\delta_{ij}$ at mesh nodes with $\delta_{ij}$ the Kronecker Symbol. Substitute these expressions into \eqref{FSproblem} and choose $v(x)\in\{ N_i(x) \}_{i=1}^{N_{\textrm{FS}}}$. The discretisation in one spatial dimension becomes
\begin{subequations}
\label{eq:semidiscrete}
\begin{align}
M' \frac{d}{dt}\eta_h &= -A_x^{\partial_x\tilde{\phi}_h} \eta_h + M' \tilde{w}_h + A_x^{\tilde{w}_h\partial_x\eta_h}  \eta_h, \\
M' \frac{d}{dt}\tilde{\phi}_h &= -M' g \eta_h - \frac{1}{2} \left[ A_x^{\partial_x\tilde{\phi}_h} \tilde{\phi}_h +  M^{\tilde{w}_h} \tilde{w}_h -  A_x^{\tilde{w}_h^2\partial_x\eta_h} \right ] \eta_h,
\end{align}
\end{subequations}
where the following global matrices have been introduced
\begin{align}
M_{ij}' \equiv \int_{\mathcal{T}_h^{\textrm{FS}}}N_iN_j dx, \quad M^b_{ij} \equiv \int_{\mathcal{T}_h^{\textrm{FS}}}b(x) N_iN_j dx, \quad (A_x^{b})_{ij} \equiv \int_{\mathcal{T}_h^{\textrm{FS}}}b(x) N_i \frac{d}{d x}N_j dx.
\end{align}
The gradients of the free surface state variables are recovered as described in Section \ref{gradrecover}. Temporal integration of \eqref{eq:semidiscrete} is performed using an explicit fourth-order Runge-Kutta method. 

{\bf Remark}: The free surface equations \eqref{FSproblem} contain strongly nonlinear terms, up to fourth order. The discretisation of these terms calls for proper treatment to avoid aliasing effects. This will be addressed in Sections \ref{sec:remquaderrors} and \ref{sec:artificial-spectral-filtering}.

\subsubsection{Spatial discretisation of the Laplace problem}

Consider the discretisation of the governing equations for the $\sigma$-transformed Laplace problem \eqref{TransformedLaplace2}. We seek to construct a linear system of the form
\begin{align}
\label{eq:laplaceproblemdiscrete}
\mathcal{L} \Phi_h = {\bf b}, \quad \mathcal{L}\in\mathbb{R}^{n\times n}, \quad \Phi_h, {\bf b}\in\mathbb{R}^{n}
\end{align}
where $n$ is the total degrees of freedom in the discretisation.

The starting point is the weak formulation of the symmetric formulation \eqref{TransformedLaplace2} that can be expressed as: find $\Phi\in V$ such that 
\begin{align}
\int_{\mathcal{T}_h} \nabla^c\cdot(K \nabla^c\Phi)  v d\boldsymbol{x} = \oint_{\partial\mathcal{T}_h} v {\bf n}\cdot (K\nabla^c\Phi) d\boldsymbol{x}  - \int_{\mathcal{T}_h}  (K \nabla^c\Phi)  \cdot \nabla^c v d\boldsymbol{x} = 0,
\end{align}
for all $v\in V$ where the boundary integrals vanish at domain boundaries where impermeable walls are assumed. 
The discrete system operator is defined via domain decomposition as
\begin{align}
\mathcal{L}_{ij} \equiv -\int_{\mathcal{T}_h} (K \nabla^c N_j)  \cdot \nabla^c N_i d\boldsymbol{x}  = -\sum_{k=1}^{N_{el}}   \int_{\mathcal{T}_h^k} (K \nabla^c N_j)  \cdot \nabla^c N_i d\boldsymbol{x} .
\end{align}
The elemental integrals are approximated through the change of variable
\begin{align}
&\int_{\mathcal{T}_h^k} (K \nabla^c N_j)  \cdot \nabla^c N_i d\boldsymbol{x}  = \int_{\mathcal{T}_r} |\mathcal{J}^k| (K \nabla^c N_j ) \cdot \nabla^c N_i d\boldsymbol{r} 
\end{align}
where $\mathcal{J}^k$ is the Jacobian of the affine mapping $\chi^k:\mathcal{T}_h^k\to\mathcal{T}_{r}$ and $\mathcal{T}_r$ is the computational reference element. The global assembly of this operator preserves symmetry, and the resulting linear system is then modified to impose the Dirichlet boundary conditions  \eqref{eq:FSeqs} at the free surface.
Typically, $\mathcal{L}$ is a large and sparse operator with a narrow band structure of non-zero elements in two space dimensions obtained by a proper permutation of the rows and numbering of nodes to reduce fill-in in the factorisation of Gaussian Elimination procedures.
By using a symmetric reverse Cuthill-Mackee permutation, the bandwidth of the sparse matrix is minimised and the system can be efficiently solved by a sparse direct Gaussian elimination procedure. In 2D this leads to optimal and scalable $\mathcal{O}(n)$ work effort and is used in this work. For sparse symmetric positive definite systems such as Eq. \eqref{eq:laplaceproblemdiscrete}, the iterative preconditioned conjugate gradient solver is an attractive choice when system sizes become large \cite{CaiEtAl1998} as would be expected for typical applications in three space dimensions, since convergence is guaranteed and memory footprint is minimal. 

\subsubsection{A generic technique for gradient recovery}
\label{gradrecover}

The gradients of the globally piece-wise continuous basis functions will be discontinuous across element interfaces in the classical sense. To guarantee global continuity of derivatives a gradient recovery procedure can be used. 
Several gradient recovery techniques are reviewed in \cite{HintonCampbell1974,HawkenTownsendWebster1991,SriramEtAl2010}. In this work, a global gradient recovery technique is used within our spectral element framework. 

The global approximation of components of the horizontal first derivative as a $C^0$ function is expressed as
\begin{align}
\label{eq:representationU}
u_h=\partial_x \phi_h = \sum_{i=1}^{n} u_i N_i(x).
\end{align}
By a global Galerkin projection 
\begin{align}
\int_{\mathcal{T}_h} u_h v(x) dx &= \int_{\mathcal{T}_h} \partial_x \phi_h v(x) dx
= \sum_{k=1}^{N_k}  \int_{\mathcal{T}_h^k}  \left( \sum_{j=1}^n \phi_j \frac{d}{dx} N_j \right)N_i dx,
\end{align}
we can generate the linear systems of equations 
\begin{align}
\mathcal{M} \boldsymbol{u}_h = \mathcal{S}_x {\boldsymbol \phi}_h, \quad \mathcal{S}_x \equiv \int_{\mathcal{T}_h} \partial_x N_j N_i dx,
\end{align}
to recover the coefficients ${\bf u}_h$ of the expansion   \eqref{eq:representationU}. This procedure is similar to the one described by \cite{HawkenTownsendWebster1991} and used in other models, cf.  \cite{WuTaylor1994,RobertsonSherwin1999,TurnbullEtAl2003}.

{\bf Remark}: In the FEM model described in \cite{Westhuis2001} the gradient recovery step is related to stability of the numerics. It was found that a global projection method may lead to instability when using FEM. As we shall see in the numerical experiments in Section \ref{sec:numexp} we do not reach the same empirical conclusion.

Using this gradient recovering technique and after the solution of  \eqref{eq:laplaceproblem}, it is possible to estimate the vertical free surface velocity $\tilde{w}=\partial_z\sigma \partial_\sigma \phi|_{z=\eta}$ to obtain closure in the free surface problem \eqref{FSphi}.
The weak formulation is
\begin{align}
\int_{\mathcal{T}_h} w v(x)d x &= \int_{\mathcal{T}_h} \partial_z \phi v(x)d x = \int_{\mathcal{T}_h} \partial_z\sigma \partial_\sigma \Phi v(x)d x.
\label{eq:velocpot}
\end{align}
From this we can construct a linear system of the form
\begin{align}
\mathcal{M} {\bf w}_h = \mathcal{S}_z{\boldsymbol \Phi}_h, \quad \mathcal{M}, \; \mathcal{S}_z\in\mathbb{R}^{n\times n}, \quad {\bf w}_h, \; {\boldsymbol \Phi}_h \in \mathbb{R}^{n}
\label{eq:velocpot2}
\end{align}
where the discrete operators takes the form
\begin{align}
\mathcal{M}_{ij} \equiv \int_{\mathcal{T}_h} N_i N_j d\boldsymbol{x} , \quad \mathcal{S}_z\equiv \int_{\mathcal{T}_h} \partial_z\sigma \partial_\sigma N_iN_j d\boldsymbol{x} 
\end{align}
From the solution of  \eqref{eq:velocpot2} we obtain the vertical free surface velocities
\begin{align}
\tilde{w}_h=({\bf w}_h)_i, \quad i\in S^{\textrm{FS}}
\end{align}
where $S^{\textrm{FS}}$ denotes an index set consisting of global numbers for the free surface nodes.

\subsubsection{Nodal prismatic Lagrange finite elements in two space dimensions}

We consider first construction of elements in two space dimensions, needed for representing the solutions of the governing equations. The expansions can be based on quadrilateral elements in the time-constant computational domain. These elements are formed by a subdivision of the horizontal free surface plane with possible irregular sized non-overlapping elements. Each of these elements can then be extended in the vertical from the surface to the bottom to form the quadrilateral elements. A spectrally accurate multivariate hierarchical polynomial expansion can be constructed by a tensor product of one-dimensional basis functions.
%
This results in a quadrilateral element $\mathcal{T}_{p}=\{ (r,t)\in\mathbb{R}^2:-1<r,t<1  \}$. 
We introduce the element basis functions  
\begin{align}
\label{eq_basis}
\tilde{\varphi}_{\bf k} ({\bf r}) = \tilde{P}_{k_1}^{(0,0)}(r)\tilde{P}_{k_2}^{(0,0)}(t),
\end{align}
where $\tilde{P}_n^{(\alpha,\beta)}(\xi)$ is the $n$'th order orthonormal Jacobi polynomial on the interval $\xi\in[-1,1]$ with orthogonality property
\begin{align}
\int_{-1}^1\tilde{P}_m^{(\alpha,\beta)}(\xi)\tilde{P}_n^{(\alpha,\beta)}(\xi)(1-\xi)^\alpha(1+\xi)^\beta d\xi = \delta_{mn}, \; \tilde{P}_n^{(\alpha,\beta)} = \frac{P_n^{(\alpha,\beta)} }{ ||P_n^{(\alpha,\beta)}||_{L_w^2([-1,1])}}.
\end{align}
These polynomials can be evaluated efficiently using a simple recurrence relation \cite{HGG07,Kopriva2009}. The basis  \eqref{eq_basis} have arbitrary polynomial orders $k_1$ in the horizontal plane and $k_2$ in the vertical plane. This makes it possible to tune the orders of the approximations to balance accuracy and efficiency needs in simulations.

For interpolating polynomials, the Unisolvence Theorem guarantees a unique connection between the hierarchical polynomial (modal) expansion and the corresponding Lagrange polynomial (nodal) expansion. 
Thus, for all $x_i^k$,  $i=1,...,{N_p}$, and $\forall k$,  and for each element $k=1,...,N_k$, we have 
\begin{align}
\label{eq:duality}
f_h^k(x_i,t) = \sum_{n=1}^{N_p}\hat{f}_n(t)\tilde{\varphi}_n(\chi^k(x_i))=\sum_{n=1}^{N_p}f_n(t)l_n(\chi^k(x_i)), 
\end{align}
which defines a relationship between modal and nodal coefficients in the form \cite{HW08}
\begin{align}
{\bf f}_h = \mathcal{V} \hat{\bf f}, \quad \mathcal{V}_{ij} =\tilde{\varphi}_j({\bf r}_i), \quad {\bf f}_h=(f_1,...,f_{N_p})^T, \quad \hat{\bf f}=(\hat{f}_1,...,\hat{f}_{N_p})^T,
\end{align}
where a non-singular generalised Vandermonde matrix $\mathcal{V}$ has been introduced.
This connection can be exploited together with the duality in polynomial representation to define the local Lagrange basis functions with property $l_i(x_j)=\delta_{ij}$ and their derivatives
\begin{align}
\label{eq:derivative}
l_i({\bf r})= \sum_{j=1}^{N_p} (\mathcal{V}^T)_{ij}^{-1} \tilde{\varphi}_j({\bf r}) , \;\;
\partial_rl_i({\bf r})= \sum_{j=1}^{N_p} (\mathcal{V}^T)_{ij}^{-1} \partial_r \tilde{\varphi}_j({\bf r}) = \sum_{j=1}^{N_p} (\mathcal{V}^T)_{ij}^{-1} (\mathcal{V}_r)_{ji}  l_j({\bf r}),
\end{align}
and the matrices defined in terms of the first derivatives of the modal basis functions
\begin{align}
(\mathcal{V}_r)_{ij} = \partial_r\tilde{\varphi}_j({\bf r}_i), \quad (\mathcal{V}_t)_{ij} =\partial_t\tilde{\varphi}_j({\bf r}_i).
\end{align}
For use with integration on the local elements, the local nodal mass matrix is introduced
\begin{align}
\label{eq:massmatrix}
\mathcal{M}_{ij} = \int_{\mathcal{T}_r} l_i({\bf r})l_j({\bf r}) d{\bf r} = (\mathcal{V}^T)^{-1} \left[ \int_{\mathcal{T}_r} \tilde{\varphi}_i({\bf r}) \tilde{\varphi}_j({\bf r}) d{\bf r} \right] \mathcal{V}^{-1}  = (\mathcal{VV}^T)^{-1}_{ij} \;,
\end{align}
where orthonormality of the basis functions is exploited to avoid the use of discrete quadrature rules in the constructions, leaving the implementations {\em quadrature-free}.

By evaluating expressions such as  \eqref{eq:derivative} at the chosen inter element node distribution we obtain the generic arbitrary-order elemental operators
\begin{align}
\mathcal{D}_r = \mathcal{V}_r \mathcal{V}^{-1}, \quad \mathcal{D}_t = \mathcal{V}_t \mathcal{V}^{-1}, \quad \mathcal{D}_\sigma = 2\mathcal{D}_t,
\end{align}
that can be used in connection with the chain rule to define derivatives in physical space
\begin{align}
\frac{\partial}{\partial x} = \frac{\partial r}{\partial x} \frac{\partial }{\partial r}, \quad \frac{\partial}{\partial z} = \frac{\partial \sigma}{\partial z} \frac{\partial }{\partial \sigma},
\end{align}
with discrete counterparts based on collocation expressed in the form
\begin{align}
\mathcal{D}_x = \textrm{diag}[{\bf r}_{\bf x}] \mathcal{D}_r, \quad \mathcal{D}_z = \textrm{diag}[{\bf \sigma}_{\bf z}] \mathcal{D}_\sigma.
\end{align}
The choice of the nodal distribution on the simplexes in two horizontal dimensions can be based on explicit construction of the nodal distribution set $\{ {\bf r}_i \}_{i=1}^{N_p^{2D}}$ using, e.g., a blend and warp procedure \cite{War06}. This is used to define the high-order nodal basis functions \cite{WarburtonSherwinKarniadakis1999}. 


\subsection{Generalised local element matrices via quadrature-free matrix-based operations}

All global operators can be assembled from generalised local elemental operators. Consider global integrals in the general form
\begin{align}
\int_{\mathcal{T}_h} f g d x.
\end{align}
Approximate each of the integrands using finite element basis functions of the form
\begin{align}
f_h = \sum_i f_i l_i, \quad g_h = \sum_j f_j l_j.
\end{align}
The global integrals can be reduced to local integrals through domain decomposition
\begin{align}
\int_{\mathcal{T}_h} f g d x = \sum_k \int_{\mathcal{T}_h^k} f g d x,
\end{align}
and by inserting the approximated integrands, we obtain local integrals of the form
\begin{align}
\int_{\mathcal{T}_h^k} f_h g_h d x = \int_{\mathcal{T}_r} |\mathcal{J}^k| f_h^k g_h^k d {\bf r}.
\end{align}
Expand the approximate integrands using nodal expansions such that
\begin{align}
\label{eq:localintegrals}
\int_{\mathcal{T}_r} |\mathcal{J}^k| f_h g_h d {\bf r} &= 
 \sum_{i} \sum_j  |\mathcal{J}^k| (f_h^k)_i (g_h^k)_j \int_{\mathcal{T}_r}  l_i l_j d {\bf r} 
= \sum_i \sum_j |\mathcal{J}^k| (f_h^k)_i (g_h^k)_j \mathcal{M}_{ij}.
\end{align}
In the special cases, where $f\in C^q(\mathcal{T}_r)$ and $g\in C^p(\mathcal{T}_r)$ are differential operators, e.g.
\begin{align}
f = \frac{\partial^q u}{\partial r^q}, \quad g = \frac{\partial^p v}{\partial s^p}, \quad p,q\in \mathbb{N}_0
\end{align}
the local integrals in \eqref{eq:localintegrals} can be evaluated by exploiting that for nodal differentiation matrices $\mathcal{D}^{(q)}=(\mathcal{D})^q$, cf. \cite{CHQZ2006}. 
An implementation of the local elemental operator following the derivation just outlined, enables generic implementation in a single framework that has support for $hp$-adaptivity to balance accuracy and cost in computations. Furthermore, the integration of the integrands approximated using polynomial basis functions are without quadrature errors when quadrature-free matrix-based operations are carried out using evaluations on sufficiently fine meshes as described next. 

\subsection{Removal of quadrature errors via higher-order numerical integration}
\label{sec:remquaderrors}

Numerical integration of nonlinear terms is handled via super-collocation \cite{KK03} and is expressed in the general form
\begin{align}
\int_{\mathcal{T}_h} f_H g_H d x \simeq \int_{\mathcal{T}_h} (\mathcal{I}_{H}^hf_H)(\mathcal{I}_{H}^h g_H) d x,
\end{align}
where we have introduced the interpolation operator $\mathcal{I}_H^h$ that maps the representation of a function on coarse mesh (subscript $H$) to that of a fine mesh (subscript $h$).
Exploiting the result in \eqref{eq:localintegrals}, the integration (projection) of the polynomial representation can be done without quadrature errors via mass matrix based operations for integration of the form 
\begin{align}
\int_{\mathcal{T}_h} f_Hg_H d x \simeq
f_H^T \mathcal{M}_H g_H \simeq
\int_{\mathcal{T}_h} (\mathcal{I}_{H}^hf_H)(\mathcal{I}_{H}^h g_H) d x \simeq 
(\mathcal{I}_{H}^hf_H)^T \mathcal{M}_h (\mathcal{I}_{H}^h g_H),
\end{align}
where $M_H\in\mathbb{R}^{m_H\times m_H}$ and $M_h\in\mathbb{R}^{m_h\times m_h}$. The computational cost of this integration step has complexities $\mathcal{O}(m_H^2)$ and $\mathcal{O}(m_h^2)$ due to the involved matrix operation. The relation between size of the coarse and fine variables is $m_h=2m_H$ for exact integration of the polynomial representation of the coarse space basis functions when quartic terms are present. Therefore, the increase in cost is about four times for these operations. 

\subsection{Dealing with aliasing by spectral filtering}\label{sec:artificial-spectral-filtering}

The solution of governing equations that contain strongly nonlinear terms may pose a challenge for maintaining both accuracy and numerical stability in time when simulations are marginally resolved. In such cases, numerical instability may be related to aliasing effects. Aliasing results from evaluating interpolated products of functions, which when represented with insufficient or marginal resolution introduces errors in the functional representation.  Possible remedies \cite{KK03,Mengaldo201556} focus on increasing resolution or introducing a proper stabilisation strategy to artificially dissipate such errors. While the former is simple, it is generally not considered feasible for large-scale systems or for long-time integration. 

In this work, we employ a spectral filtering strategy exploiting the duality in the local element representation, cf.  \eqref{eq:duality}. 
On the $k$'th element the filtered local solution can be expressed as (assuming an order $P$ expansion in one space dimension here)
\begin{align}
\bar{\eta}_h^k(t,x) = \sum_{i=0}^{P} \sigma(i) \eta_i(t) \varphi_i(\chi^k(x)).
\end{align}
The filtering is applied only for the time-dependent free surface variables $\eta$ and $\tilde{\phi}$. An exponential filter  \cite{HK08,HW08} can be used with cut-off low-pass filter index $i_c$, to only affect the highest modes $i>i_c$, such that
\begin{align}
\sigma(i) = \exp (-\alpha((i - i_c)/(1-i_c))^s), \quad i_c\leq i \leq P
\end{align}
For example, by choosing the parameters $(\alpha,s)=(0.0513,0)$ a mild damping is achieved which gently removes five percent of the energy from only the highest mode in the basis. 
The filtering is done on a per element basis using a matrix-vector product
\begin{align}
\hat{\tilde{\eta}}_h^k =\mathcal{F} \tilde{\eta}_h^k, \quad \mathcal{F}=\mathcal{V}\mathcal{S}\mathcal{V}^{-1}\in\mathbb{R}^{N_p \times N_p}, \quad \mathcal{S}_{ij}= \delta_{ij} \sigma_i, \quad \mathcal{V}_{ij} = \hat{\varphi}_j({\boldsymbol r_i}),
\end{align}
which has a work complexity of $\mathcal{O}(N_p^2)$ with total work effort proportional to $\mathcal{O}(N_kN_p^2)$. This local filtering matrix $\mathcal{F}$ is constructed and used on all elements and applied repeatedly as necessary. We use the  model basis in one space dimension \cite{KS99}
\begin{align}
\hat{\varphi}_0(r) &= \tfrac{1}{2}(1-r), \;\;  \hat{\varphi}_1(r) = \tfrac{1}{2}(1+r), \;\; \hat{\varphi}_{n+1}(r) = \tfrac{1}{4}(1-r)(1+r)P_{n-1}^{(1,1)}, \;\; n=1,..,P-1
\end{align}
to avoid introducing interface jumps or affect the mean through only filtering the higher modes ($n>1$).  If the filter is applied gently, i.e. by removing very little 'energy' from the highest modes only, spectral accuracy can be recovered. Excessive filtering may reduce the convergence rate of the method albeit improve stability. 

\section{Numerical properties of the model}

\subsection{Temporal stability}

For general schemes, the connection between nonlinear instability and explicit time-integration is important to understand \cite{DiasBridges2006}. To solve the governing equations efficiently, an explicit time integration method is preferred. 
Explicit schemes come with conditional stability in the form of the global CFL condition 
$\Delta t \leq \frac{C_1}{\max_i |\lambda_i|}$ 
that dictates an upper bound for the choice of stable time step sizes $\Delta t$ where the constant $C_1=\mathcal{O}(1)$. 
It is well-known that a main challenge for many high-order Spectral Element Methods is an unattractive scaling of the modulus of the eigenvalues of the discrete operators of the form 
$\max_i |\lambda_i| \sim C_2 P^{2\gamma}$
where $P$ is expansion order and $\gamma$ is the highest order of differentiation operator in the evolution equations. Typically, the constant $C_2$ is dependent on the minimum mesh size for an element in the mesh. This property may pose a severe problem in the accurate local representation of geometric features with small elements. 

\begin{figure}[!htb]
\begin{minipage}{0.5\textwidth}
\begin{center}
(a) Maximum eigenvalues \\
\includegraphics[height=5.0cm]{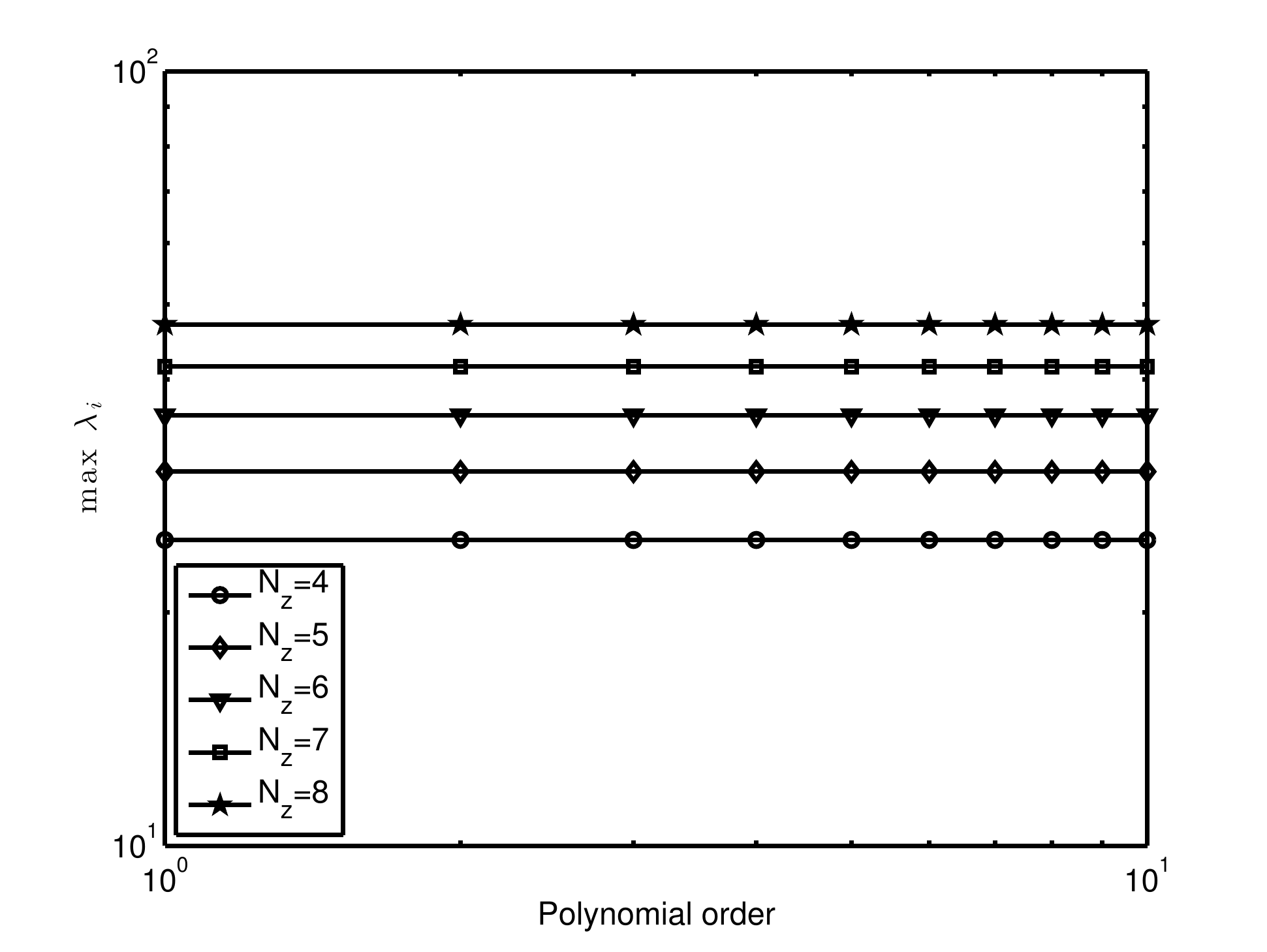}
\end{center}
\end{minipage}
\hspace{0.2cm}
\begin{minipage}{0.5\textwidth}
\begin{center}
(b) Computed eigenspectrum  \\
\includegraphics[height=5.0cm]{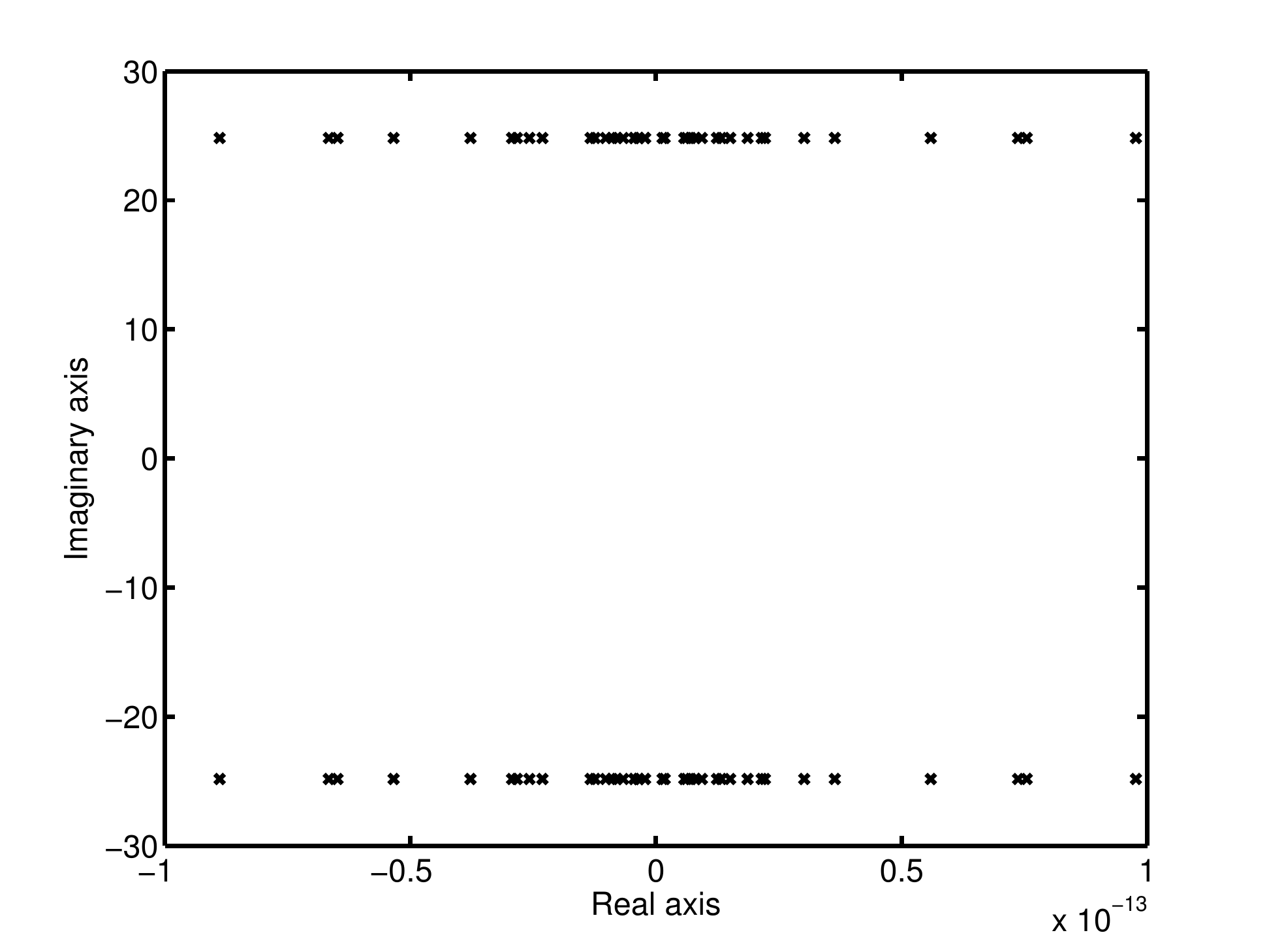}
\end{center}
\end{minipage}
\caption{Computed eigenvalues of the Jacobian matrix of the semi-discretised linearised equations on a constant bottom when system is discretised using SEM. (a) Lack of growth of maximum eigenvalues with expansion order illustrated for varying number of nodes $N_z$ in the vertical direction. (b) Example of computed eigenspectrum of purely imaginary eigenvalues to within machine precision for $N_z=4$ and $N_x=4$. Five elements in horizontal direction are used in all results.}
\label{fig:bounded}
\end{figure}

As discussed and shown in \cite{EGNL13} for a linear finite difference scheme of the same governing equations, the stable time step size for explicit schemes has an upper bound given by the CFL condition  
$\Delta t \leq C(N_z) \sqrt{\frac{g}{h}}$, 
where $h$ is here still water depth and $g$ the gravitational acceleration constant. 
This bound is only dependent on the scale of the physics (still water depth) and the resolution chosen in the vertical. 
This result is demonstrated and shown in Figure \ref{fig:bounded}, following \cite{EBL08} via discretisation and numerical eigenvalue analysis of the linearised system.
This property is similar to the property inherent in several other wave models as described in the works \cite{RobertsonSherwin1999,ENG06,EHBM06,EHBW08,EGNL13,EskilssonEngsigKarup2014}. 
This property  implies that the CFL condition is tamed \cite{WarburtonHagstrom2008} in the sense that the time step size is not dictated by the numerics but only the physics (depth). 
Along the same line of the experiments described in \cite{EGNL13}, we find that there are only small changes in the CFL properties for nonlinear problems and hence small elements in the mesh do not impose any severe time step size restriction. 

\subsection{Linear accuracy and dispersion properties}

The accuracy of the numerical model is compared to the theoretical solution to the system of equations that arise when the system is subject to the assumption of small-amplitude waves (assume $H/L<1/20$ where wave height $H$ relative to wave length $L$). 
The theoretical solution for linear progressive monochromatic waves in one space dimension is given by \cite{SJ01}
\begin{align}
\eta(x,t) &= \frac{H}{2}\cos(\omega t - kx), \quad \phi(x,z,t) = -\frac{Hc_s}{2} \frac{\cosh(k(z+h))}{\sinh(kh) } \sin(\omega t-kx),
\end{align}
with linear dispersion relation from Stokes theory $c_s=\omega/k=\sqrt{\frac{g}{k}\tanh(kh)}$, where $k=2\pi/L$ is the wave number, $g$ is the gravitational acceleration (assumed as 9.81 $m/s^2$) and $\omega=2\pi/T$ is the angular velocity with $T$ being the wave period.

\begin{figure}[!htb]
\centering
\begin{minipage}{5.6cm}
\begin{center}
(a) $h$- and $p$-refinement \\
\includegraphics[height=4cm]{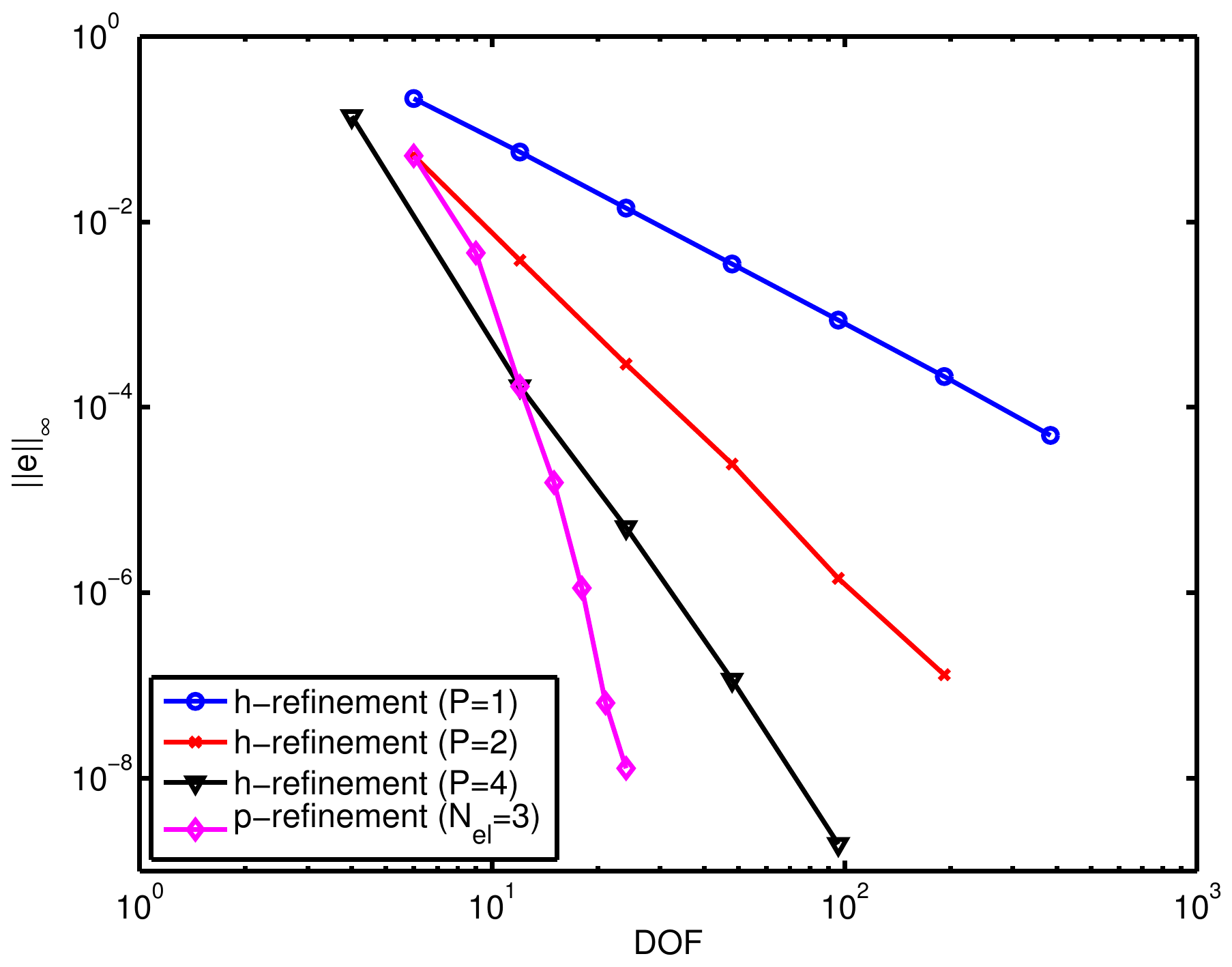}
\end{center}
\end{minipage}
\begin{minipage}{5.6cm}
\begin{center}
(b) Optimised accuracy in time \\
\includegraphics[height=4cm]{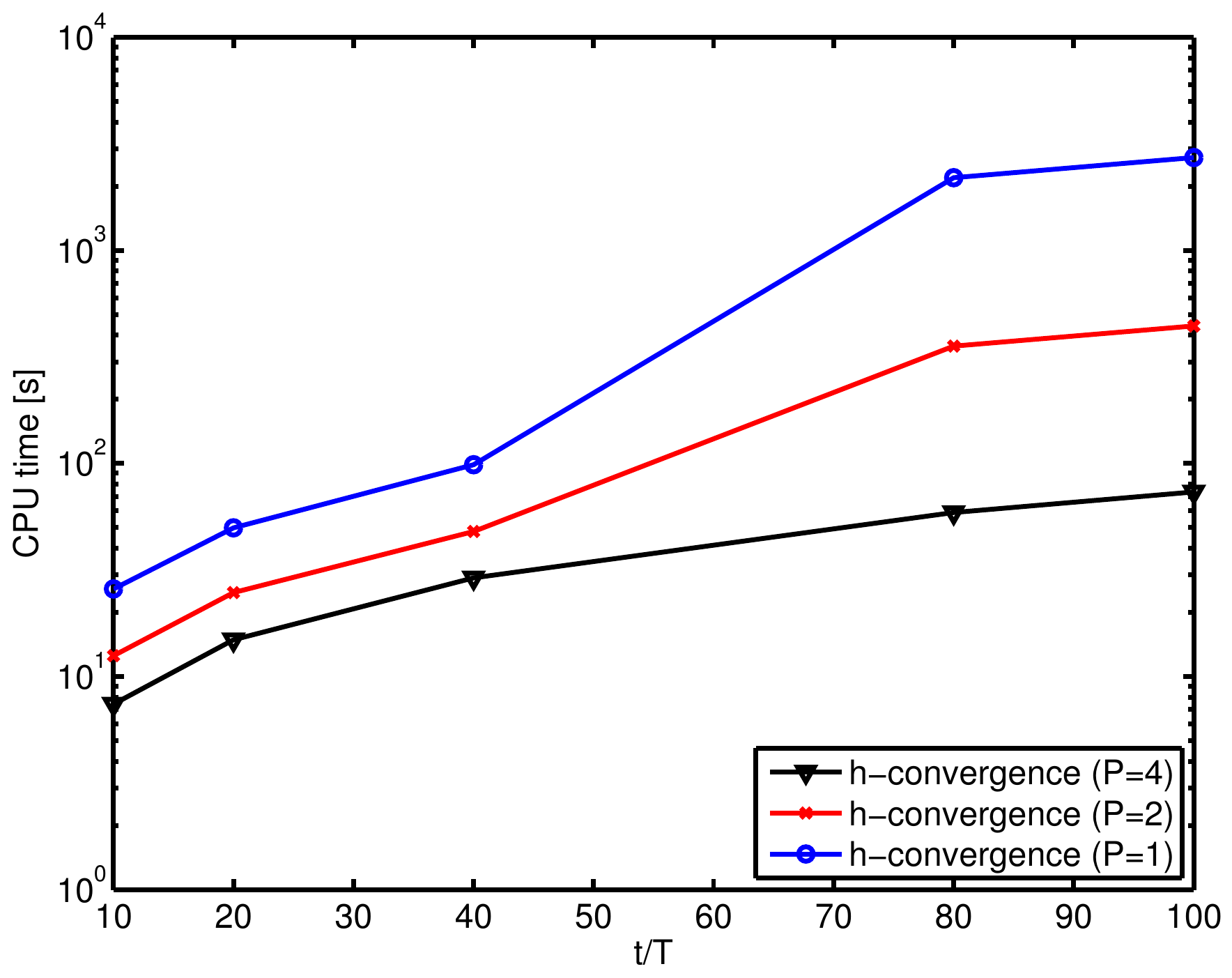}
\end{center}
\end{minipage}
\begin{minipage}{5.6cm}
\begin{center}
(c) Optimised spatial resolution \\
\includegraphics[height=4cm]{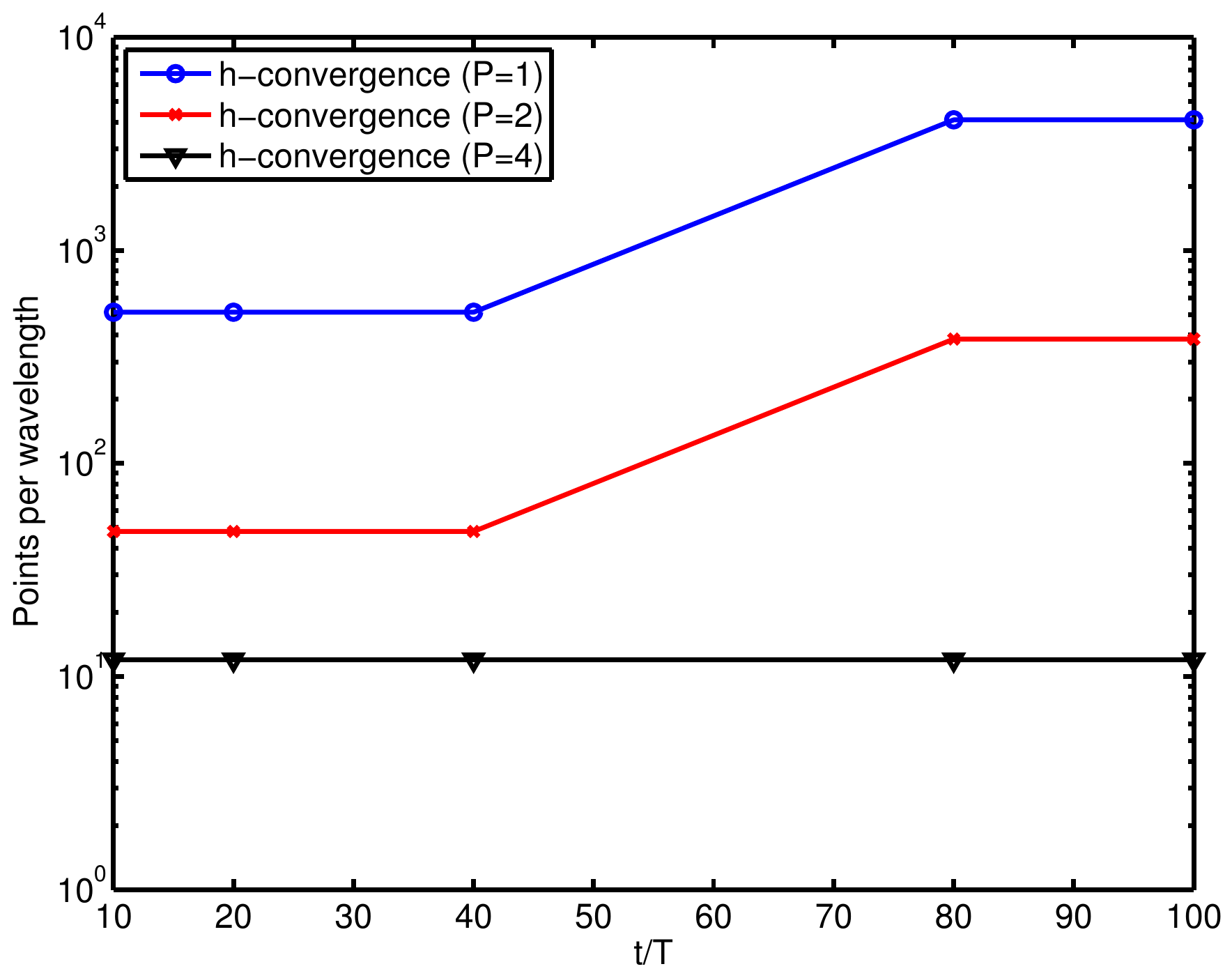}
\end{center}
\end{minipage}
\begin{minipage}{5.6cm}
\begin{center}
(d) Speedup \\
\includegraphics[height=4.25cm]{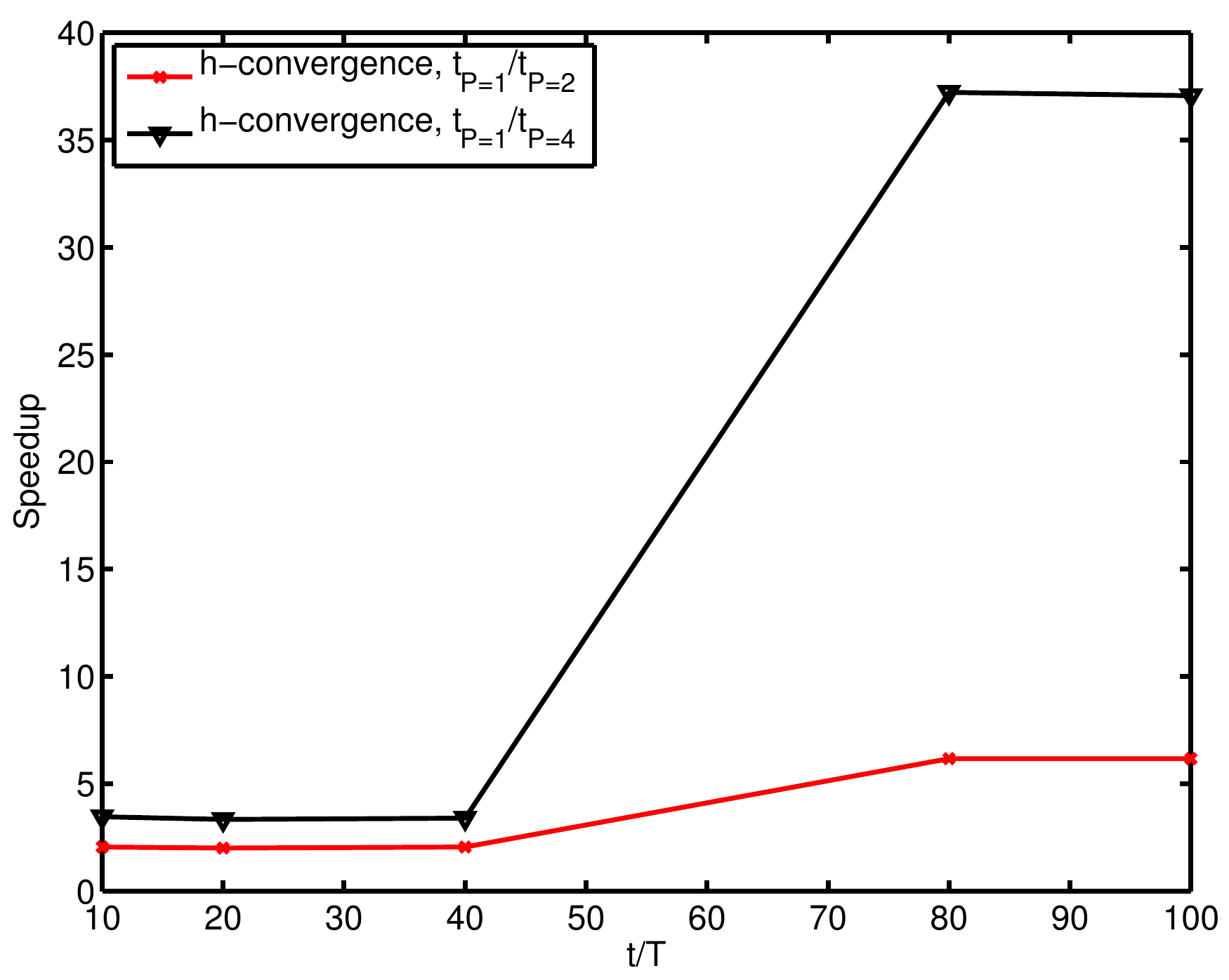}
\end{center}
\end{minipage}
\caption{(a) Convergence tests measuring absolute errors in amplitude. (b) Optimised accuracy in time for a fixed relative error in amplitude of $1\%$ (engineering accuracy). (c) Number of points per wave length for optimised accuracy in time. A uniform mesh is used. (d) Optimised speedup based on results of (b) and relative to second order results ($P=1$ curve). Results are for $kh=1$ with small-amplitude waves of one wave length.}
\label{fig:convresults}
\end{figure}

In Figure \ref{fig:convresults} (a) results of $h$- and $p$-refinement strategies are presented and highlight an important advantage of the SEM. When solutions are smooth, it is possible to use high-order basis functions to improve the cost-efficiency of the method by allowing fewer degrees of freedom to be used to achieve the same level of accuracy of lower order methods. 
Most previous works have focused exclusively on classical FEM methods based on piece-wise linear approximations, which has a convergence rate that matches the curve for linear ($P=1$) basis functions.

The decisive criterion for choosing between different numerical strategies is to understand what is the amount of work (cost) to achieve a given level of accuracy. 
This is illustrated via results obtained via a semi-optimal brute-force procedure in Figures \ref{fig:convresults} (b) and (c) where the work effort has been minimised\footnote{Tests done using a sequential proof-of-concept code on a laptop with a 2,3 GHz Intel Core I7 processor and 16 GB 1600 MHz DDR3 RAM.} with respect to time to be as numerically efficient for a fixed accuracy level of 1$\%$ error in surface elevation. With no significant dissipation in the scheme, this implies that the error measure mainly numerical dispersion error. 
The results in Figure \ref{fig:convresults} (d) show that for our current sequential proof-of-concept implementation there is speedup of approximately x2 for short time and close to x6 for longer times, achievable by switching from a ($P=1$)-scheme to a $(P=2)$-scheme. These gains can be improved further to close to x3 for short time and close to x37 for longer times, achievable by switching from a ($P=1$)-scheme to a ($P=4$)-scheme. 
As expected (cf. \cite{KO72}), the longer a simulation the more the possible gain in terms of numerical efficiency, and for even higher orders, there are additional gains although they end up being marginal for the modest accuracy requirement chosen. 

\begin{figure}[!htb]
\centering
\begin{minipage}{5.6cm}
\begin{center}
(a) High-order elements ($x$ and $\sigma$) \\
\includegraphics[height=4.6cm]{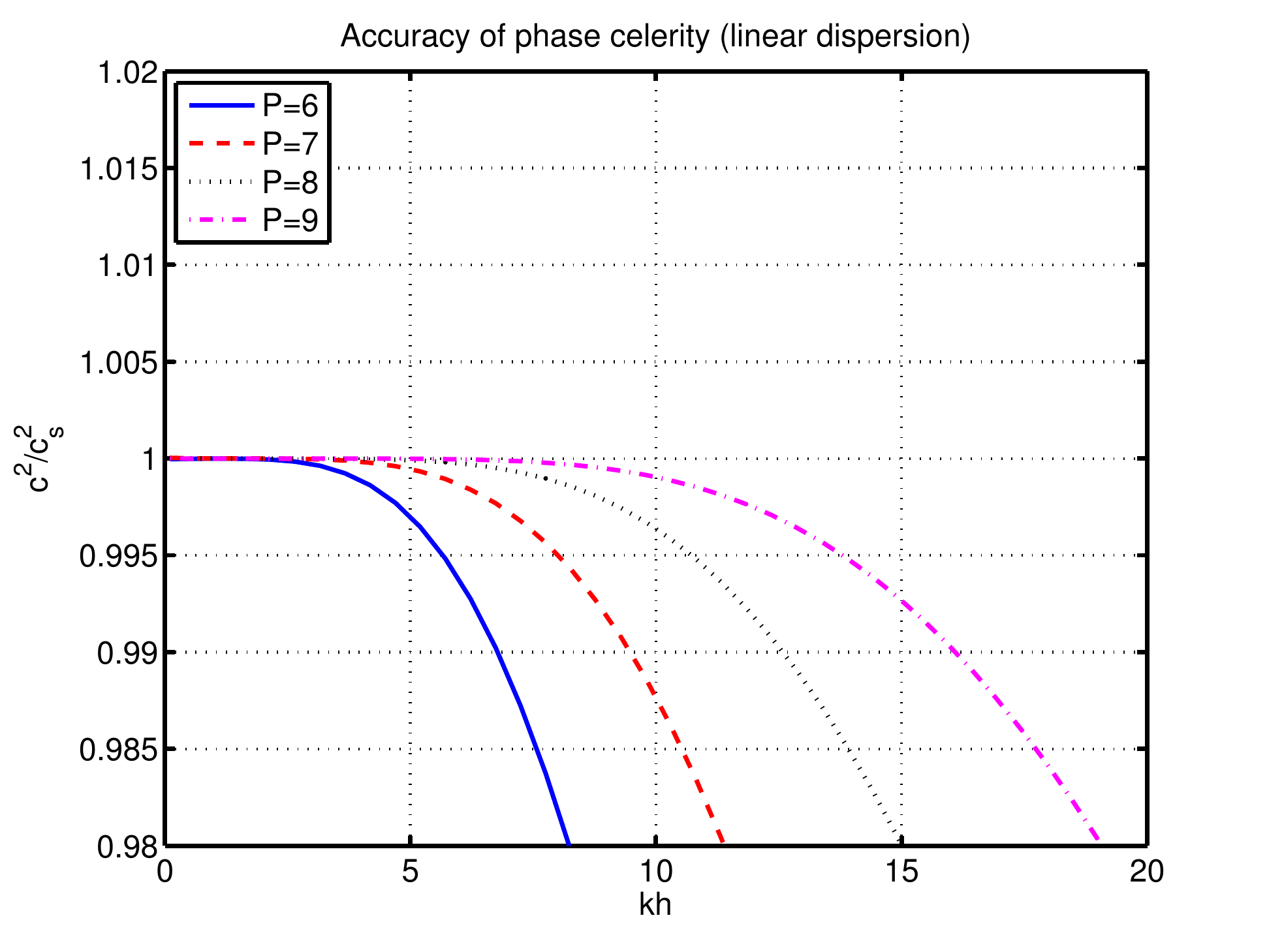}
\end{center}
\end{minipage}
\begin{minipage}{5.6cm}
\begin{center}
(b) High-order ($x$), low-order $(\sigma)$\\
\includegraphics[height=4.6cm]{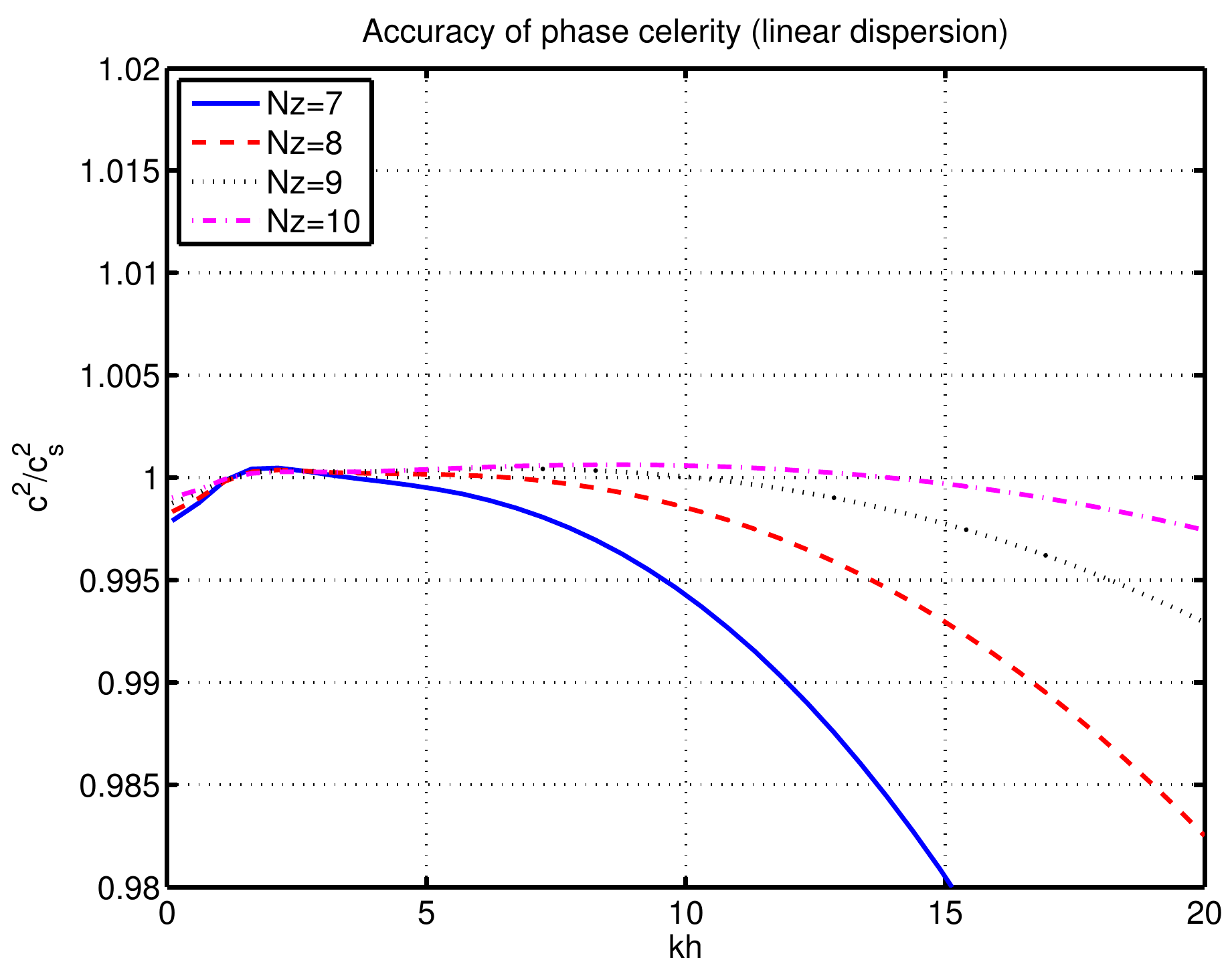}
\end{center}
\end{minipage}
\caption{Linear dispersion curves to within 2\% accuracy for (a) vertical polynomial orders $P$ where the number of vertical points are $N_z=P+1$ and (b) cosine-clustered element sizes with elements in the vertical using a local vertical polynomial basis of order $P=1$ and horizontal resolution high enough to have no impact on the dispersion curves.
The application range is given in terms of the dimensionless dispersion parameter $kh$ and increases with spatial resolution in the vertical measured in terms of $N_z$ nodes.}
\label{fig:lindispersion}
\end{figure}

Clustering mesh nodes more densely closer to the free surface can improve accuracy in linear dispersion without increasing the CPU time significantly \cite{MaWuTaylor2001,BinghamZhang2007,EBL08}. 
This is confirmed by results presented in Figure \ref{fig:lindispersion}. 
The results show that the use of a high-order method or clustered vertical distribution of low-order elements is a must for an accurate approximation of dispersion in applications where $kh$ is large (short wave length relative to depth). 
This highlights that one can tune the accuracy by proper choice of discretisation parameters. An important implication of these results is that the vertical node distribution can be used to control the range of validity of the model in terms of dispersive properties, i.e. a numerical truncation counterpart to the analytic truncation used in Boussinesq-type models. Interestingly, flexible-order finite difference solvers with cosine-clustered vertical distributions of nodes appears to be superior in linear dispersion compared to single layer SEM for equal number of vertical nodes. However, multi-layer SEM with cosine clustered elements of vertical order $P = 1$ almost match the FDM, cf. \cite[Figure 2.6, page 59]{EGNL13}. 

\begin{figure}[!htb]
\centering
\begin{minipage}{4.3cm}
\begin{center}
(a) $kh = 0.5$ \\
\includegraphics[height=3.5cm]{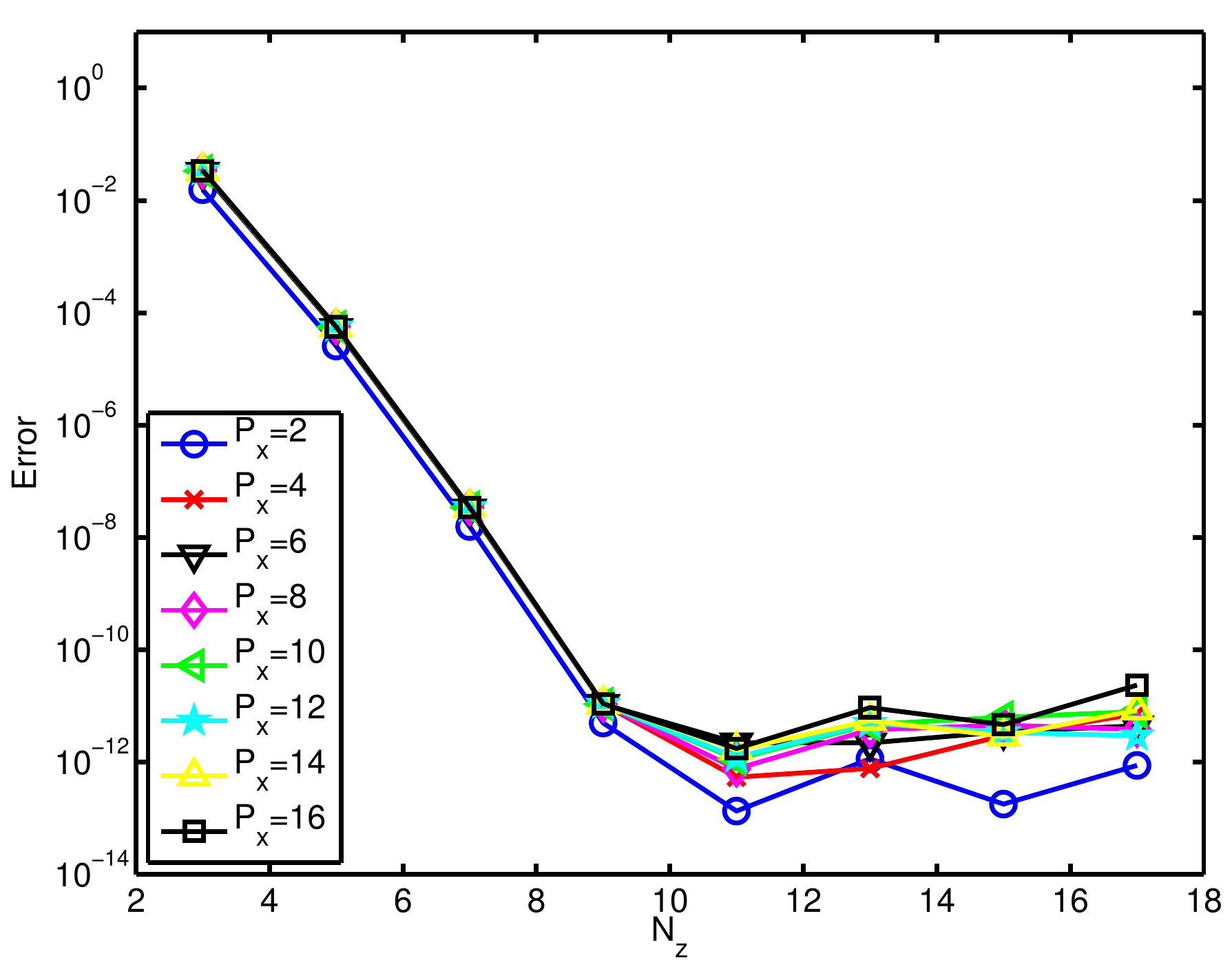} 
\end{center}
\end{minipage}
\begin{minipage}{4.3cm}
\begin{center}
(b) $kh = 2.0$\\
\includegraphics[height=3.5cm]{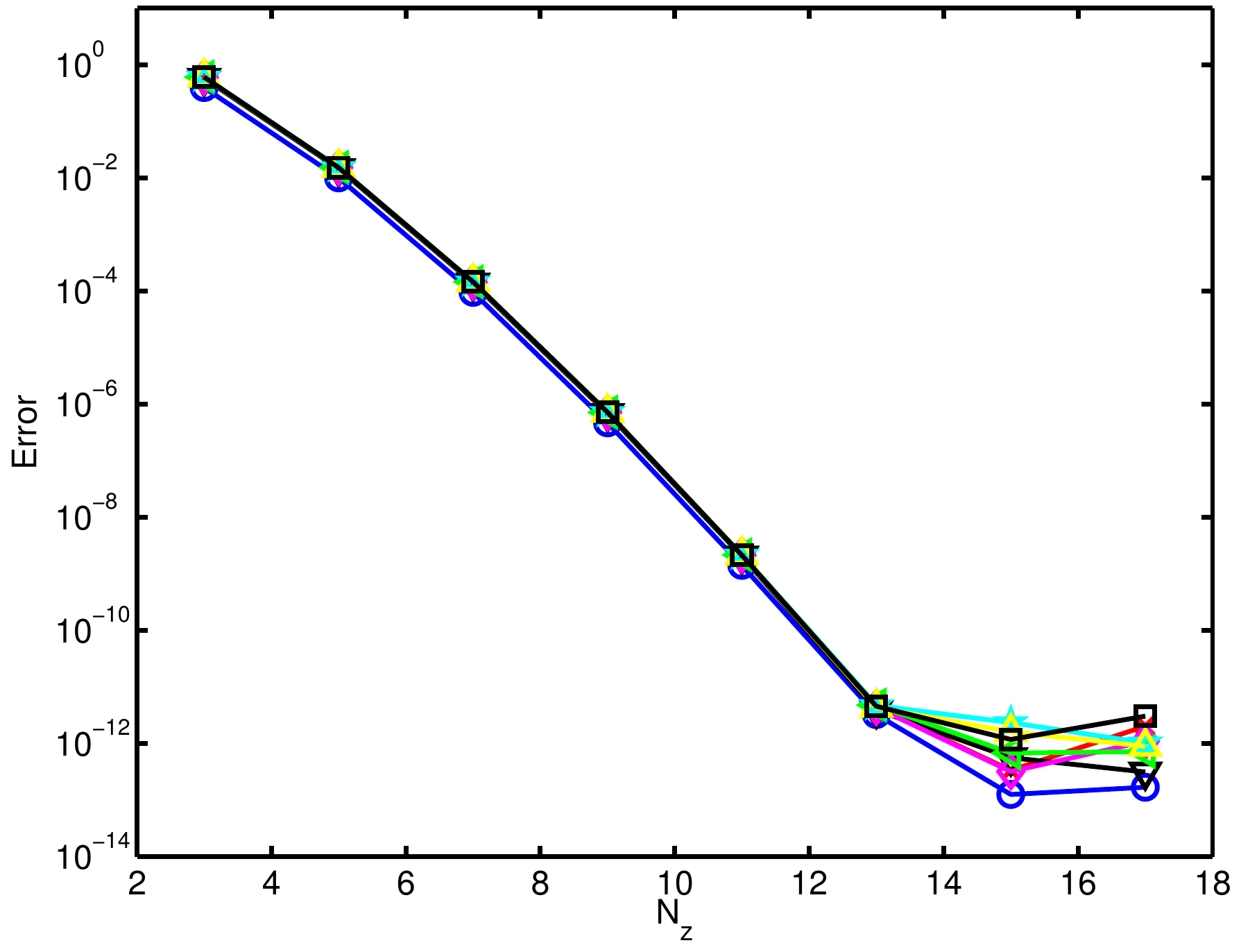} 
\end{center}
\end{minipage}
\begin{minipage}{4.3cm}
\begin{center}
(c) $kh = 4.0$ \\
\includegraphics[height=3.5cm]{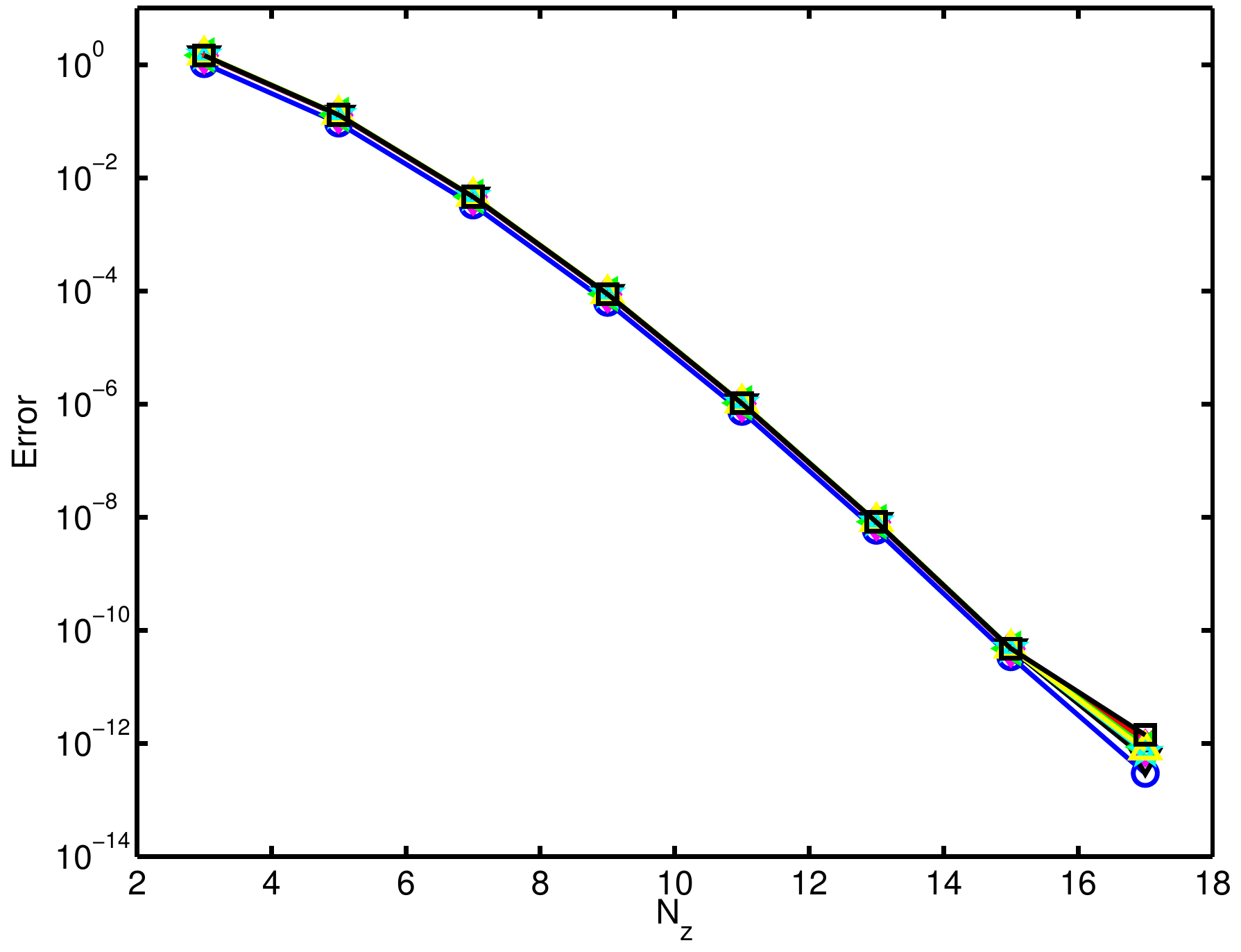} 
\end{center}
\end{minipage}
\caption{Linear accuracy for Airy waves at fixed time. $P_x=2,...,16$ used in the horizontal and one layer of elements in the vertical. One element in the horizontal, $N_k=1$.
}
\label{fig:linearaccuracy}
\end{figure}

\subsection{On nonlinear accuracy, stability and kinematics properties}

\setlength{\unitlength}{1cm}
\begin{figure}[!htb]
\centering
\begin{minipage}{4.3cm}
\begin{center}
$H/L=10\%$ (a) $kh = 0.5$ \\
\includegraphics[height=3.5cm]{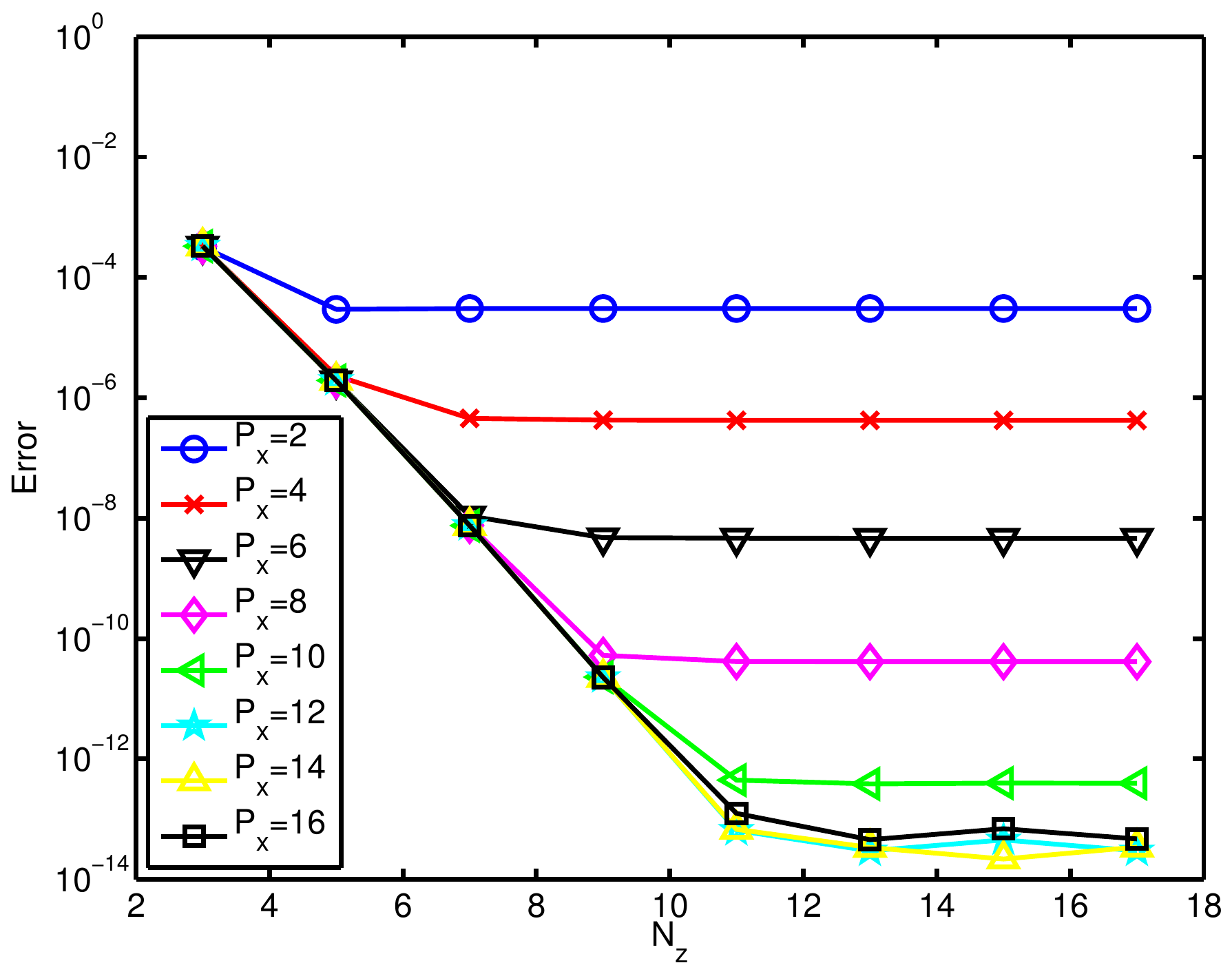} \\
$H/L=50\%$ (d) $kh = 0.5$ \\
\includegraphics[height=3.5cm]{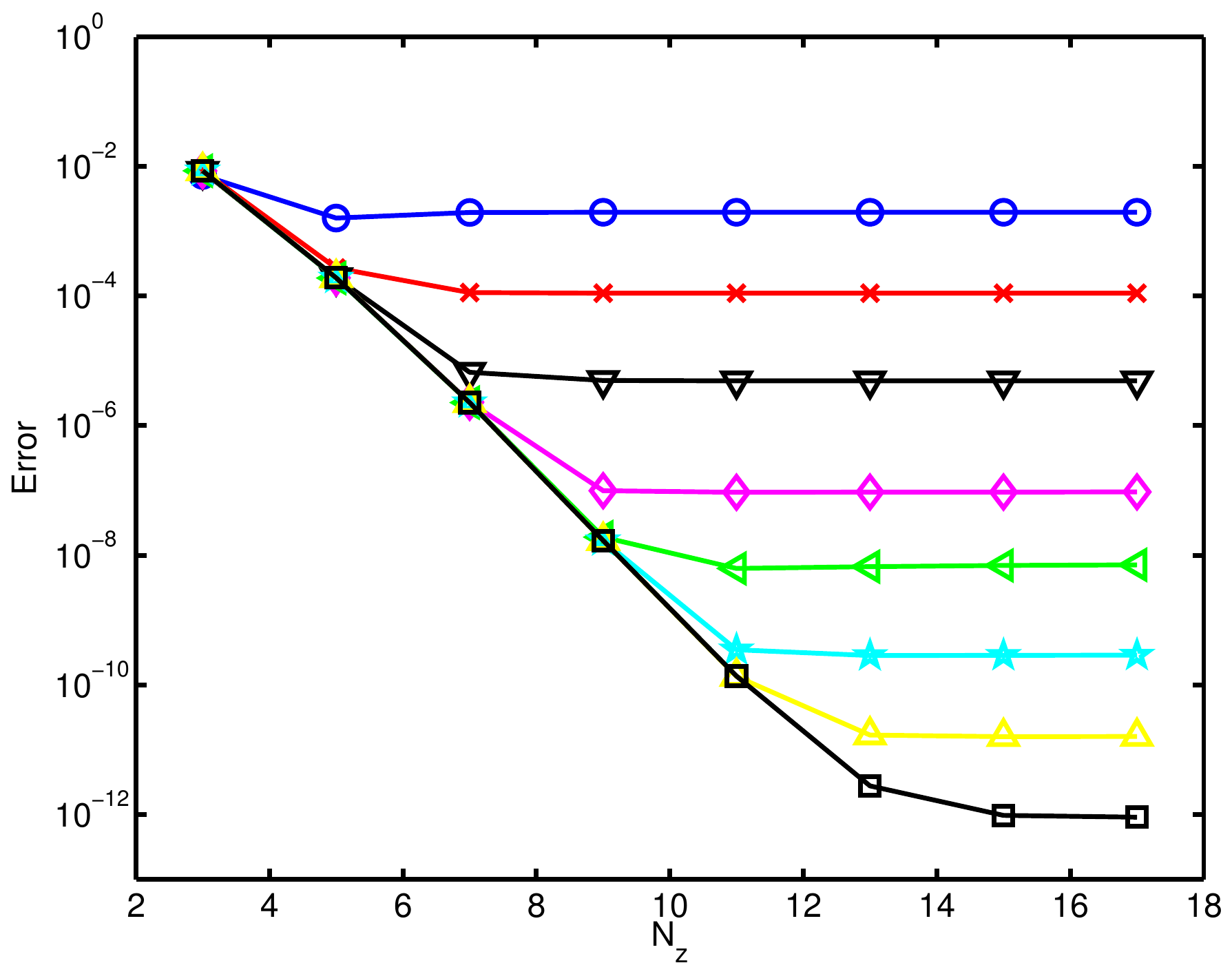} \\
$H/L=90\%$ (g) $kh = 0.5$ \\
\includegraphics[height=3.5cm]{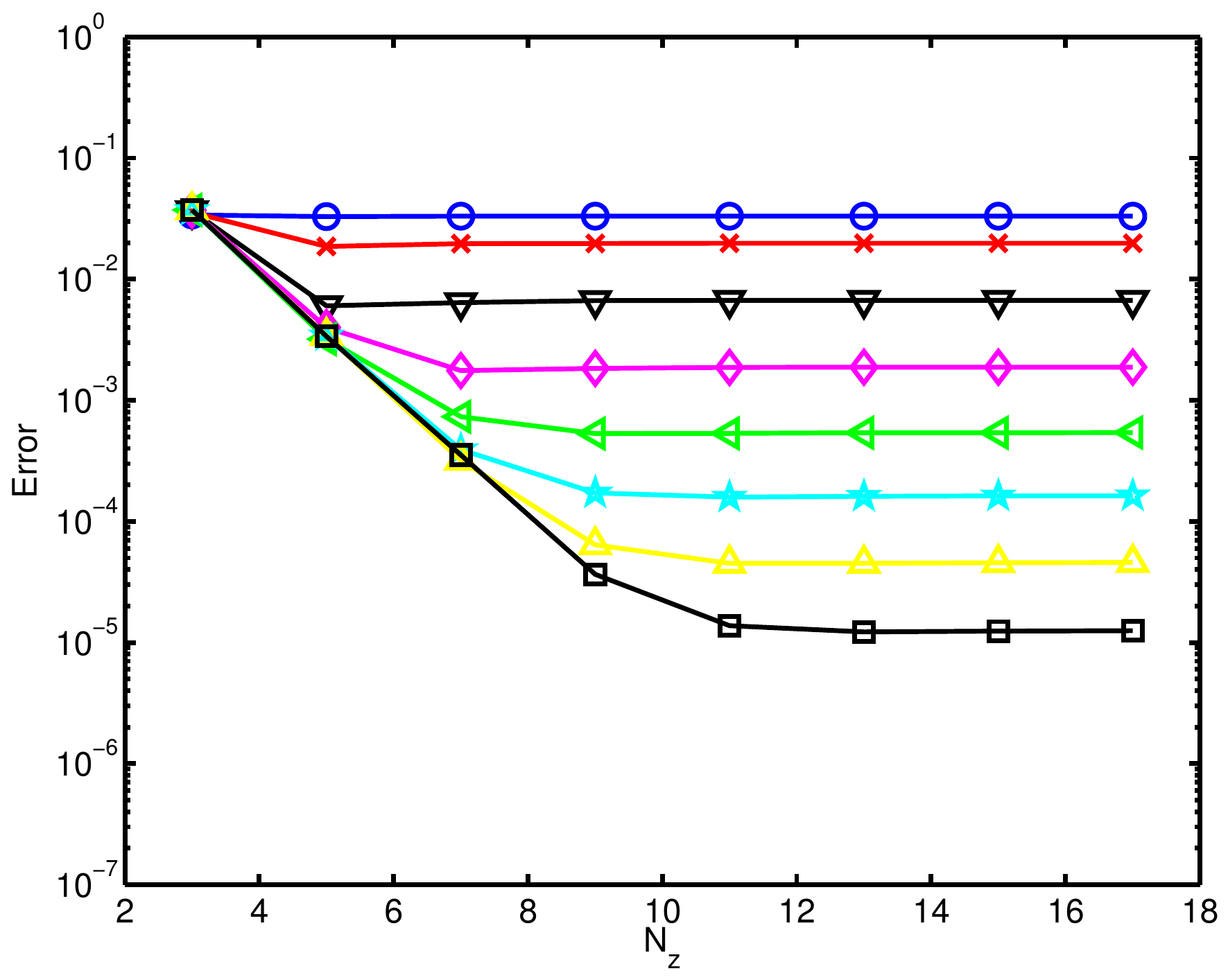}
\end{center}
\end{minipage}
\begin{minipage}{4.3cm}
\begin{center}
$H/L=10\%$  (b) $kh = 2.0$\\
\includegraphics[height=3.5cm]{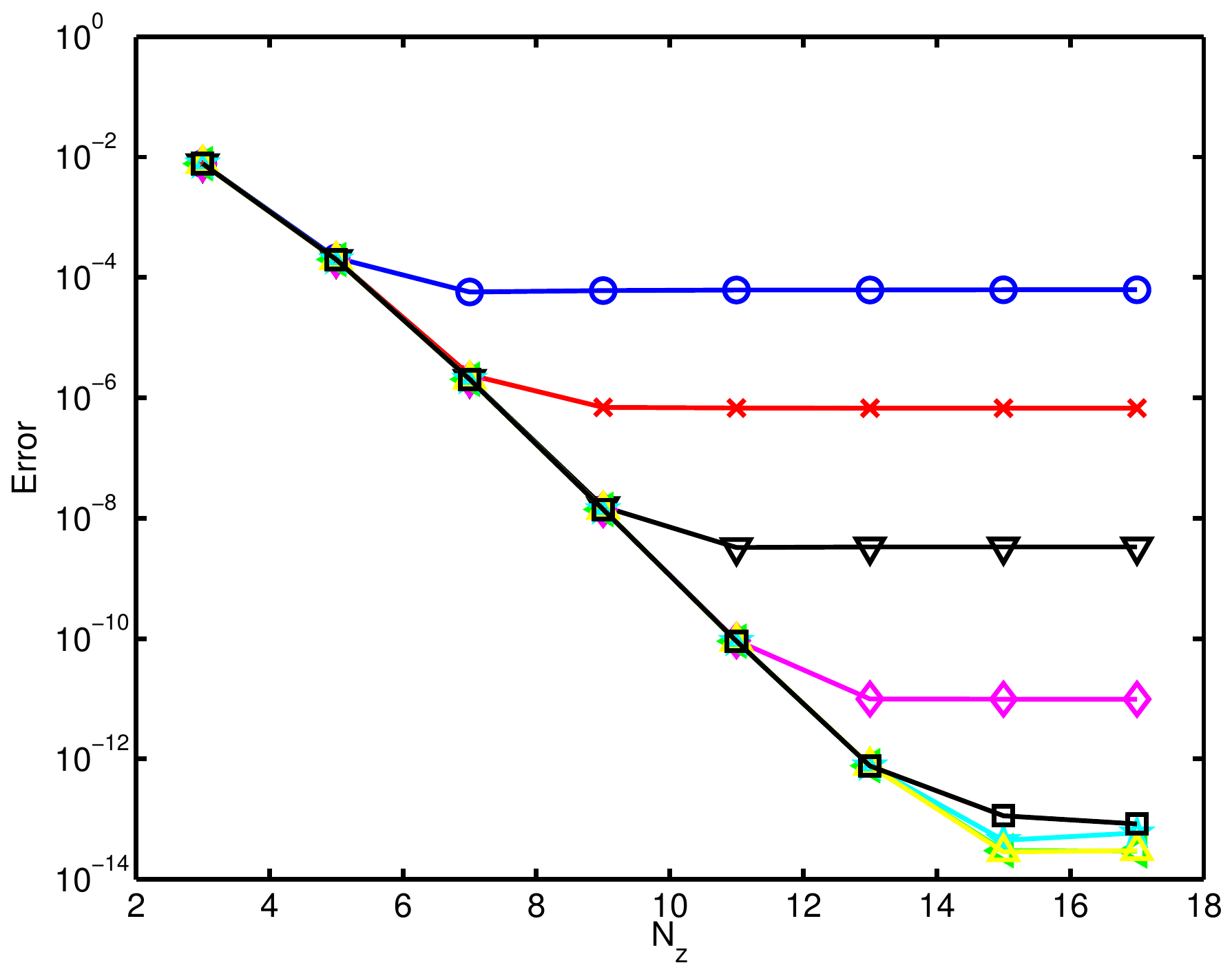} \\
$H/L=50\%$  (e) $kh = 2.0$ \\
\includegraphics[height=3.5cm]{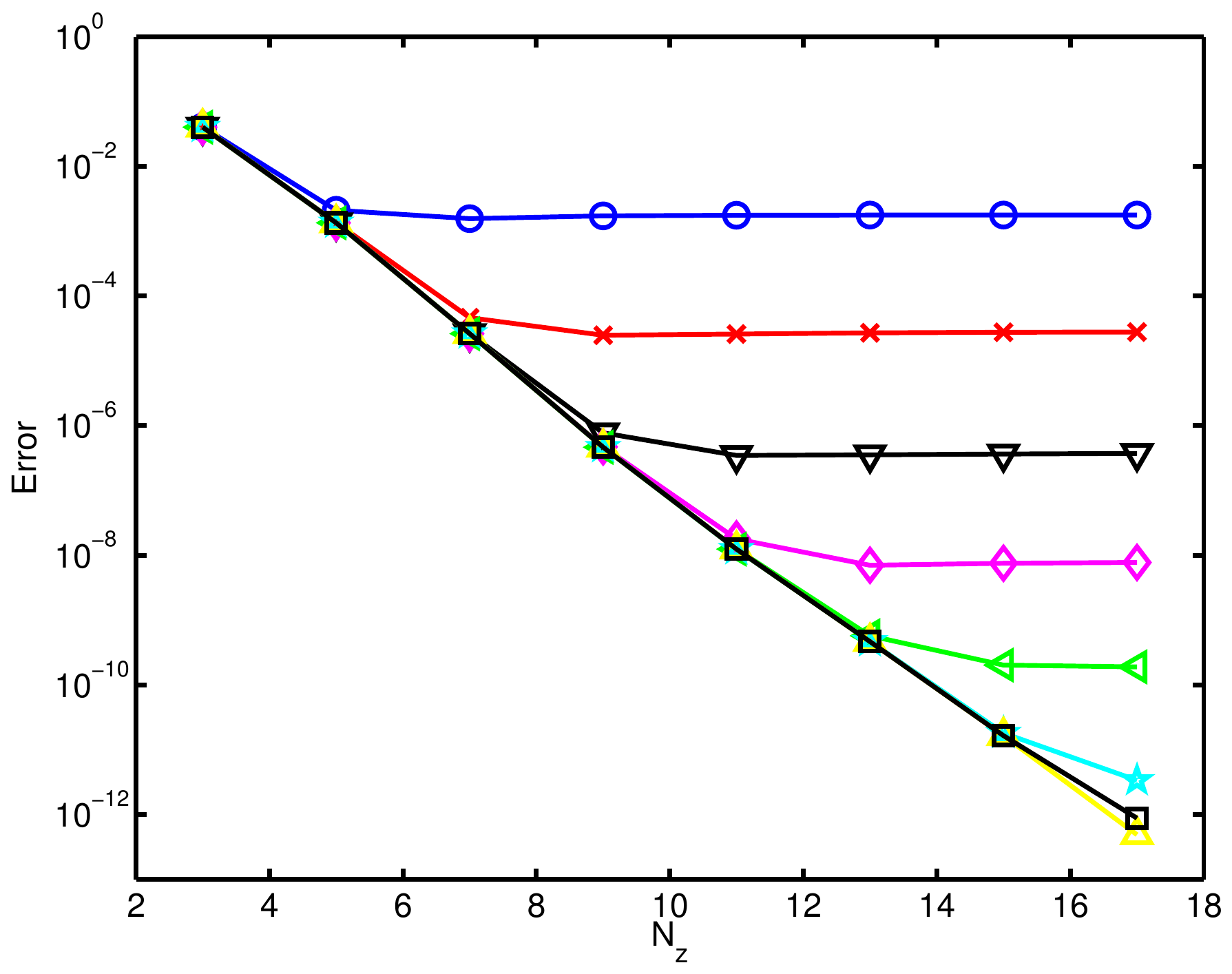} \\
$H/L=90\%$  (h) $kh = 2.0$ \\
\includegraphics[height=3.5cm]{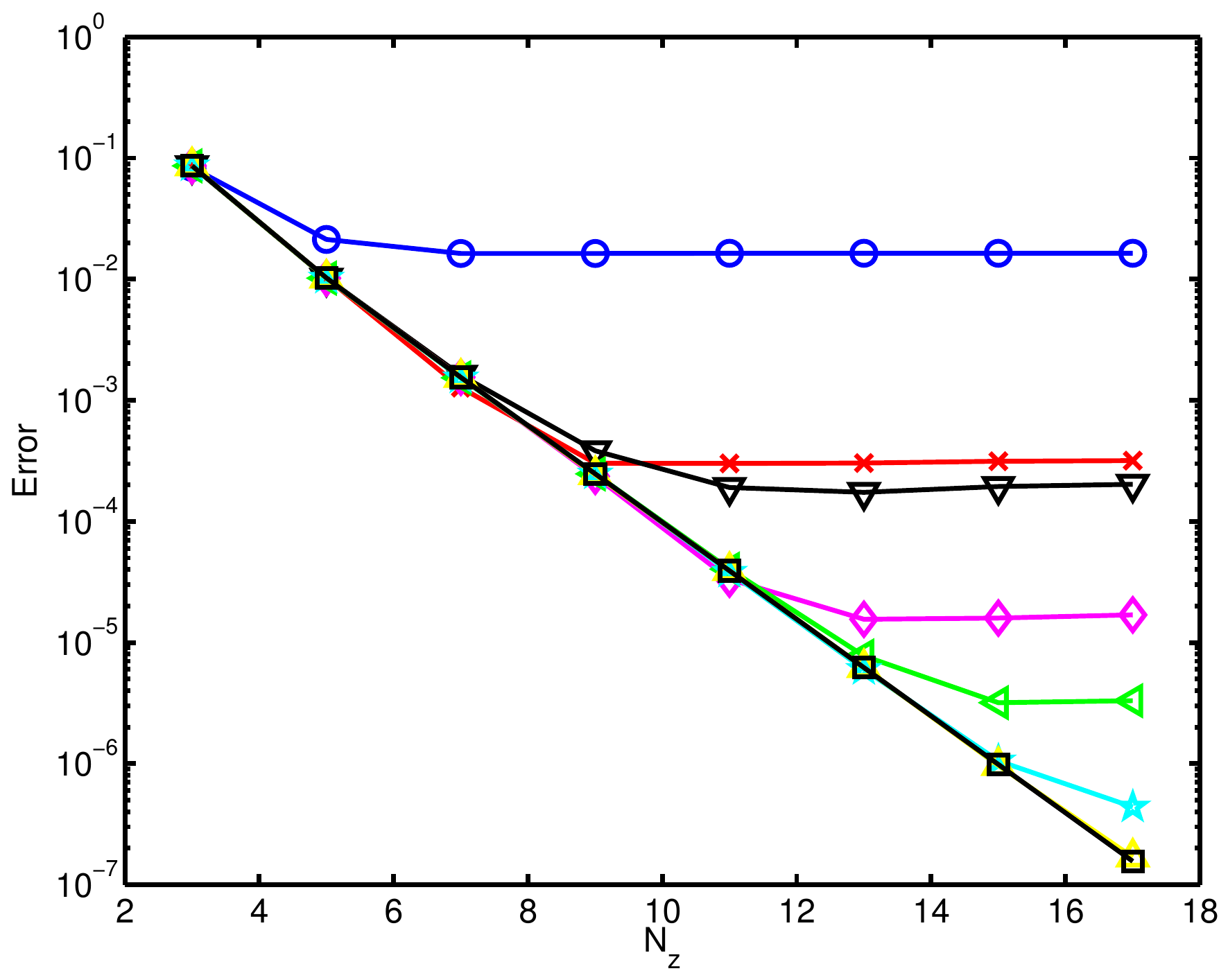}
\end{center}
\end{minipage}
\begin{minipage}{4.3cm}
\begin{center}
$H/L=10\%$  (c) $kh = 4.0$ \\
\includegraphics[height=3.5cm]{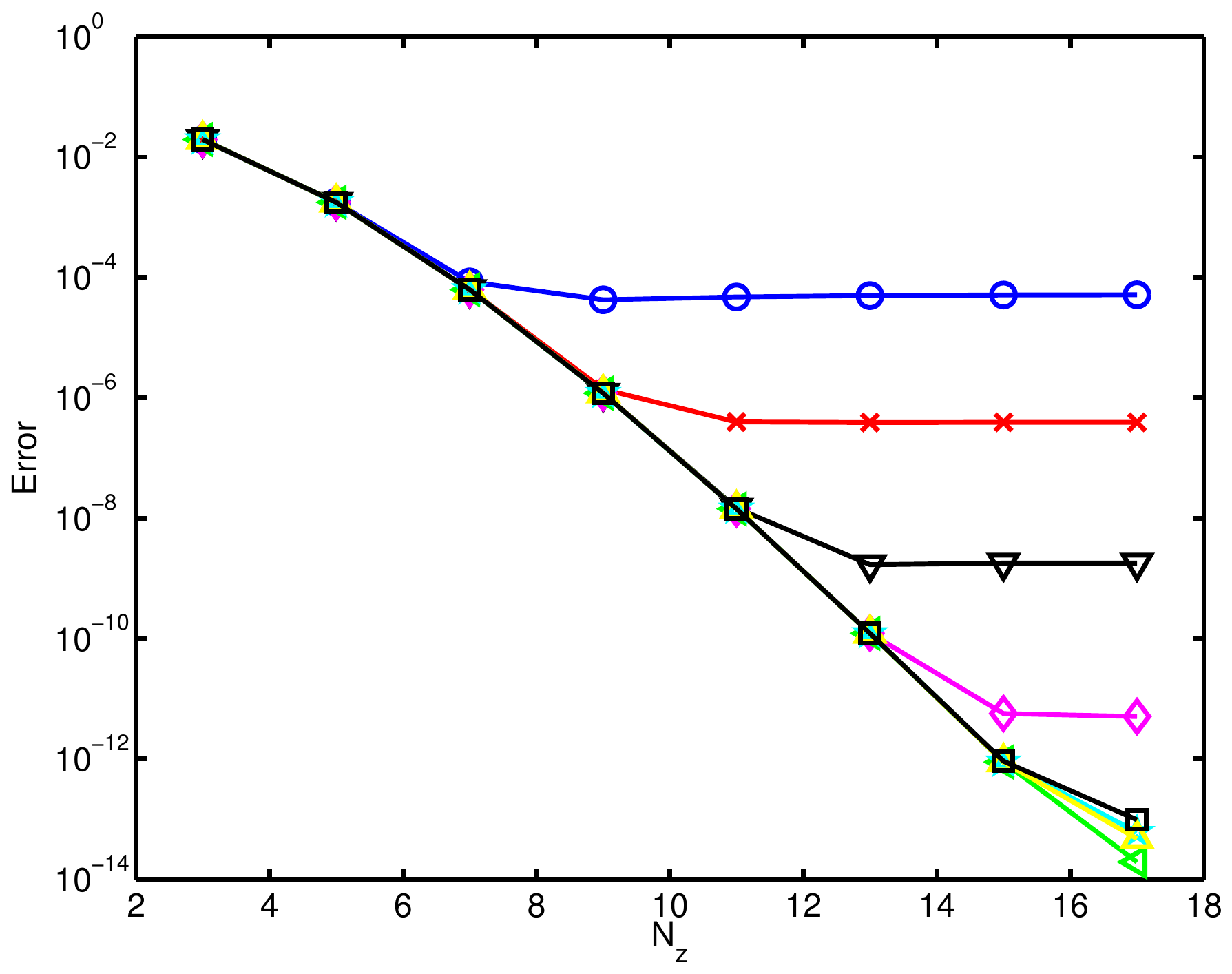} \\
$H/L=50\%$  (f) $kh = 4.0$ \\
\includegraphics[height=3.5cm]{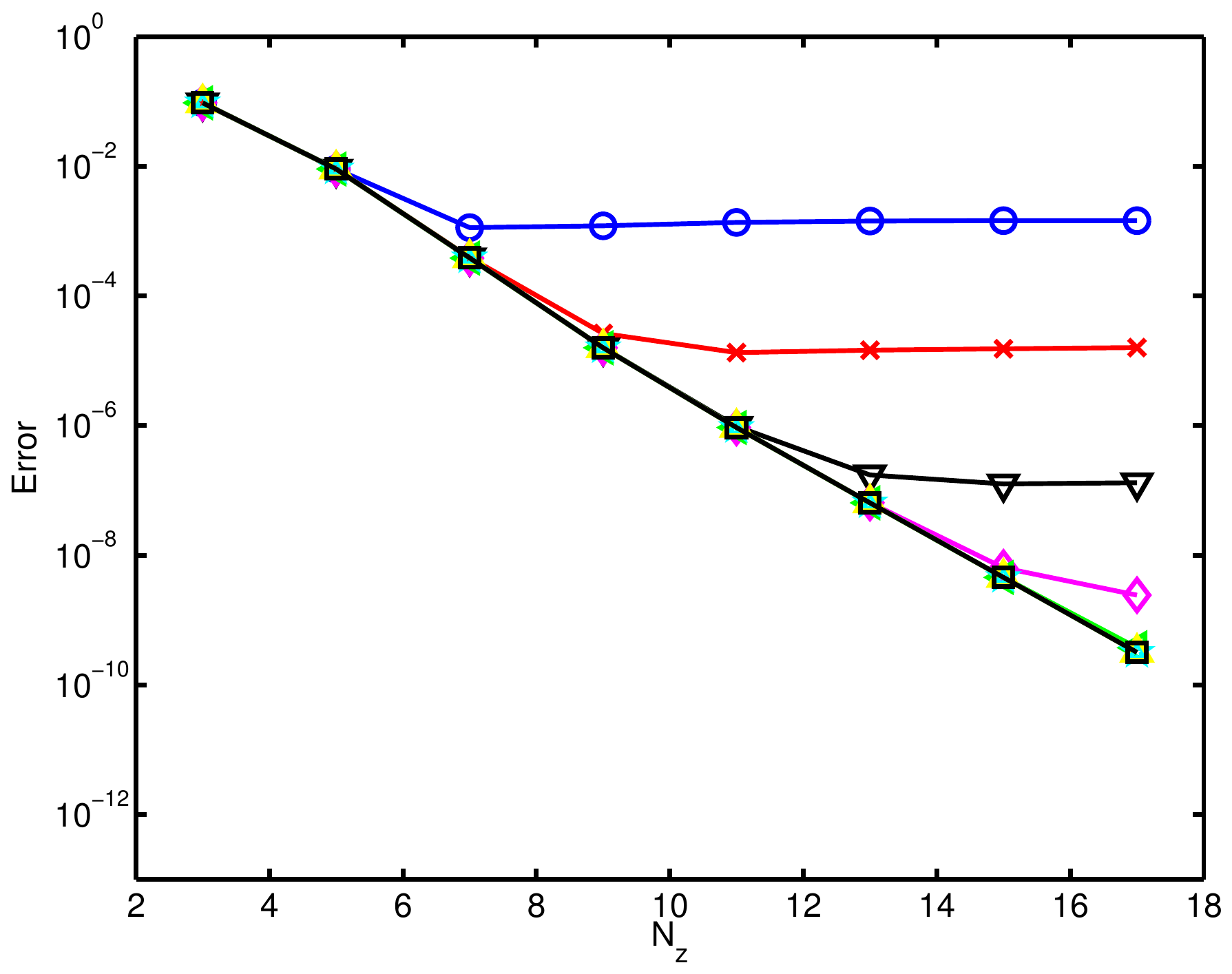} \\
$H/L=90\%$  (i) $kh = 4.0$ \\
\includegraphics[height=3.5cm]{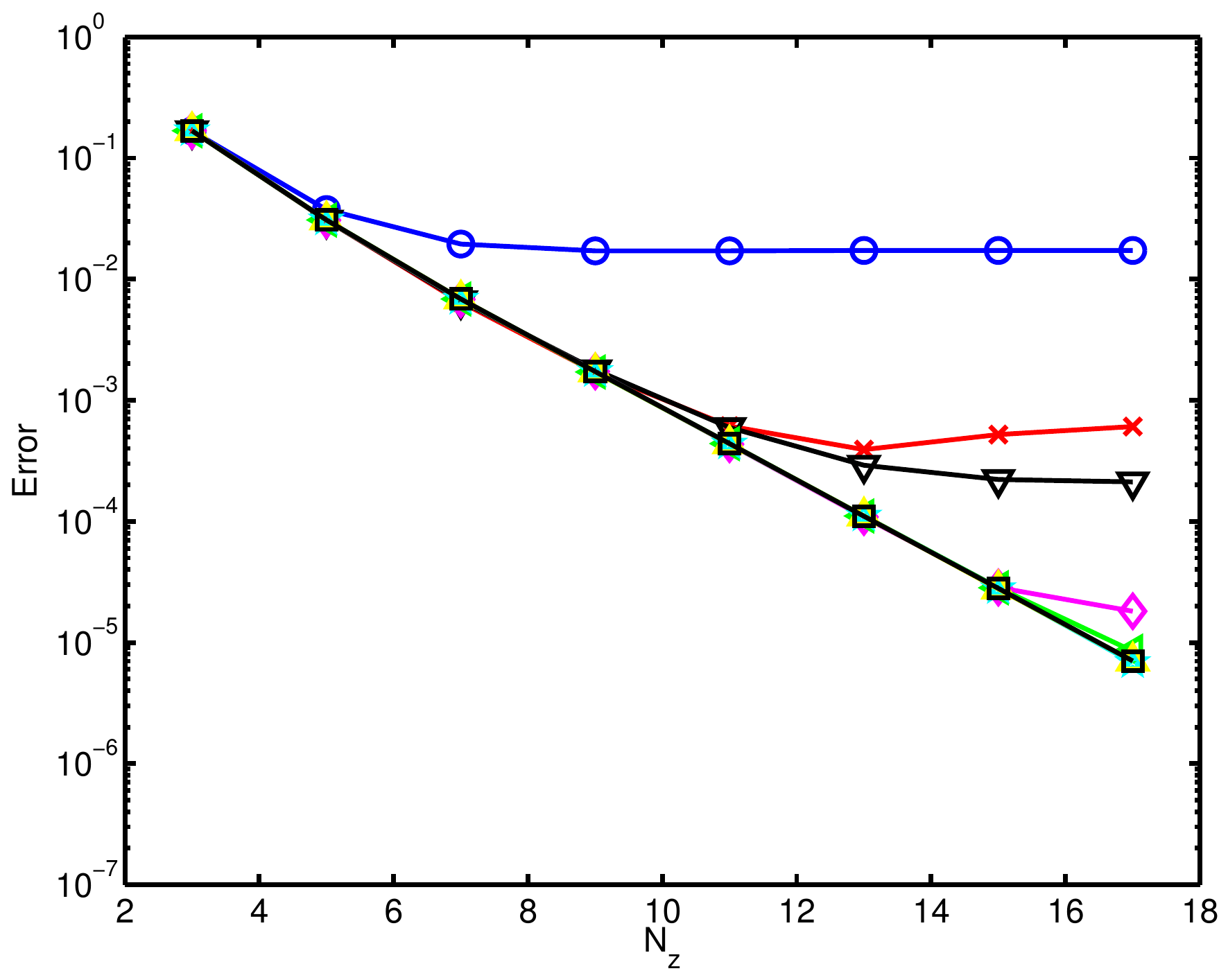}
\end{center}
\end{minipage}
\caption{Nonlinear accuracy for Stream Function Waves at fixed time. $P_x=2,...,16$ used in the horizontal and one layer of elements in the vertical. Fixed number of elements in the horizontal, $N_k=10$.
}
\label{fig:nonlinearaccuracy}
\end{figure}

\begin{figure}[!htb]
\begin{minipage}{0.95\textwidth}
\begin{center}
(a) ($kh=1,H/L=95\%$ of maximum steepness) \\
\includegraphics[height=5.0cm]{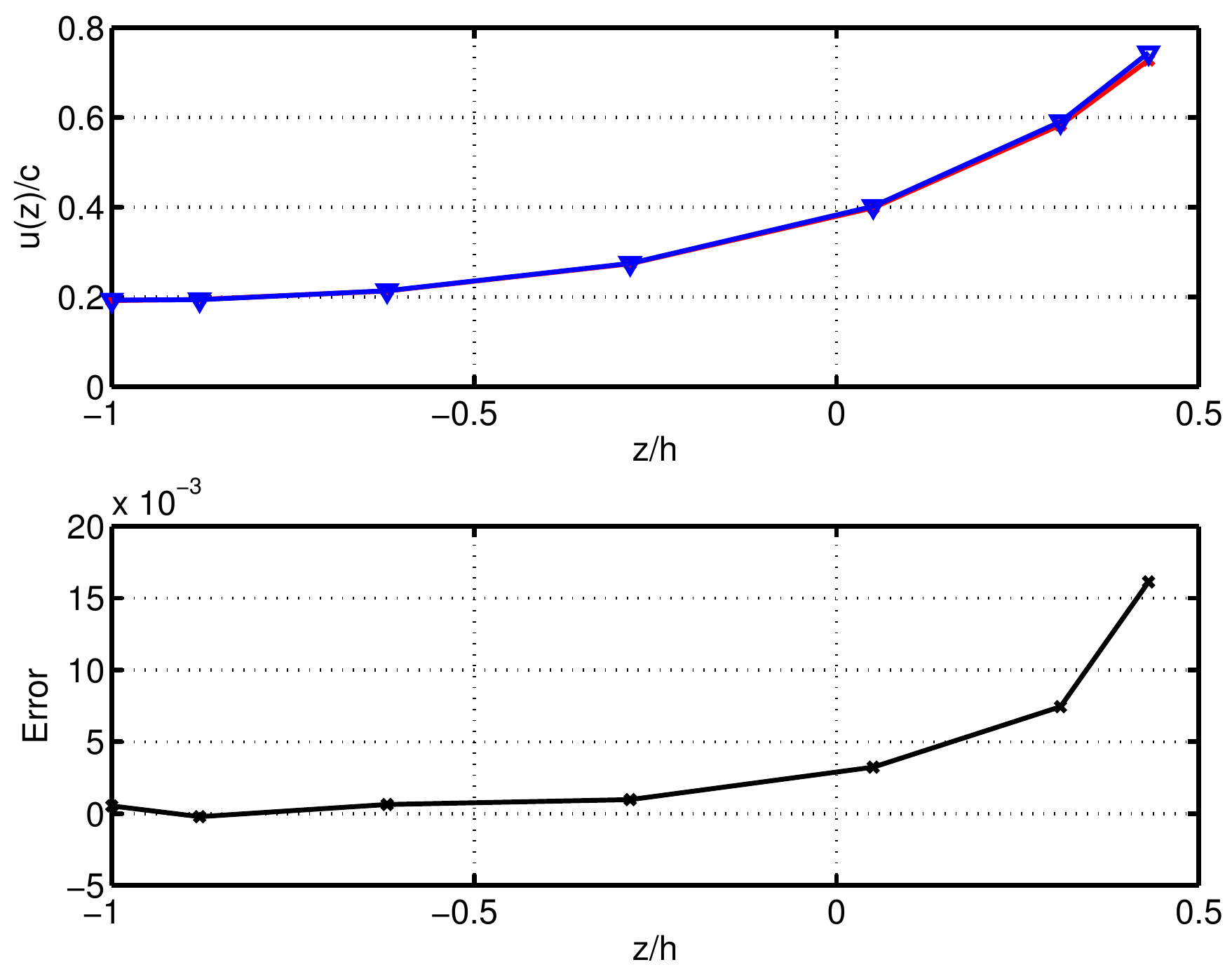}
\includegraphics[height=5.0cm]{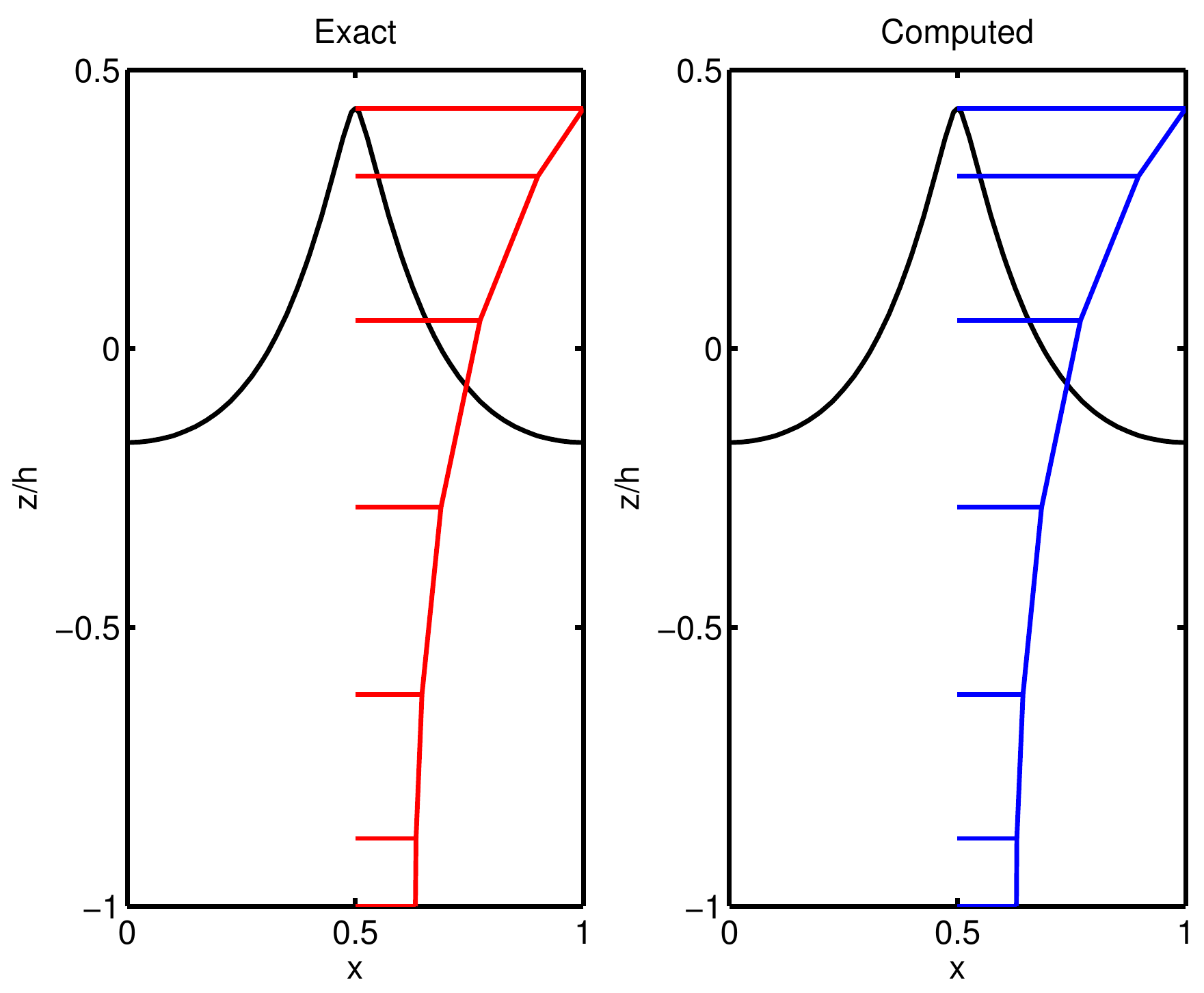}
\end{center}
\end{minipage}
\\
\begin{minipage}{0.95\textwidth}
\begin{center}
(b) ($kh=10,H/L=95\%$ of maximum steepness) \\
\includegraphics[height=5.0cm]{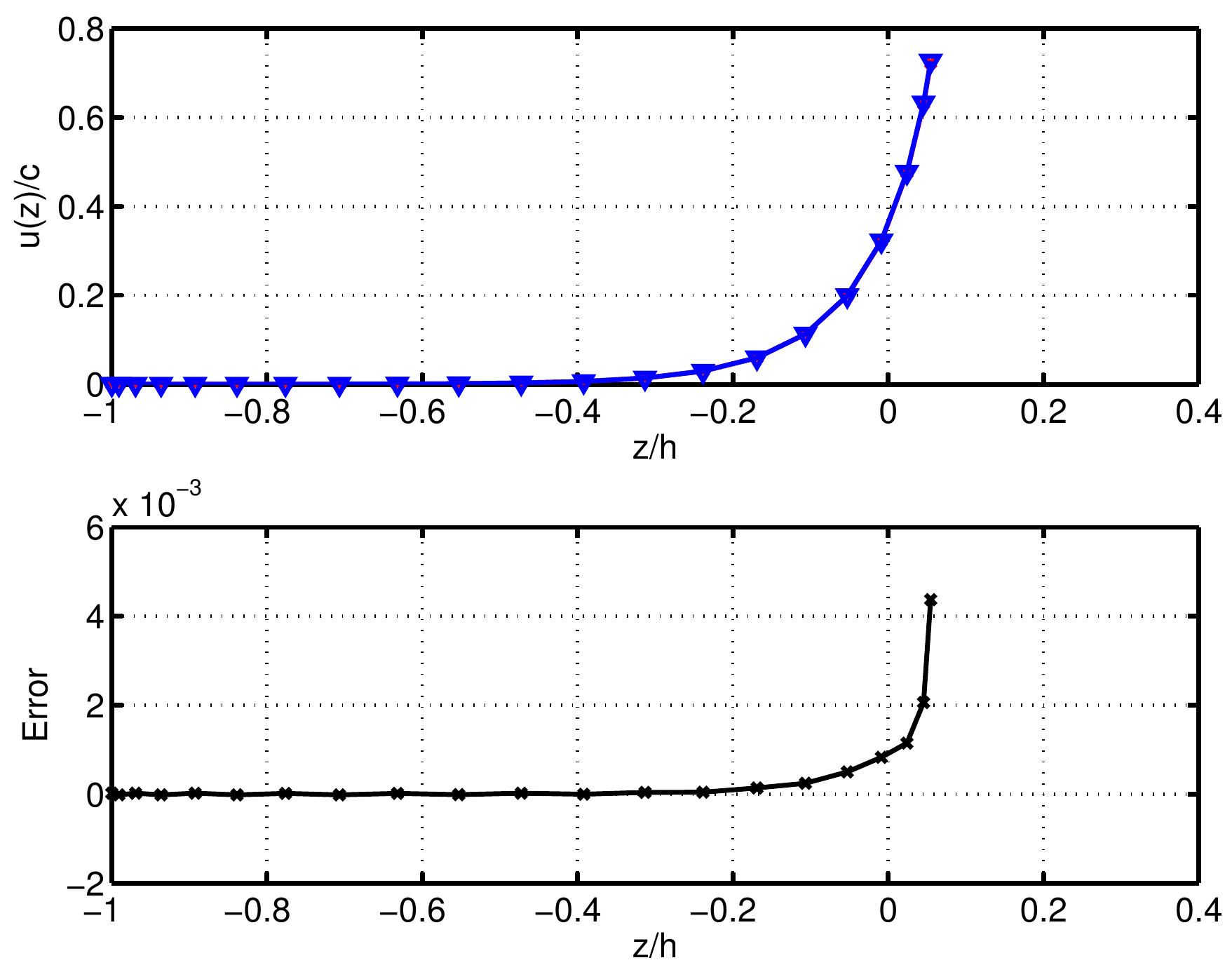}
\includegraphics[height=5.0cm]{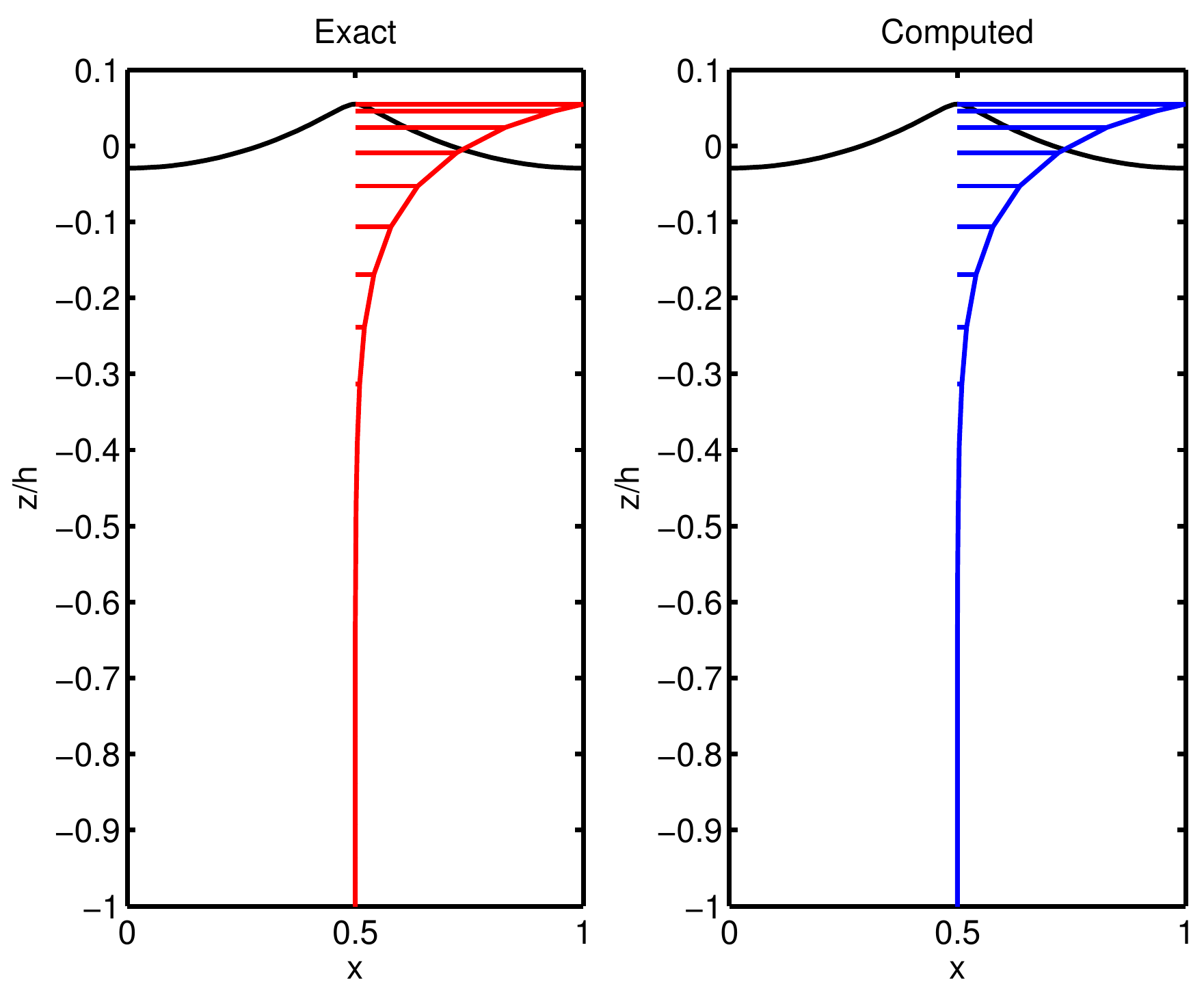}
\end{center}
\end{minipage}
\caption{Kinematic accuracy for stream function waves. (a) Intermediate water $(P_x,P_z,N_k) = (6,6,10)$ and (b) deep water $(P_x,P_z,N_k) = (6,20,10)$.}
\label{fig:kinematics}
\end{figure}

The accurate computation of kinematics is essential for load predictions in wave-structure applications, e.g. for offshore foundations of wind turbines. 
For nonlinear waves, exact stream function (SF) wave solutions of permanent form \cite{Dean1965} based on assuming a flat sea bed can be used to assess the accuracy with respect to variation in dimensionless nonlinearity ($H/L$) and dispersion ($kh$) parameters. 
This is done by solving the Laplace problem first. 
Then, from the scalar velocity potential solution we can calculate the vertical free surface velocity and compare with exact results. 
Numerical results for linear, weakly nonlinear and strongly nonlinear SF waves in combinations of shallow, intermediate and deep waters are presented in Figures  \ref{fig:linearaccuracy} and  \ref{fig:nonlinearaccuracy}. Here we use only one layer of elements in the vertical and a fixed number of elements in the horizontal direction.
Convergence is achieved by the variation of the polynomial order to achieve fast $p$-convergence. 
All tests show convergence with increasing resolution as expected. 
When depth or nonlinearity increases more resolution is required. 
Similar tests were carried out for a flexible-order finite difference model in \cite{BinghamZhang2007}. 
An immediate conclusion is that the SEM has similar resolution requirements as the corresponding finite difference solver to match the order of accuracy for nonlinear applications. This highlights that from a purely algorithmic (implementation independent) viewpoint there seems to be no significant tradeoffs in introducing geometric flexibility through this choice of discretisation.
In Figure \ref{fig:kinematics}, the accuracy of the kinematics computations is shown for intermediate depth and deep water for very steep nonlinear stream function waves. Excellent agreement is found between exact and computed results for both intermediate and deep waters, which is difficult to represent in conventional wave propagation models due to lack of resolution.

\section{Numerical experiments}
\label{sec:numexp}

We examine different test cases and benchmarks, that inspect different properties of the numerical model that serves as validation. 

\subsection{Stabilised nonlinear wave propagation of stream function waves}

To test the robustness and accuracy of the numerical method, we consider the strenuous test case of propagating nonlinear stream function waves near the theoretical limit of wave steepness and nonlinearity. 
Few numerical schemes can resolve such waves accurately. 
Results from representative numerical experiments are presented in Table \ref{tab:effstrat}. 
The table shows how a standard Galerkin scheme unsurprisingly fails to be stable for all temporal resolutions chosen. 
Improvement in stability is achieved by using higher order numerical integration (cf. Section \ref{sec:remquaderrors}) integrands in the Galerkin integrals. This is done by interpolating the integrands to a finer mesh on the reference element corresponding to a basis with polynomial basis expansion order of $2P_x$ resulting in terms of less than quartic order nonlinear to be subject to {\em over-integration}. This implies that exact integration is used for all nonlinear terms to within the order of accuracy of the numerical scheme.
This is compared to using no over-integration but instead trying to stabilise only using a {\em gentle} spectral filtering strategy that caps off the highest 5\% in the highest modes of the modal expansion. 
Both of these strategies fail to be stable. 
Instead, if the two strategies are combined, where the over-integration effectively removes any aliasing in the evaluation of the strongly nonlinear terms, the mild spectral filter dissipates just enough energy for the model to stabilise completely. 
We find that over-integration is only needed in the free-surface equations, leading to a marginal increase in computational cost, which is anyway driven by the Laplace solver.
This is highlighted in the table, where the time stepping cost is only increased by approximately 15\% when using over-integration and spectral filtering in comparison with a standard Galerkin formulation based on expansion order $P_x$. This additional cost of over-integration would be even less significant for larger simulations.

\begin{table}[!htb]
\centering
\begin{tabular}{lcccccc} \hline
{\bf Accuracy} & $t/T$ & $T/\Delta t$ &  &  & Cost/$\Delta t$\\ \hline
&  &  40 & 80 & 160 & \\ \hline
No filter + & $1$ & NaN & NaN & NaN & 1.00 \\ 
No over-integration & $10$ & NaN & NaN & NaN \\ \hline 
No filter + & $1$ & 1.3608e-03 & 7.5021e-04 & 7.6731e-04 & 1.02\\ 
Over-integration & $1.8$ & NaN & NaN & NaN \\ \hline
Filter + & $1$ & NaN & 5.6019e-04 & 8.0508e-03 & 1.12 \\ 
No over-integration & $3.1$ & NaN & NaN & 1.2705e-03 \\ 
& $10$ & NaN & NaN & 7.8374e-03 \\ 
& $28.2$ & NaN & NaN & NaN \\ \hline
Filter + & $1$ & 1.3943e-03 & 7.0651e-04 & 1.0102e-03 & 1.15 \\ 
Over-integration & $10$ & 7.4032e-03 & 4.3313e-03 & 7.0332e-03 \\ 
& $50$ & 7.2826e-02 & 5.7642e-02 & 7.5093e-02 \\ \hline
\end{tabular}
\caption{Nonlinear wave propagation of stream function waves with dispersion parameter $kh=1$ and nonlinearity parameter $H/L=0.0903=$ 90$\%$ of maximum steepness. The spatial resolution is fixed using eight elements with discretisation order $P_x=6$ in both the horizontal and vertical dimensions. Unstable simulations are indicated with 'NaN' in the table. With longer integration time, the errors tend to increase due to a difference between numerical and exact phase speed and more accuracy can be recovered by increasing the resolution. The numerical efficiency is measured as a cost per time step relative to the first strategy. The time $t$ indicates when the results were achieved relative to a full wave period $T$, to make it clear how fast the non-stabilised simulations were deemed unstable.
}
\label{tab:effstrat}
\end{table}

\subsection{Convergence tests and high-order accuracy}
\label{sec:convtests}

\begin{figure}[!htb]
\begin{center}
\begin{minipage}{4.2cm}
\centering (a) $P=3$, $H/L=10\%$ \\
\includegraphics[height=3.6cm]{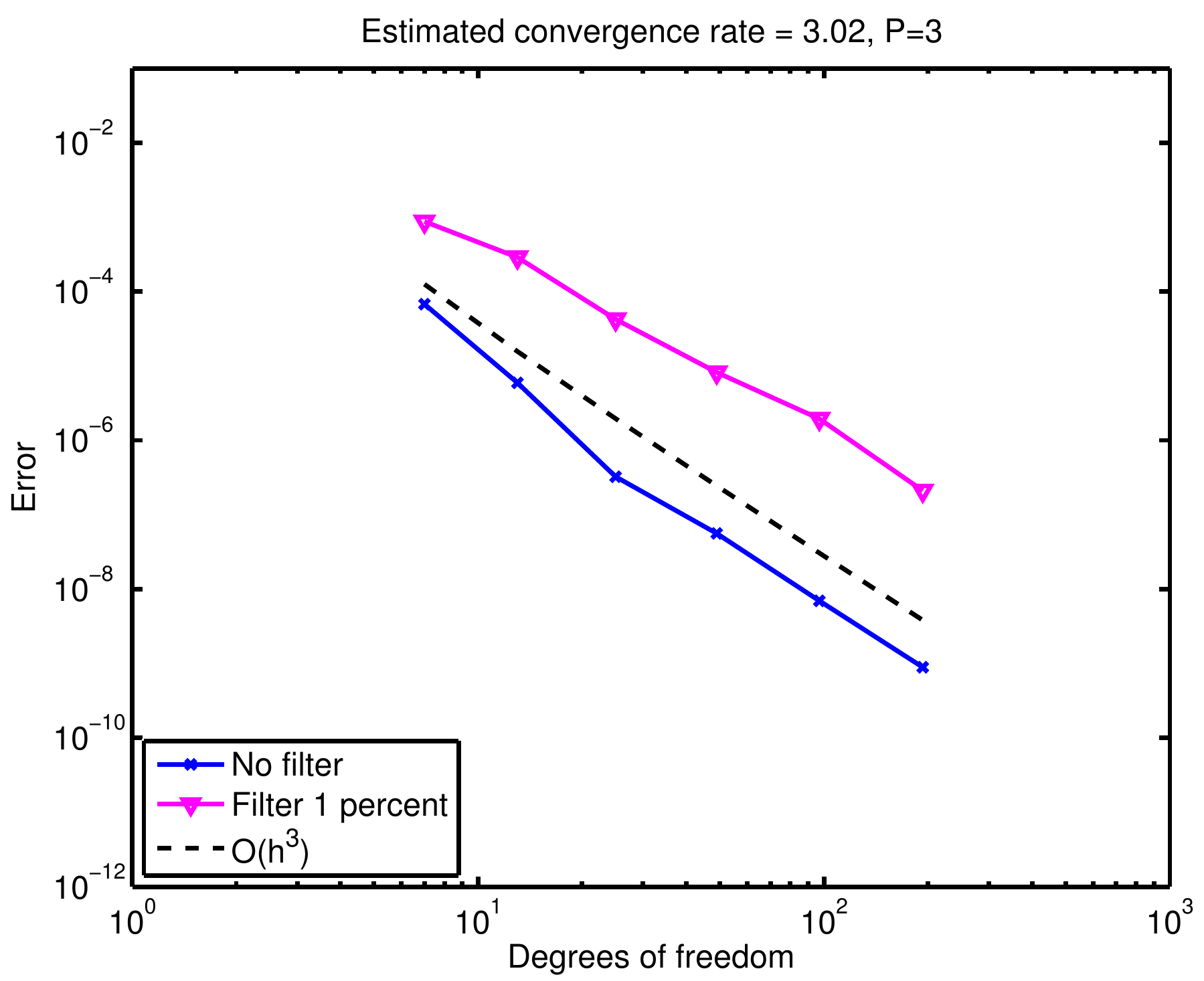} \\
\centering (d) $P=4$, $H/L=10\%$ \\
\includegraphics[height=3.6cm]{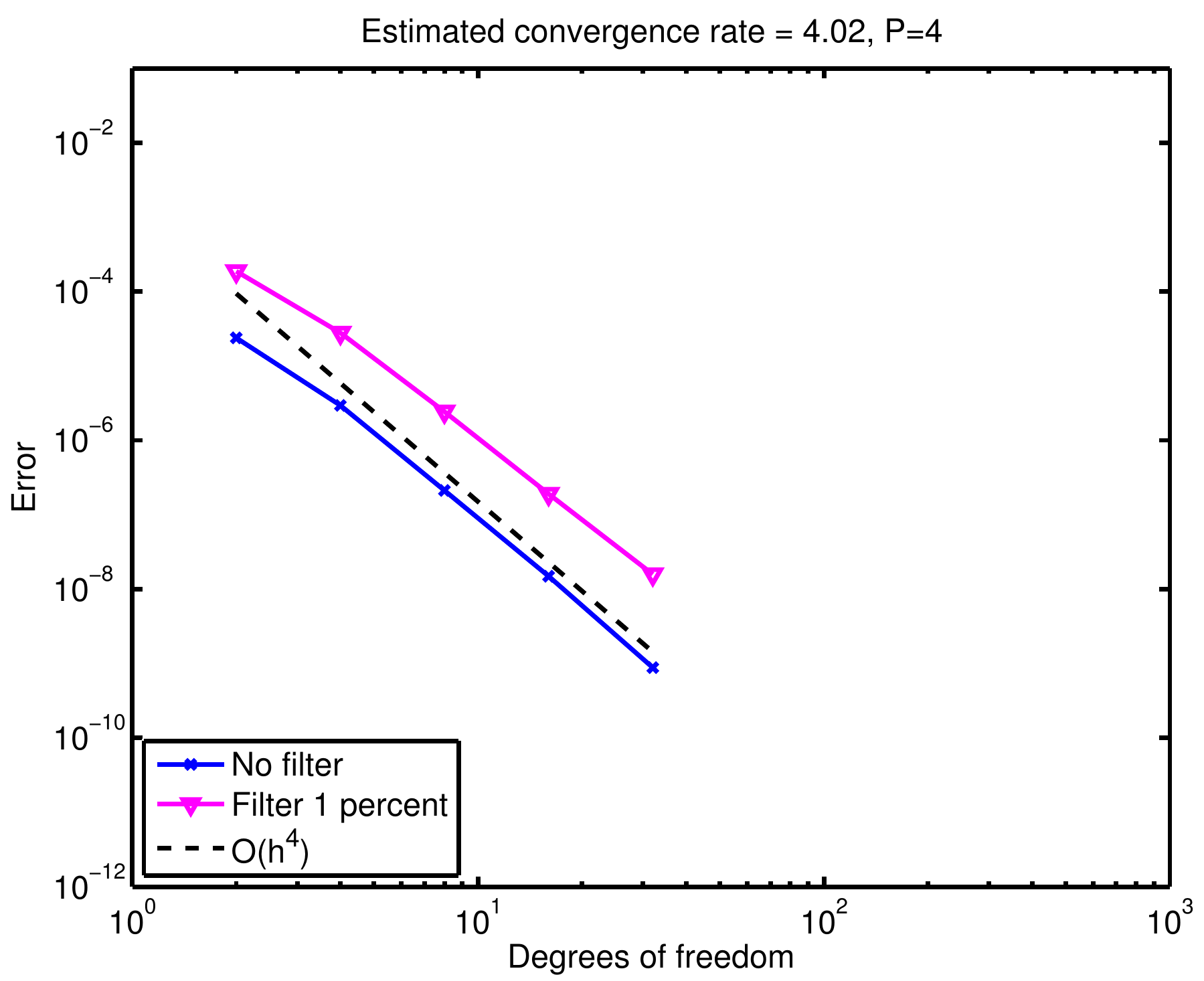} \\
\centering (g) $P=5$, $H/L=10\%$ \\
\includegraphics[height=3.6cm]{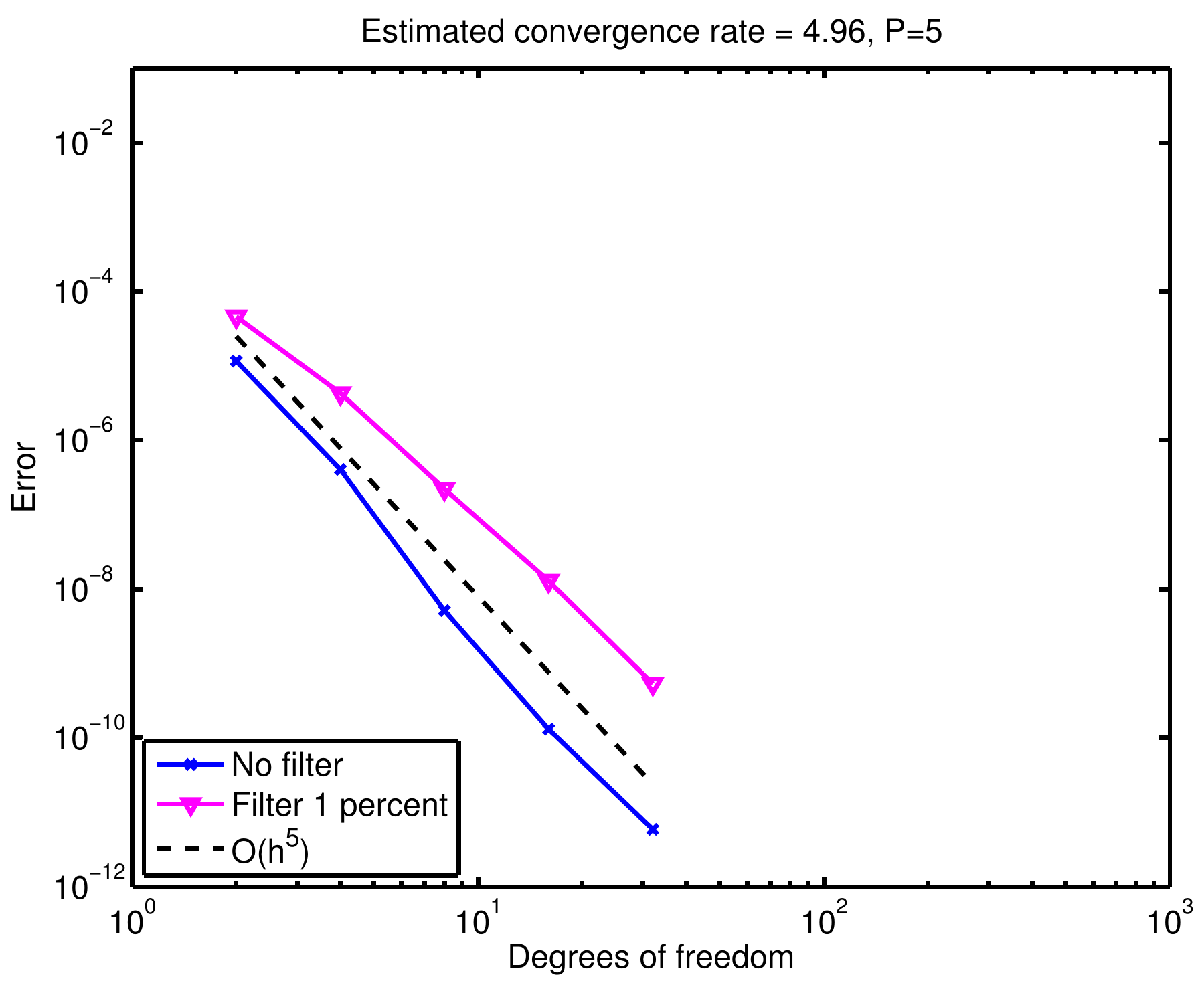} \\
\centering (j) $P=6$, $H/L=10\%$ \\
\includegraphics[height=3.6cm]{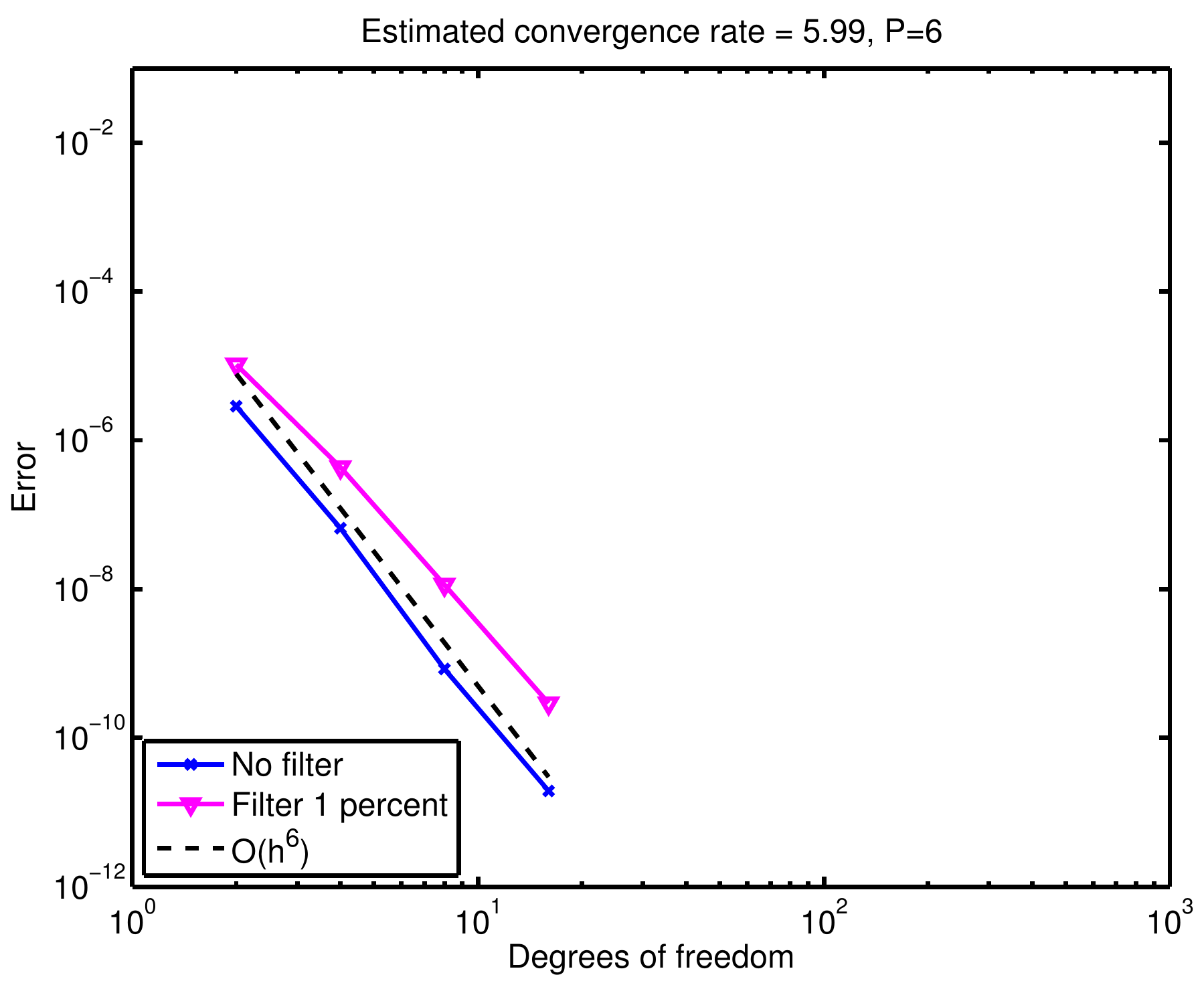}
\end{minipage}
\begin{minipage}{4.2cm}
\centering (b) $P=3$, $H/L=50\%$ \\
\includegraphics[height=3.6cm]{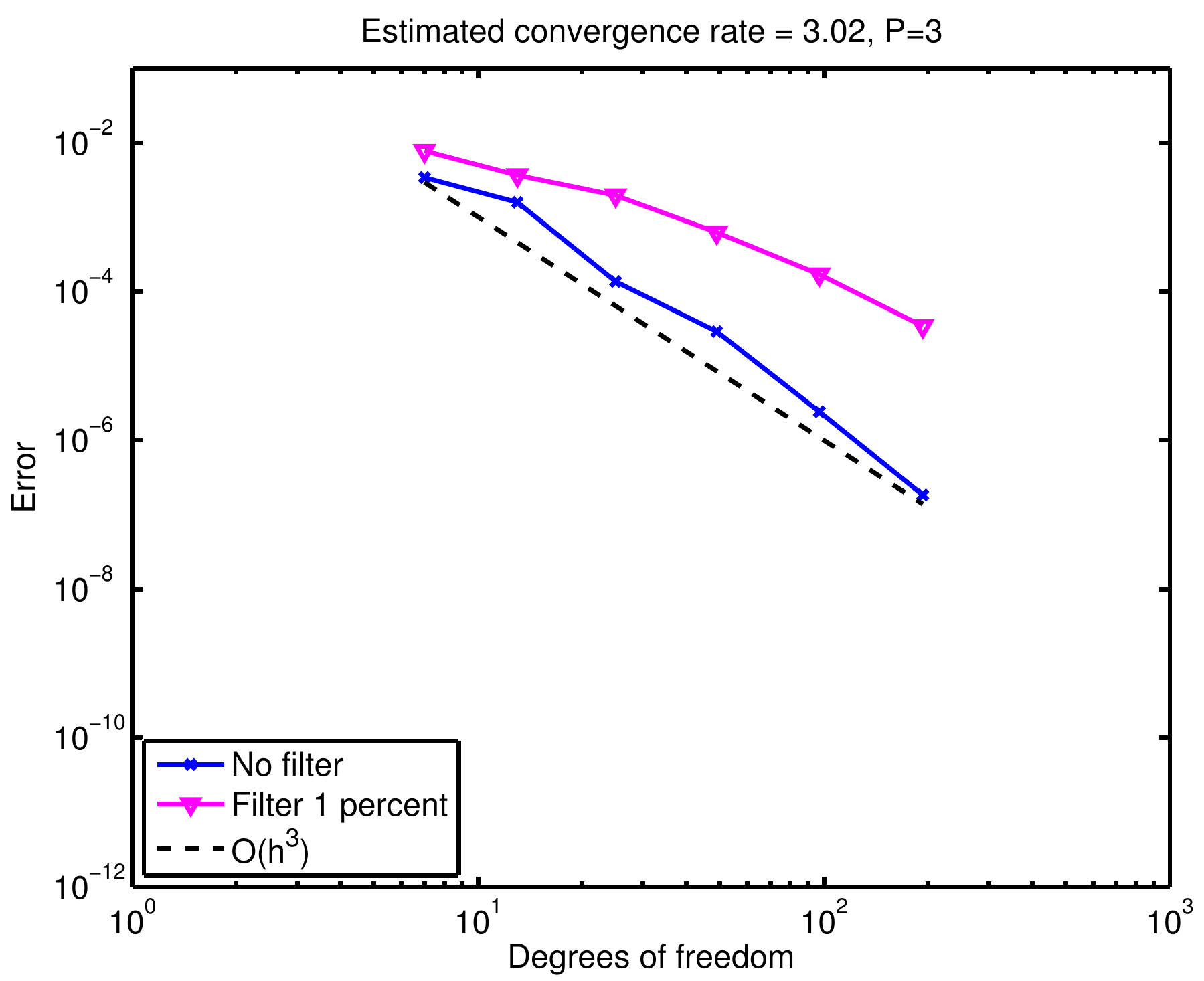} \\
\centering (e) $P=4$, $H/L=50\%$ \\
\includegraphics[height=3.6cm]{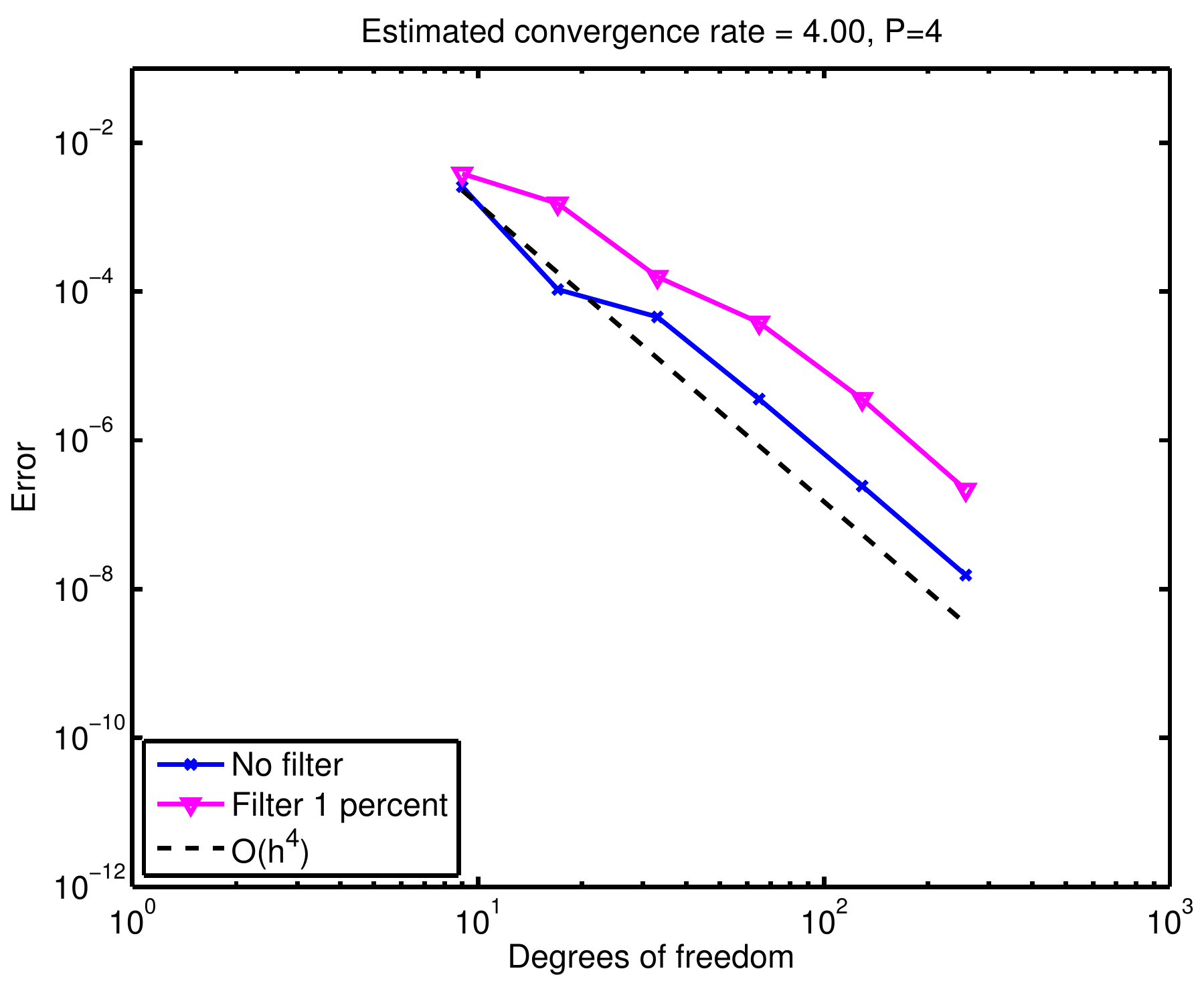} \\
\centering (h) $P=5$, $H/L=50\%$ \\
\includegraphics[height=3.6cm]{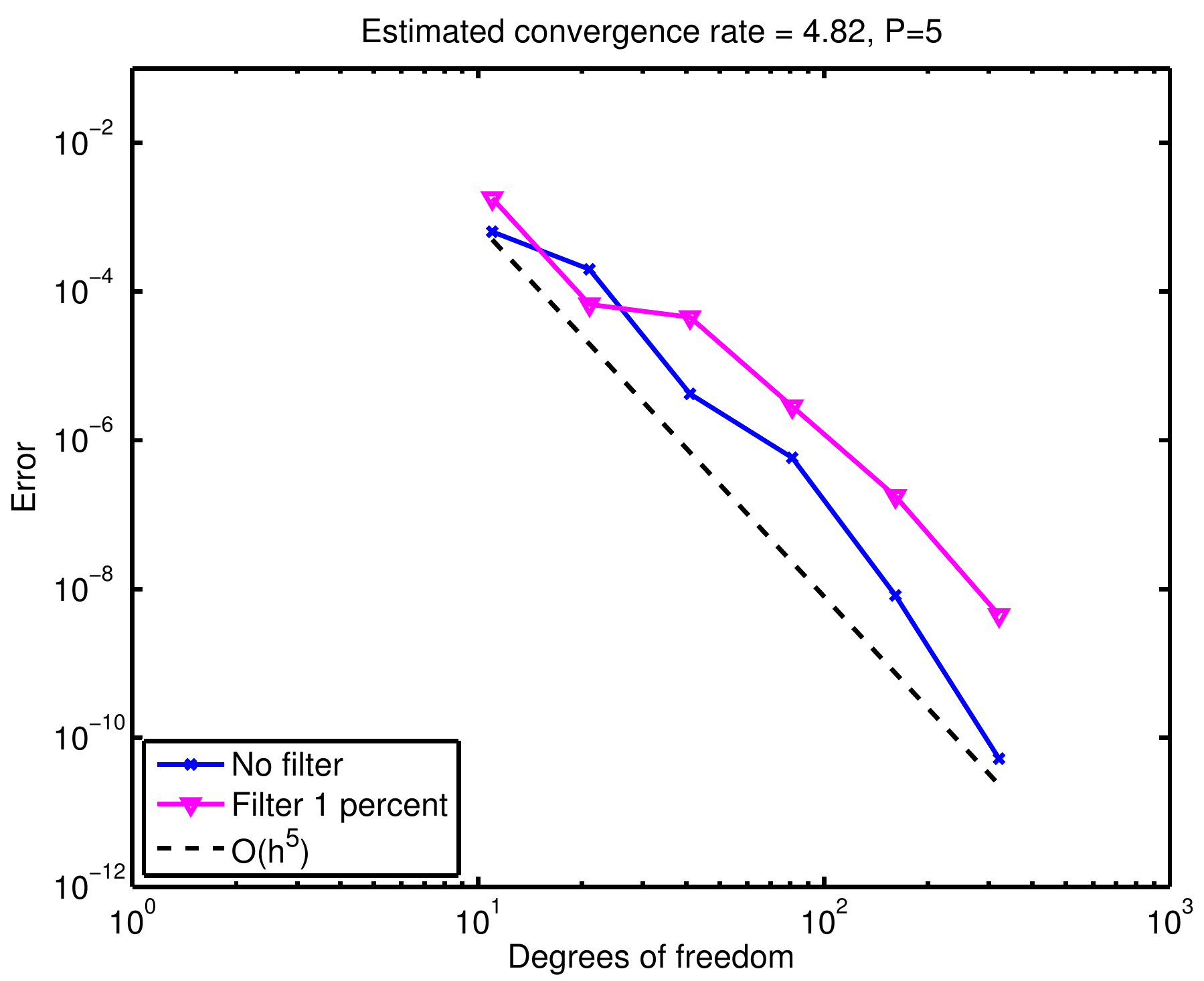} \\
\centering (k) $P=6$, $H/L=50\%$ \\
\includegraphics[height=3.6cm]{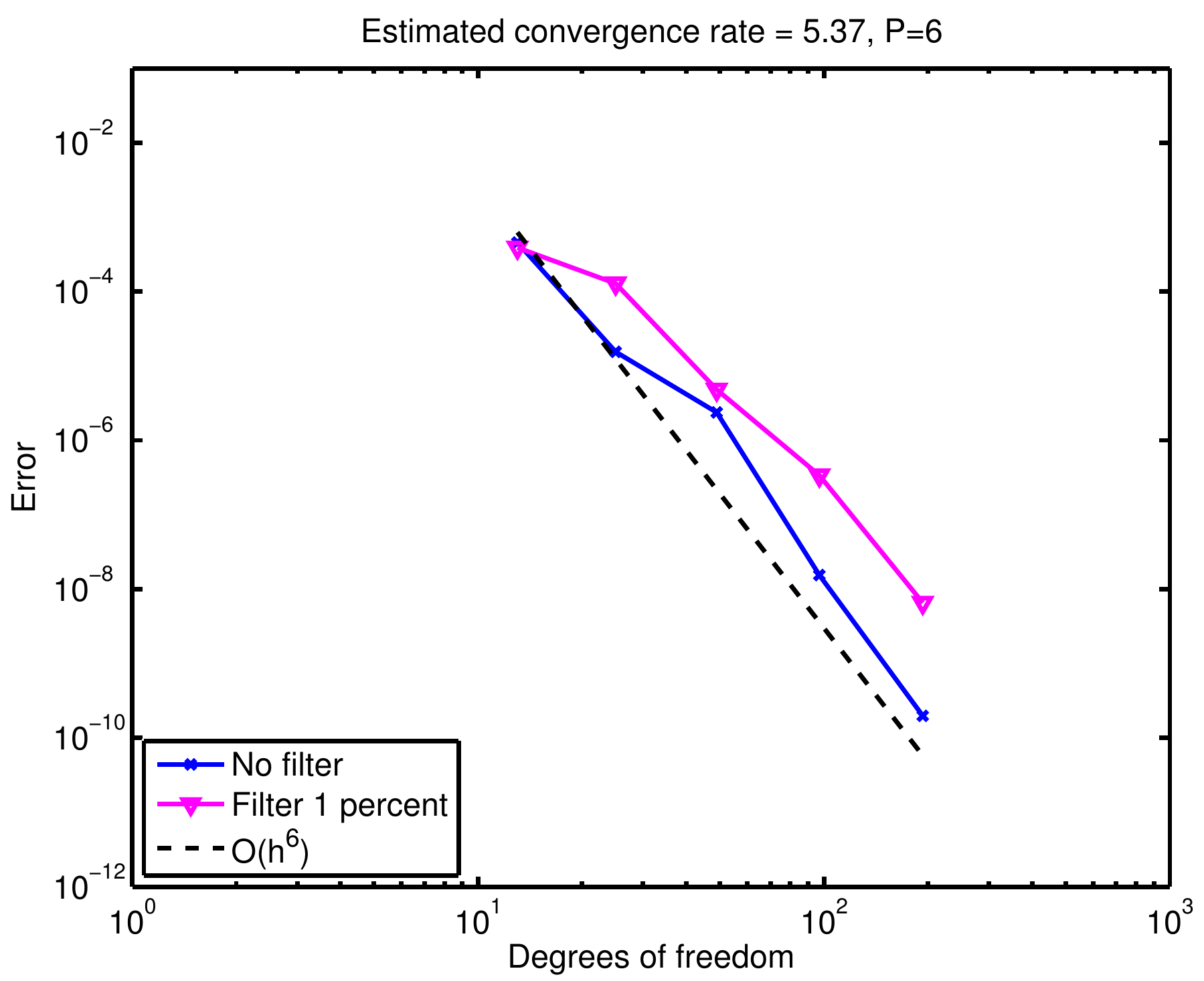}
\end{minipage}
\begin{minipage}{4.2cm}
\centering (c) $P=3$, $H/L=90\%$ \\
\includegraphics[height=3.6cm]{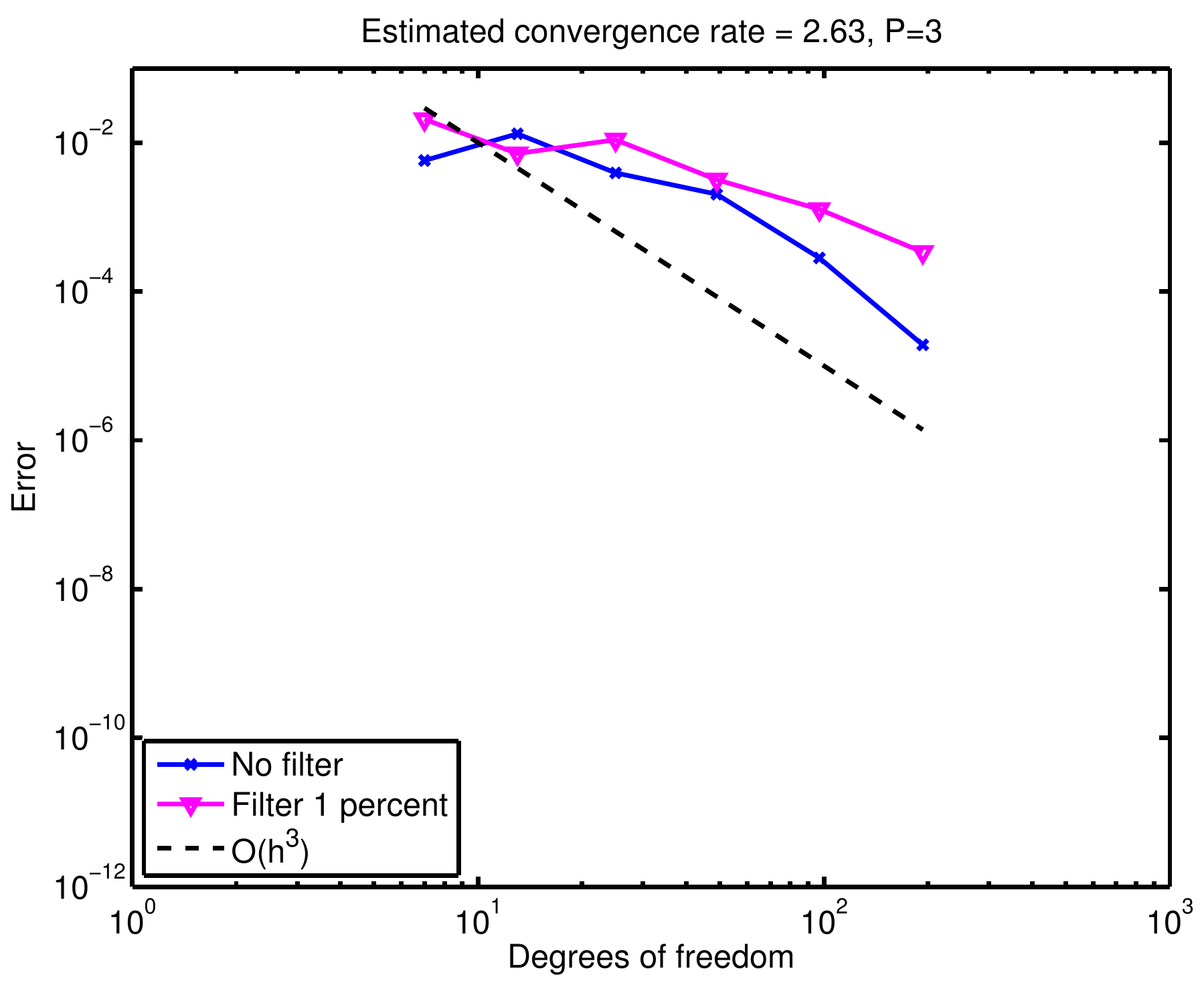} \\
\centering (f) $P=4$, $H/L=90\%$ \\
\includegraphics[height=3.6cm]{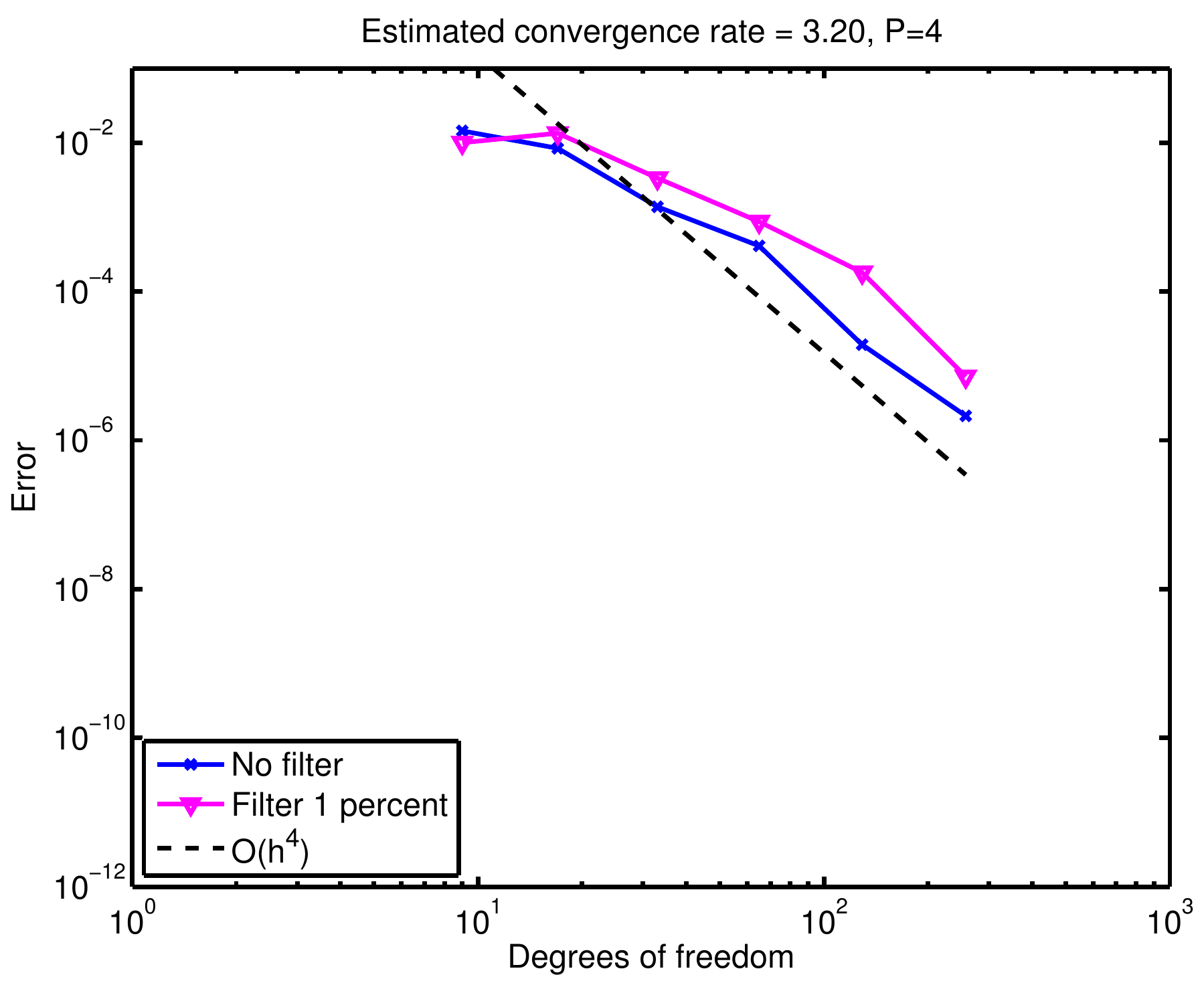} \\
\centering (i) $P=5$, $H/L=90\%$ \\
\includegraphics[height=3.6cm]{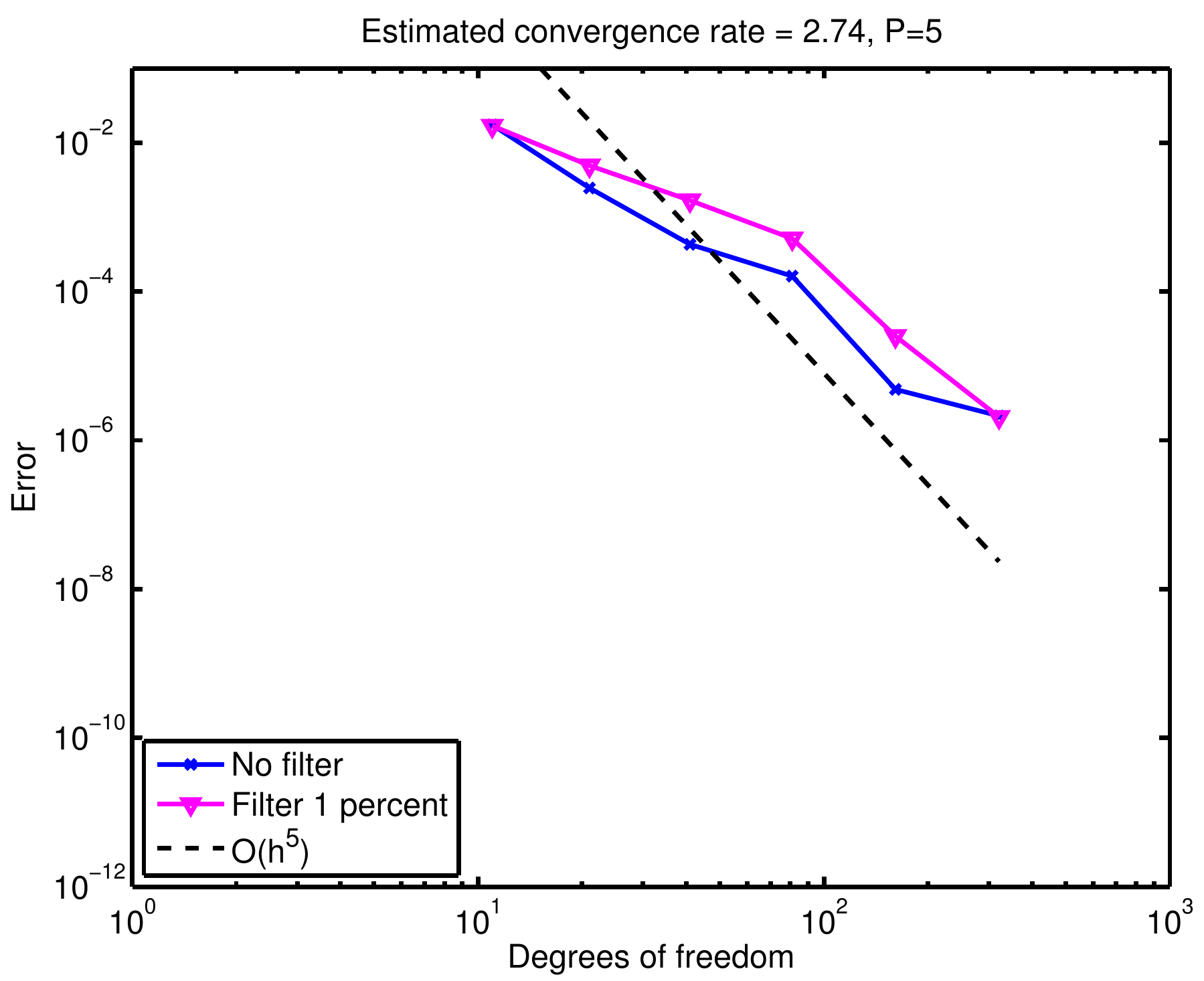} \\
\centering (l) $P=6$, $H/L=90\%$ \\
\includegraphics[height=3.6cm]{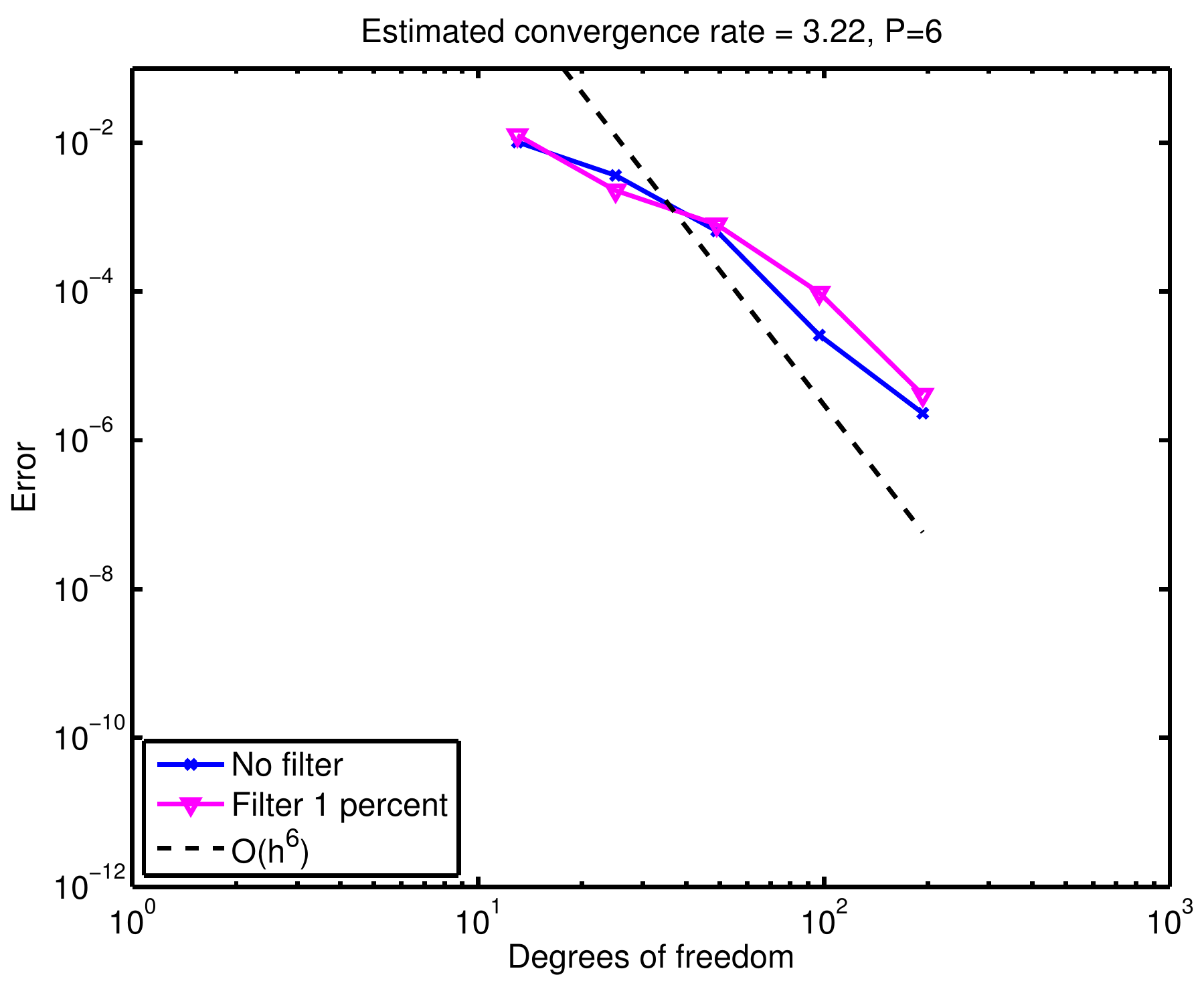}
\end{minipage}
\end{center}
\caption{Convergence tests with different expansion order $P$ in horizontal for nonlinear stream function wave solutions with parameters $kh=1$ and $H/L$ ratios of maximum wave steepness. A Galerkin scheme with over-integration is used with either no filtering or a 1\% filter applied. The time step size in all simulations is set to be small enough for spatial truncation errors to dominate. }
\label{fig:convtests}
\end{figure}

To confirm the high-order accuracy of the model and evaluate the influence of spectral filtering we have carried out convergence tests using the exact nonlinear stream function wave solutions for parameters $kh=1$ (dispersion) and $H/L=10\%$, $50\%$ and $90\%$ of maximum wave steepness (nonlinearity). 
The results for tests of the proposed Galerkin scheme with over-integration without spectral filtering and with spectral filtering using a cap of 1\% of the highest modal mode are presented in Figure \ref{fig:convtests}. 
The results confirm the high-order $\mathcal{O}(h^P)$ convergence for the spatial spectral element discretisation. 
Particularly, this is clear for the mildly nonlinear wave. 
With increasing nonlinearity more spatial resolution is required to accurately resolve all harmonic modes of the solution. 
The gentle filtering is found to reduce accuracy and has some detrimental effect on the convergence. 
For increasing order of the local basis functions, these effects become less significant.

\subsection{Harmonic generation over a submerged bar}

\begin{figure}[!htb]
\centering
\includegraphics[width=8cm]{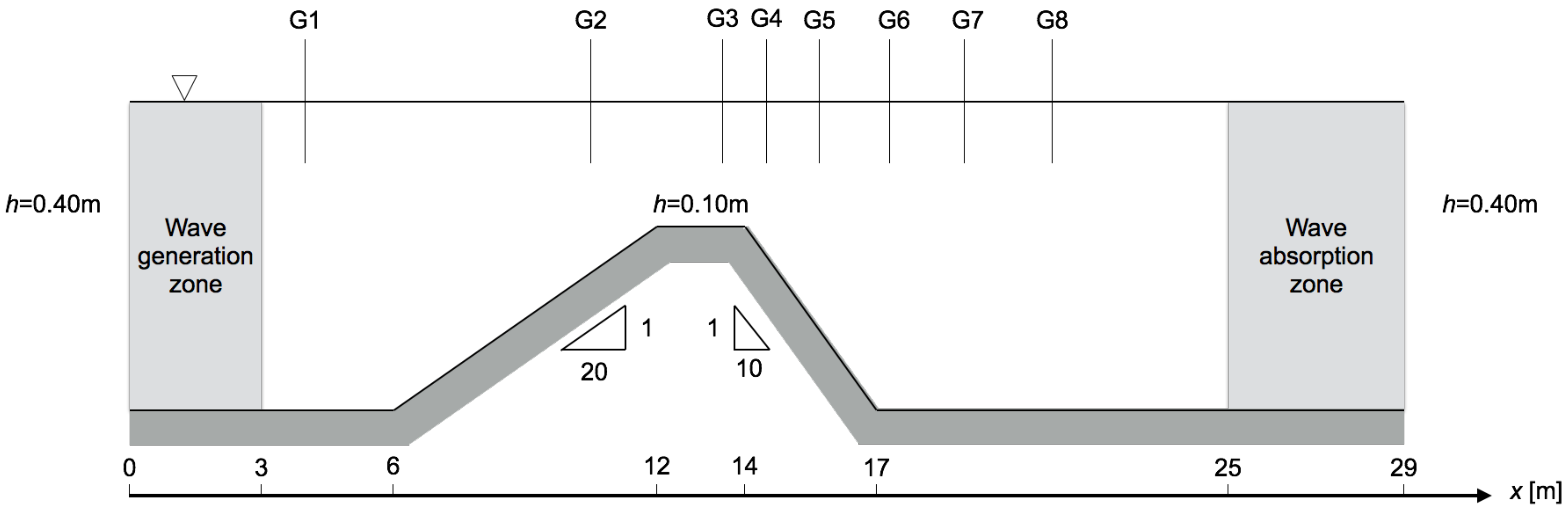}
\caption{Experimental setup of wave tank due to Beji and Battjes (1994) \cite{BB94}.}
\label{fig:bartestmesh}
\end{figure}

We consider the classical benchmark for wave transformation due to a submerged bar to test the accuracy of the SEM model. The test is often used for validation of deterministic dispersive and nonlinear wave models since it can be compared to experimental laboratory data and other known results in the literature, e.g., see \cite{GK99,MS99}. The numerical wave tank is illustrated in Fig. \ref{fig:bartestmesh}.

\begin{figure}[!htb]
\centering
\begin{minipage}{6cm}
\begin{center}
(a) $x=4$m \\
\includegraphics[width=5.2cm]{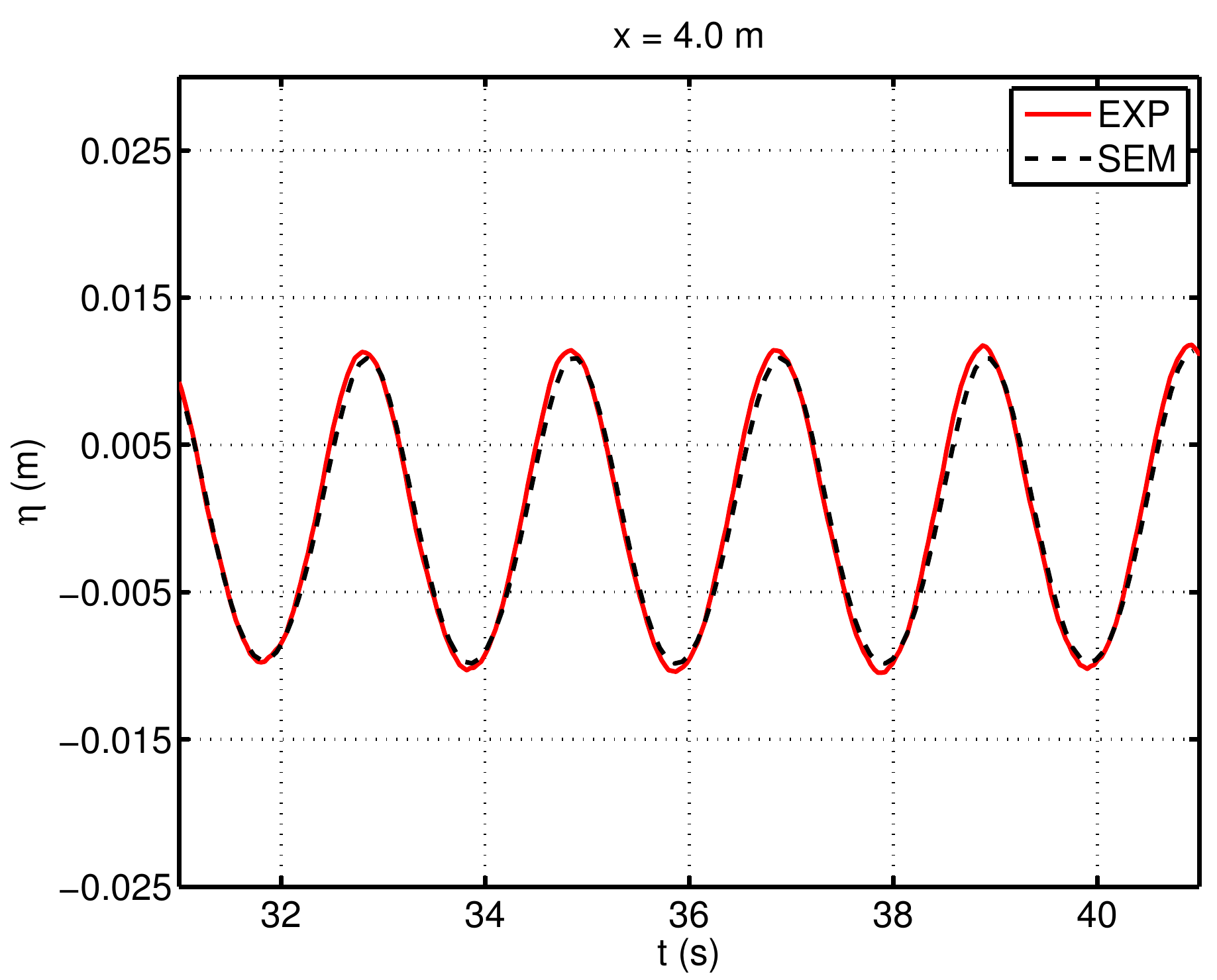}
\end{center}
\end{minipage}
\begin{minipage}{6cm}
\begin{center}
(b) $x=10.5$m \\
\includegraphics[width=5.2cm]{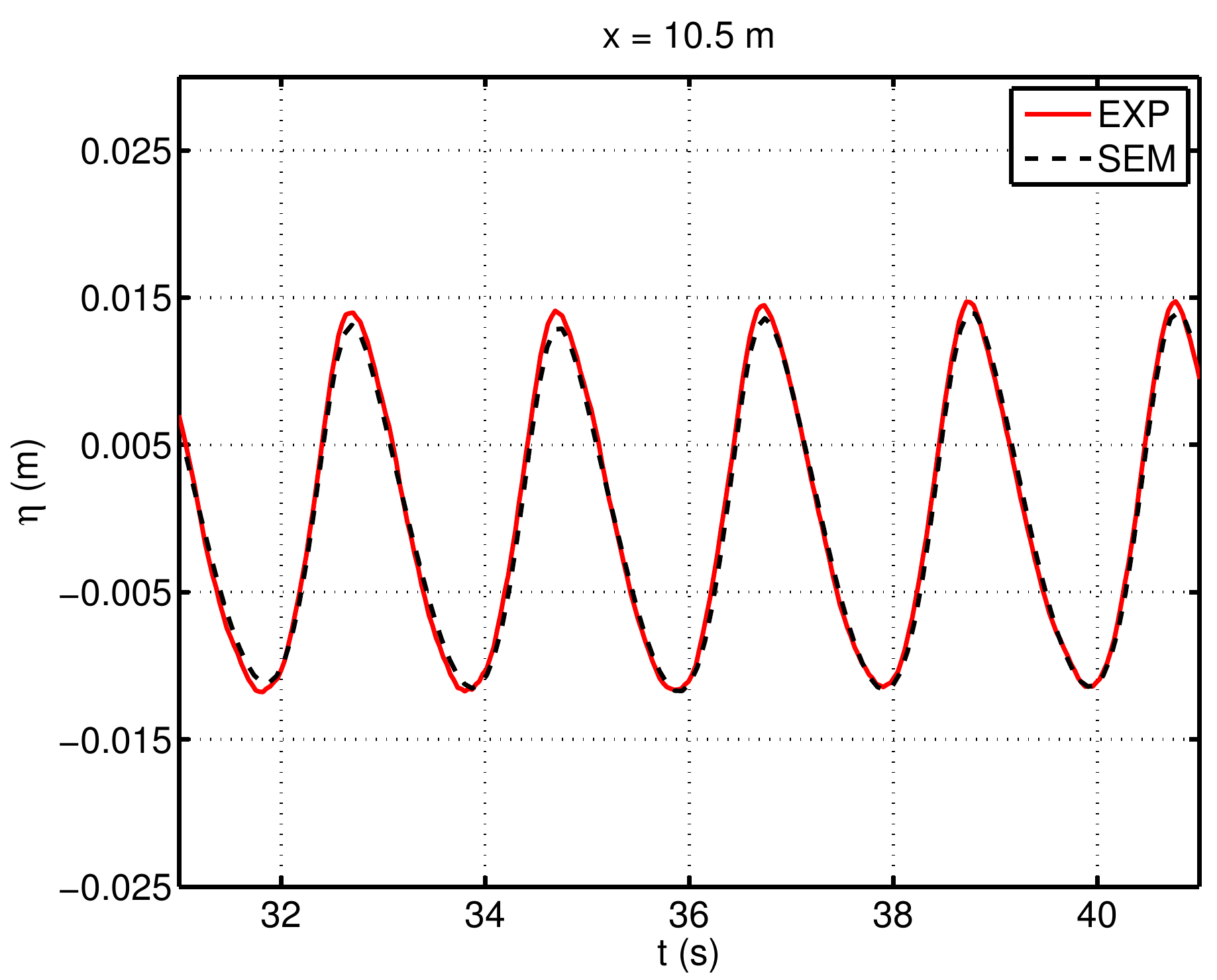}
\end{center}
\end{minipage} \\ 
\begin{minipage}{6cm}
\begin{center}
(c) $x=13.5$m \\
\includegraphics[width=5.2cm]{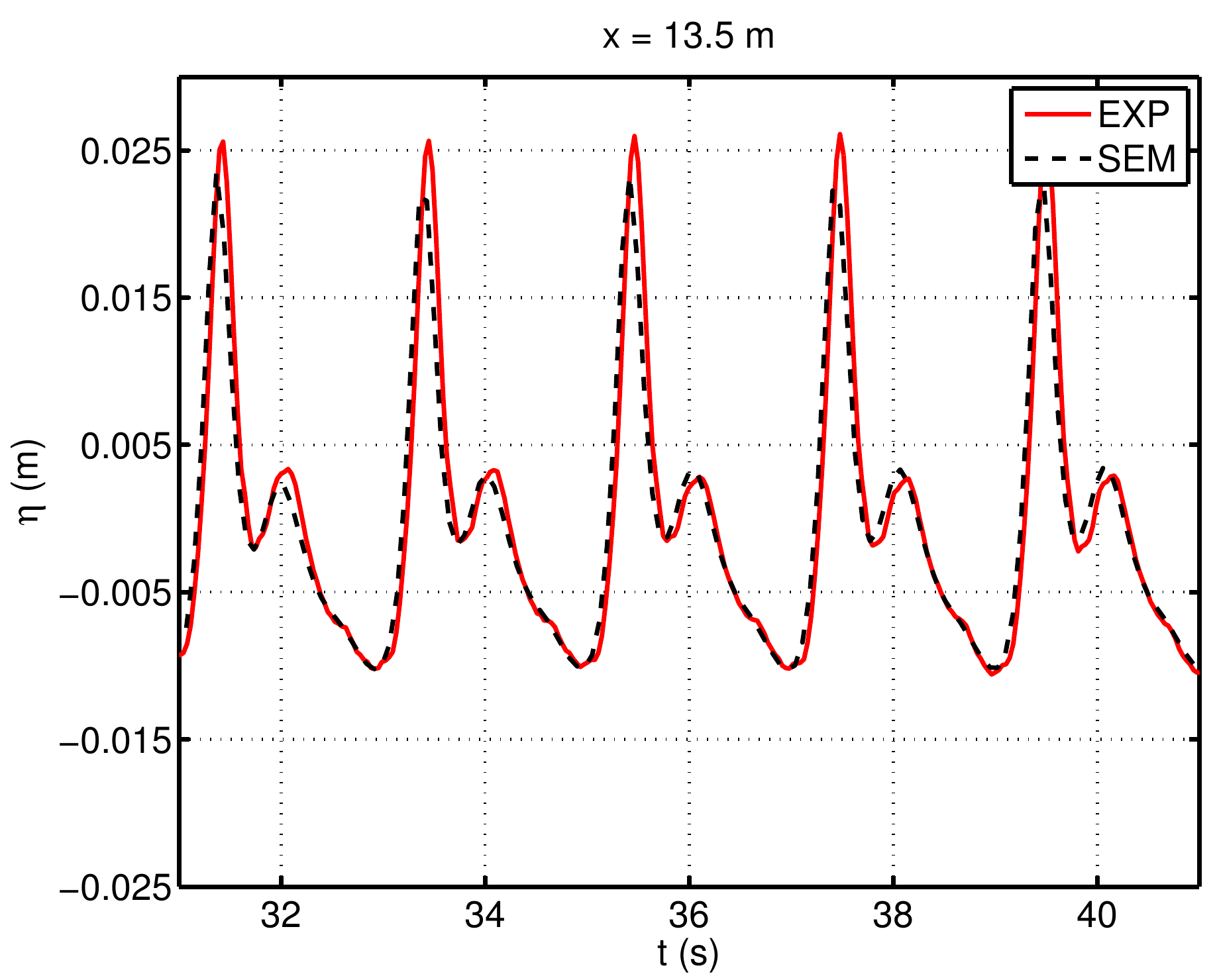}
\end{center}
\end{minipage}
\begin{minipage}{6cm}
\begin{center}
(d) $x=14.5$m \\
\includegraphics[width=5.2cm]{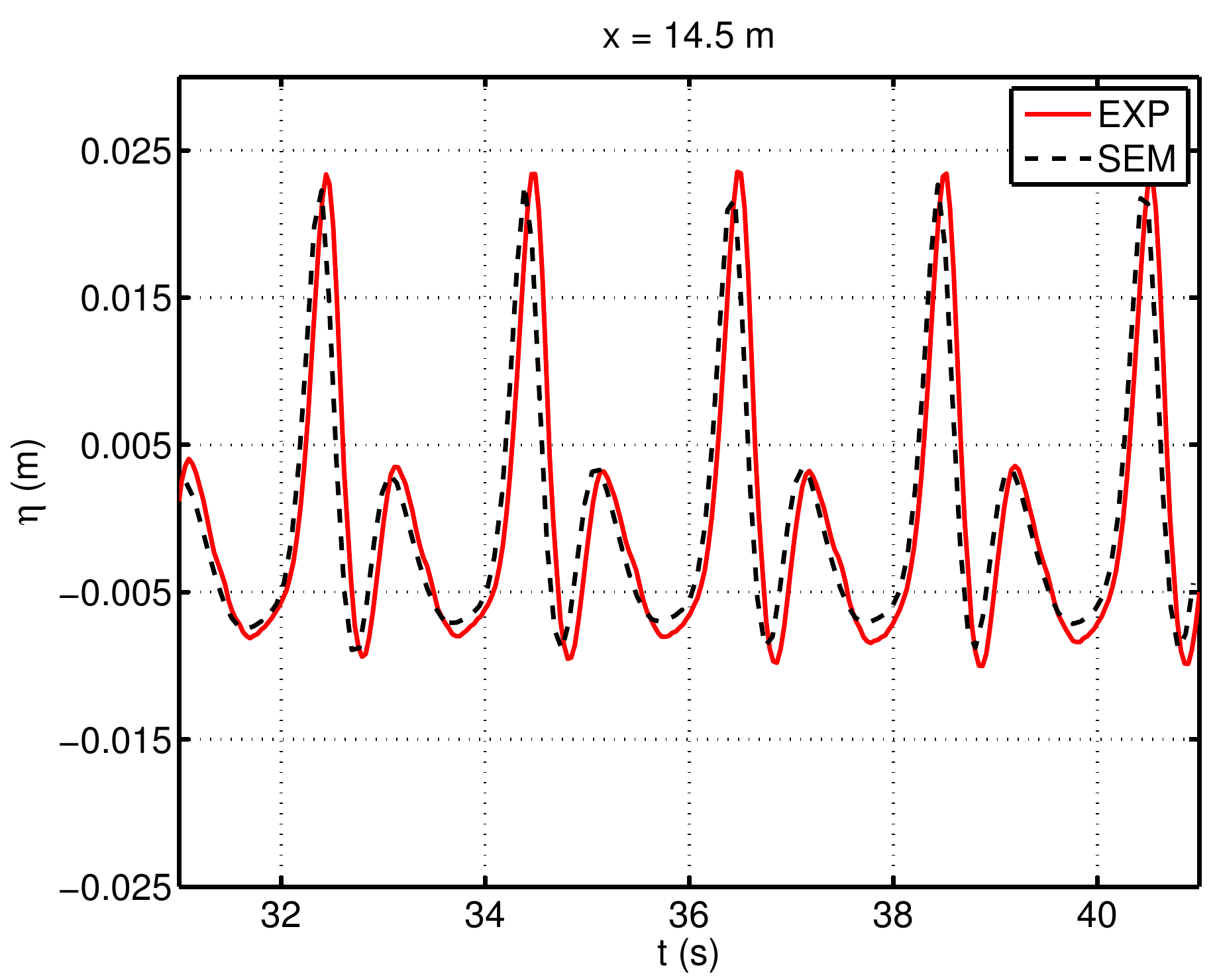}
\end{center}
\end{minipage} \\ 
\begin{minipage}{6cm}
\begin{center}
(e) $x=15.7$m \\
\includegraphics[width=5.2cm]{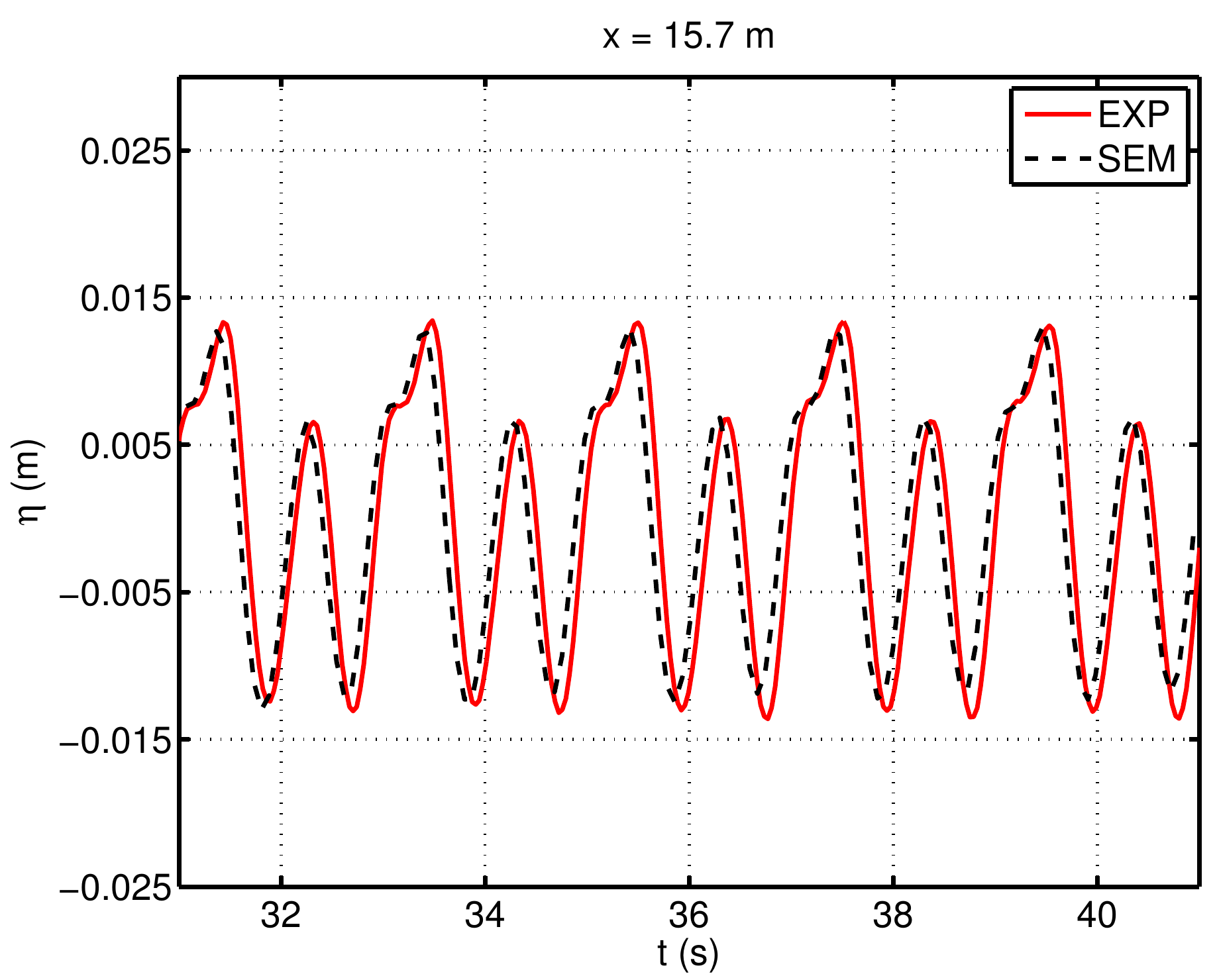}
\end{center}
\end{minipage}
\begin{minipage}{6cm}
\begin{center}
(f) $x=17.3$m \\
\includegraphics[width=5.2cm]{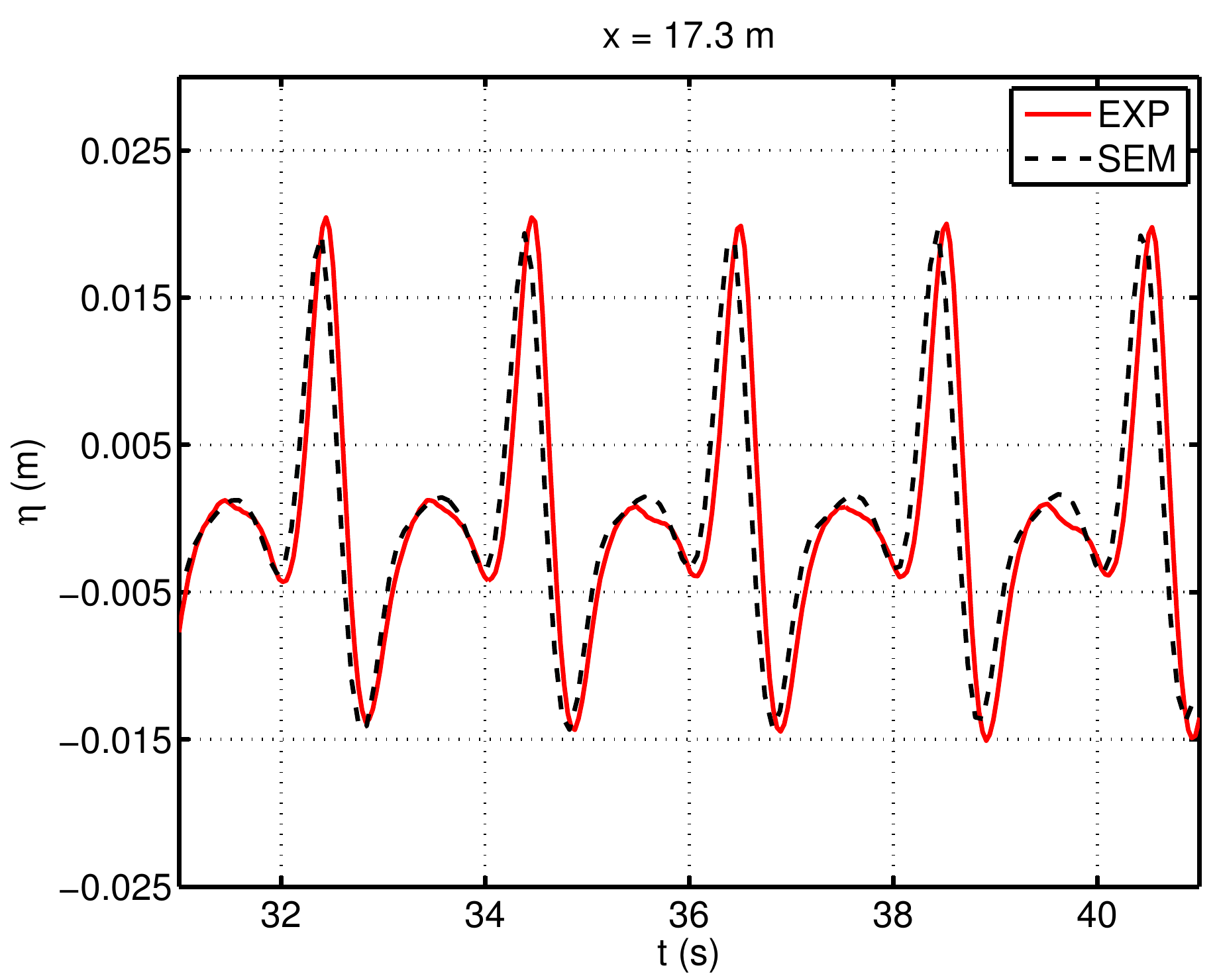}
\end{center}
\end{minipage}
\begin{minipage}{6cm}
\begin{center}
(g) $x=19.0$m \\
\includegraphics[width=5.2cm]{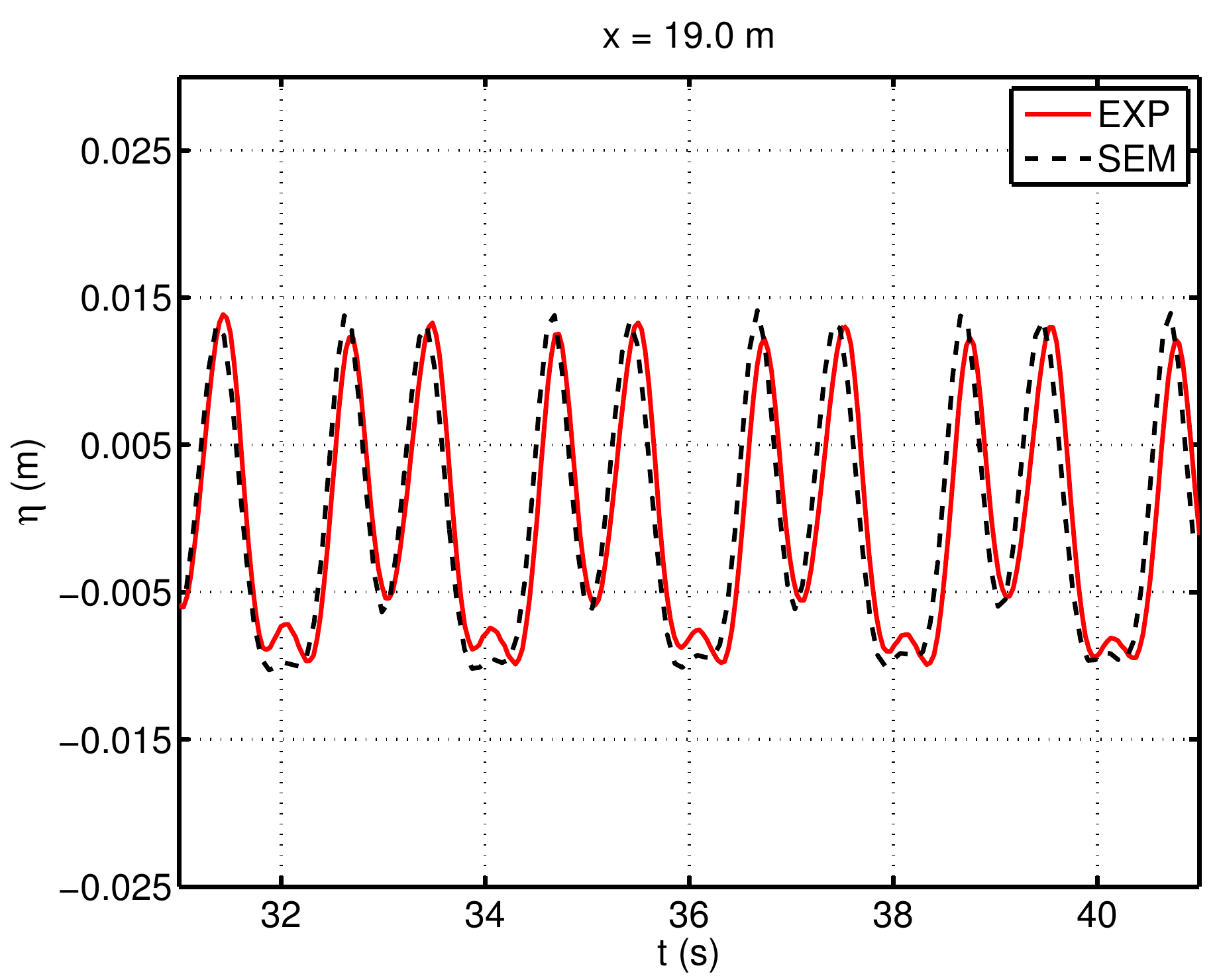}
\end{center}
\end{minipage}
\begin{minipage}{6cm}
\begin{center}
(h) $x=21.0$m \\
\includegraphics[width=5.2cm]{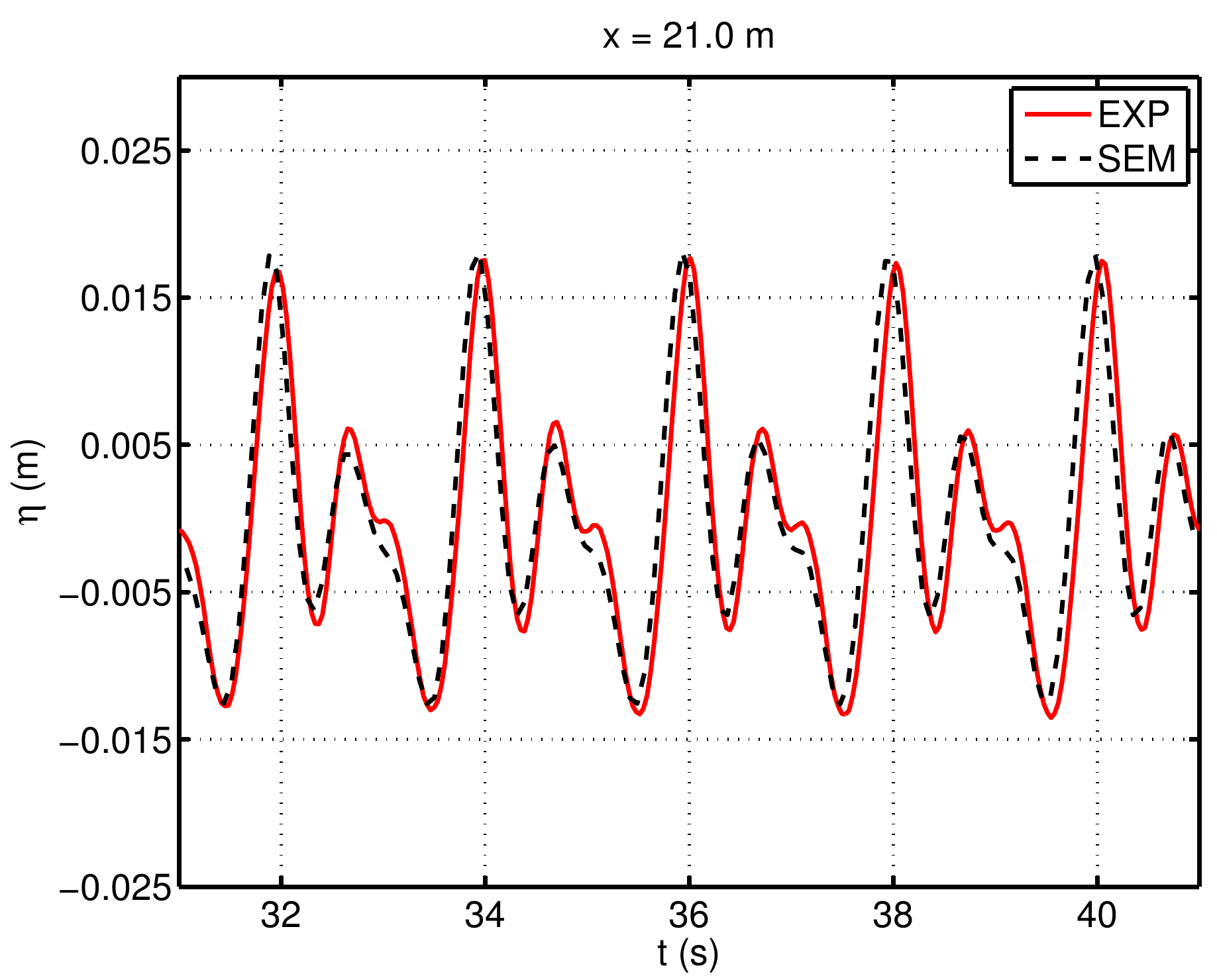}
\end{center}
\end{minipage}
\caption{Computed and measured time series of the surface elevation at different gauge locations for the bar test benchmark.}
\label{fig:bartest}
\end{figure}

The experiment was originally proposed in \cite{BB94} and subsequently an equivalent scaled experiment was carried out and described in \cite{LKK94}. We consider the setup for Case A in the original experiment. The weakly nonlinear input wave is generated in the numerical model using a regular stream function solution at undisturbed depth $0.4$ m with a wave height of $2$ cm and a wave period of $2.02$ s. The input wave is generated and propagates towards the submerged bar on a constant bottom. During the propagation over the bar, the wave will undergo a transformation resulting in a steepening and shortening of wavelengths due to nonlinear shoaling effects. At the top of the bar, the bound harmonics will be released as free harmonics (harmonic generation) decomposing the wave into shorter waves that propagate freely. Thus, to attain high accuracy in the calculations we need to use a model that can handle nonlinear wave-wave interactions and have accurate dispersion properties to capture the correct wave speed of the free harmonics after the bar. 
Taking advantage of the unstructured SEM the elements are of similar sizes but adjusted to have the interfaces positioned where the depth function has kinks in the first order gradients. 
Thus the $\sigma$-transformation is also local to the elements and entirely valid throughout the domain. 
The results presented in Figure \ref{fig:bartest} are based on a deterministic simulation where we have used 103 elements in the horizontal and one vertical layer with a multivariate basis of polynomial order $6$ in both the horizontal and vertical directions. 
A CFL condition with Courant number $C_r=0.5$ is used for defining the time step size. 
All results are found to be in excellent agreement with the experimental data with some qualitatively minor differences in phase between experimental and computed results which compare well to other published results.

\subsection{Irregular waves shoaling on a slope}

Mase and Kirby \cite{MaseKirby1992} conducted laboratory experiments of shoaling and breaking of irregular waves on a constant slope. The water depth offshore the slope is 0.47 m and the slope is 1/20. The incoming irregular waves were generated by a Pierson-Moskowitz (PM) spectrum with peak frequency of 1.0 Hz. This set of experiments has been used for testing Boussinesq models in e.g. \cite{WK95,Shietal12}. 

\begin{figure}[!htb]
\centering
\includegraphics[width=8cm]{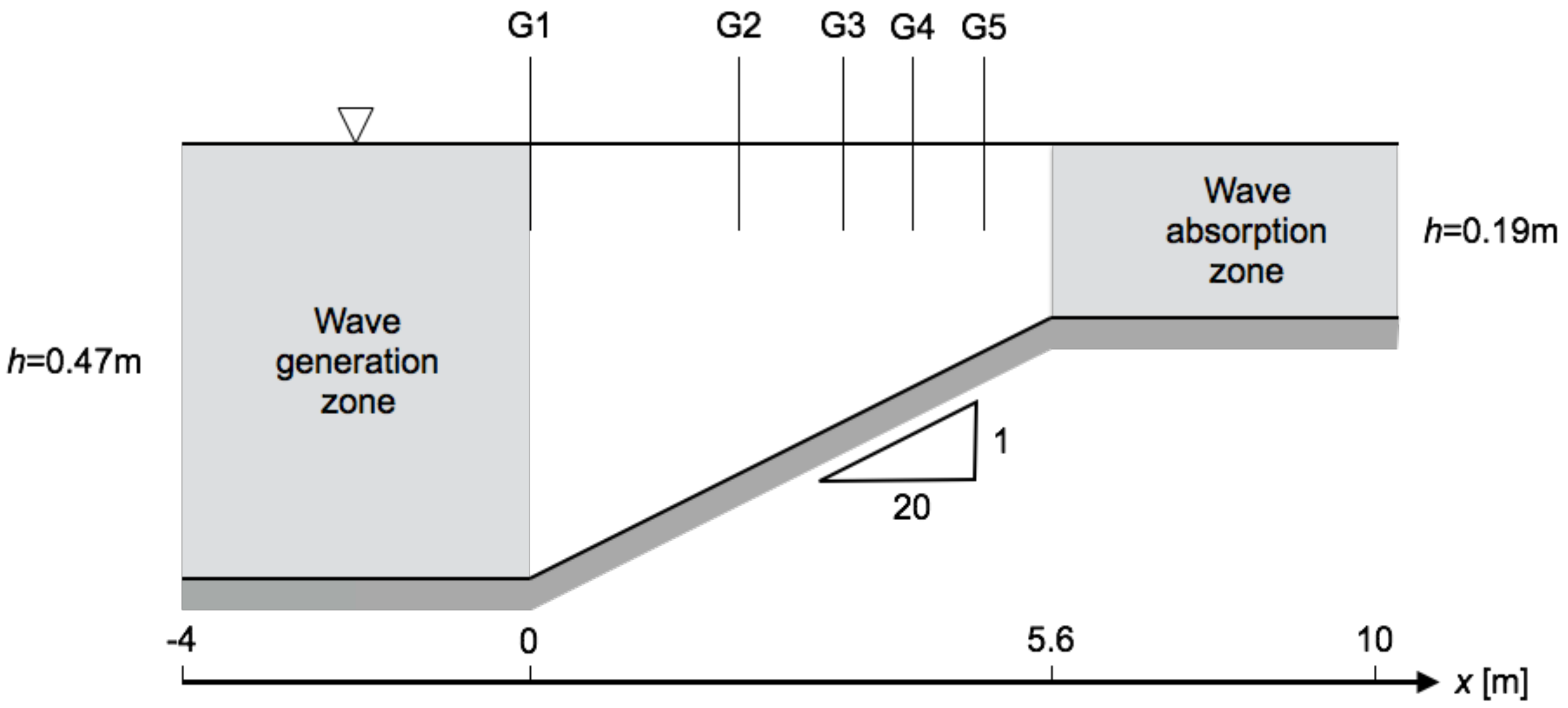}
\caption{Experimental setup of wave tank due to Mase and Kirby \cite{MaseKirby1992}.}
\label{fig:maseKirbyWaveTank}
\end{figure}

As the SEM model presently is limited to non-breaking waves we follow the approach of \cite{WK95}, i.e. the constant slope part of the wave tank is truncated and exchanged with a flat bottom at a depth such that no breaking occurs. For the SEM model we limit the still water depth to be no less than 0.19~m. The numerical wave tank is illustrated in Fig.~\ref{fig:maseKirbyWaveTank}. The wave gauges are located at $x=0,\, 2.4,\, 3.4,\, 4.4$ and $5.4$ m, respectively. Wave generation and absorption is performed with 4~m long relaxation zones. Further, as the random phase angle of the spectrum is unknown, the incoming wave train is generated in the numerical model based on a FFT of the measured waves at wave gauge 1 located at the toe of the slope. The wave kinematics is given by superposition of the linear wave modes.

\begin{figure}[!htb]
\centering
\begin{minipage}{6cm}
\begin{center}
(a) $x=2.4$m (G2) \\
\includegraphics[width=5.2cm]{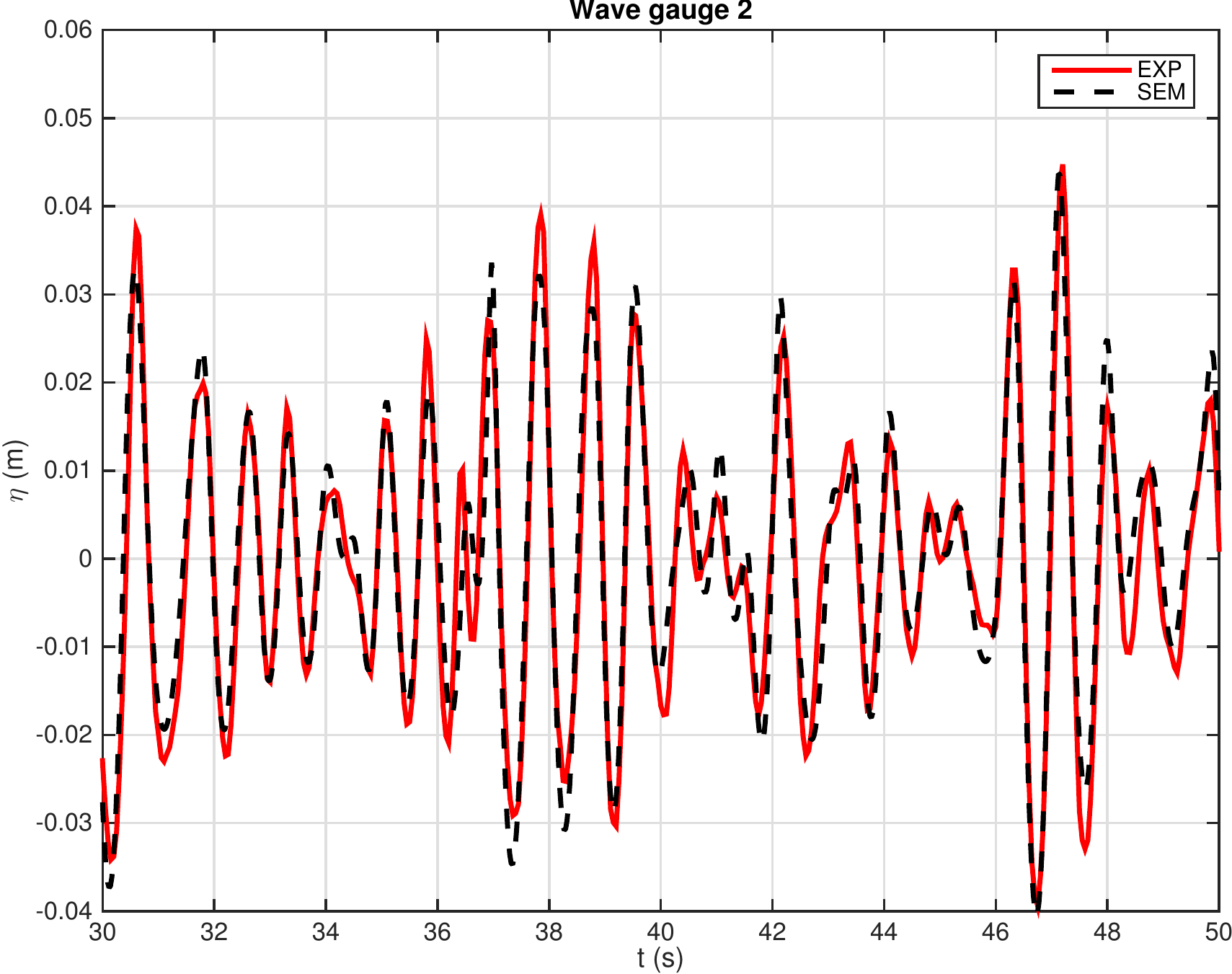}
\end{center}
\end{minipage}
\begin{minipage}{6cm}
\begin{center}
(b) $x=3.4$m (G3)\\
\includegraphics[width=5.2cm]{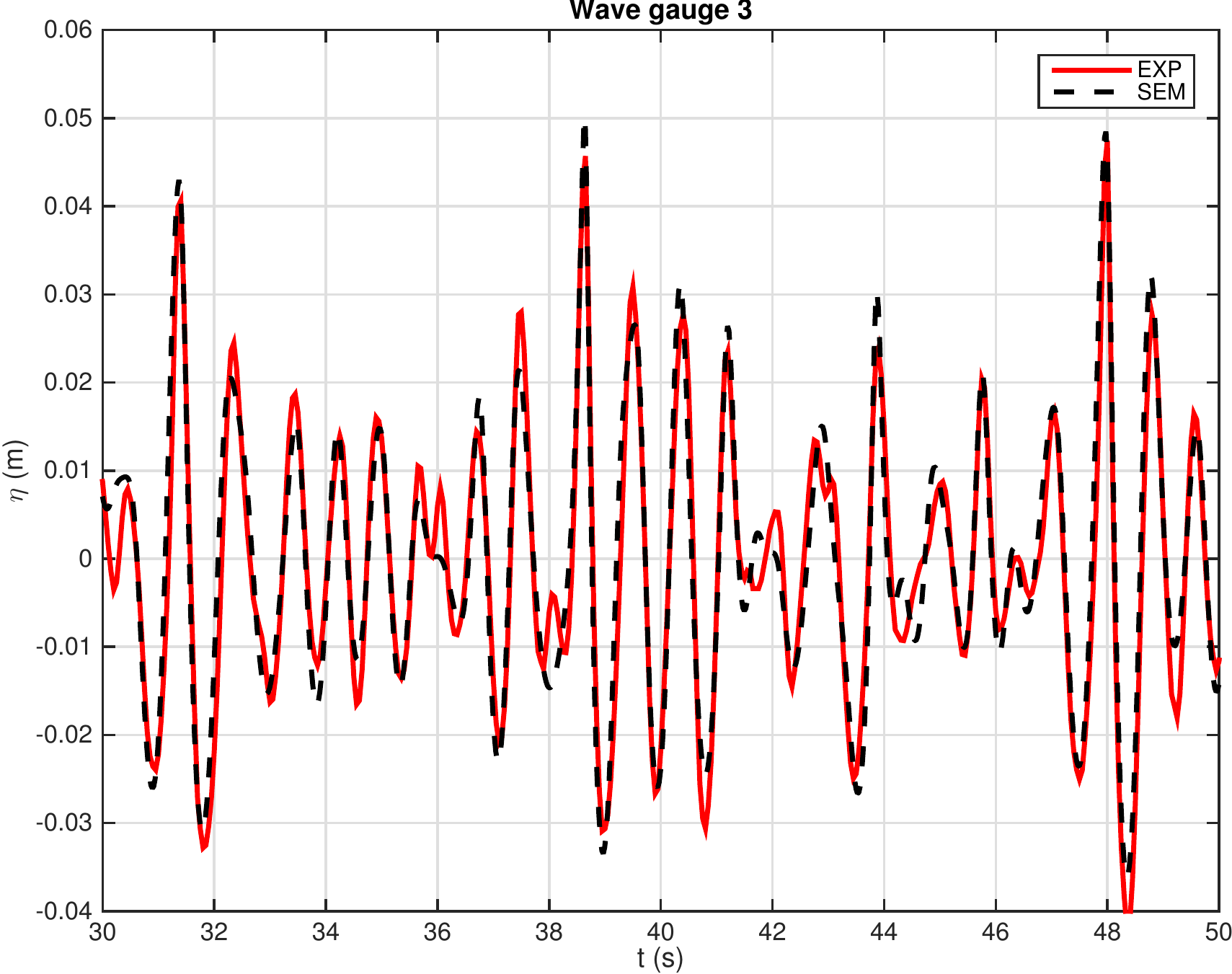}
\end{center}
\end{minipage} \\ 
\begin{minipage}{6cm}
\begin{center}
(c) $x=4.4$m (G4) \\
\includegraphics[width=5.2cm]{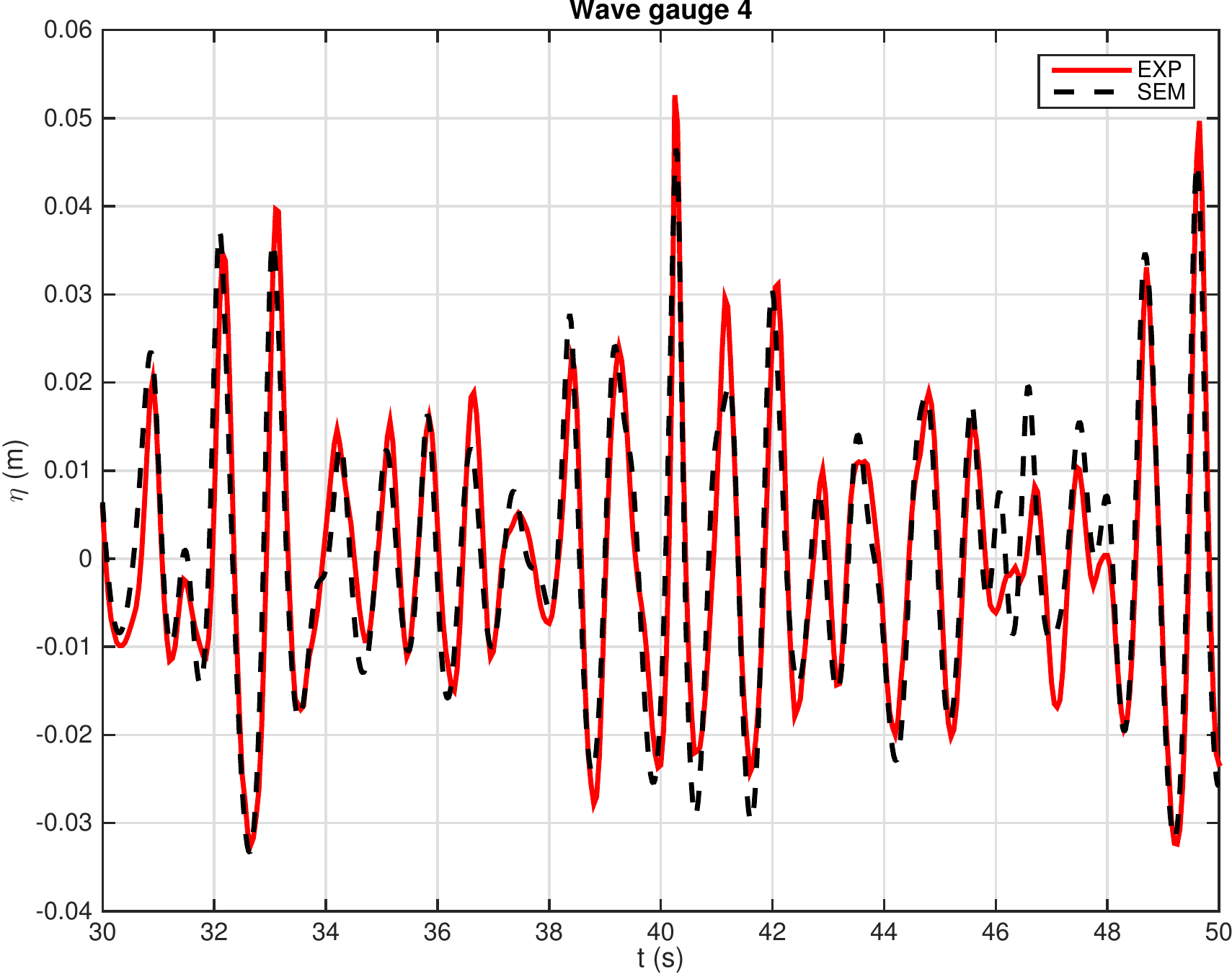}
\end{center}
\end{minipage}
\begin{minipage}{6cm}
\begin{center}
(d) $x=5.4$m (G5) \\
\includegraphics[width=5.2cm]{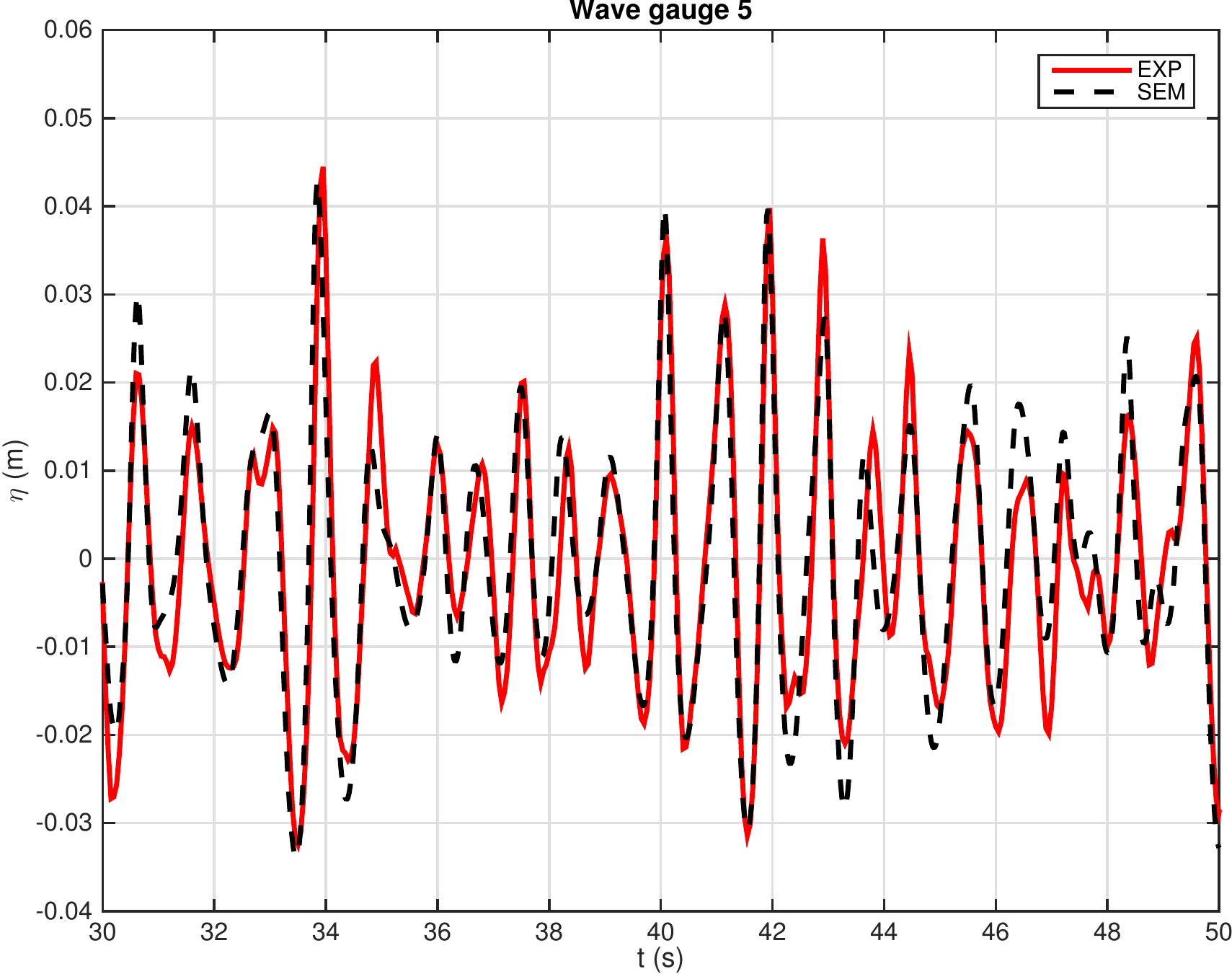}
\end{center}
\end{minipage} \\ 
\caption{Computed and measured time series of the surface elevation at different gauge locations for the irregular shoaling waves benchmark.}
\label{fig:maseKirbyResults}
\end{figure}

The computational domain is partitioned into elements of length 0.1 m with an expansion basis of polynomial order $P=5$ in the horizontal and vertical directions. 
Figure~\ref{fig:maseKirbyResults} shows the simulated and experimental free surface elevation at the different wave gauge locations. 
There is a reasonably good fit to the experimental data, with some minor differences in peak and through amplitudes. 
There are two main reasons for the discrepancy. 
First, the shorter waves of the PM spectra would require a very high polynomial order in order to be properly resolved and are thus diffused. 
Secondly, the assumption of linear superposition in the generation zone is not correct, giving that the incoming wave train does not exactly match the recorded.
 
\section{Conclusions}

We have presented a spectral element model for simulation of fully nonlinear water wave propagation. The main advantages of using the spectral element method is the opportunity for balancing high accuracy with unstructured meshes which can be adapted to geometry of arbitrary shape (sharp corners, curvilinear features, etc.) or features of the solution such as large gradients. The spectral element method's dual roads towards convergence, namely $h$- and $p$-adaptivity, allows for balancing accuracy and cost effectively.  

A two-dimensional spectral element model for fully nonlinear potential flow is implemented in a general computational framework using quadrature-free construction of local element operators. We have used arbitrary-order multivariate Lagrange (nodal) basis functions in space and an explicit fourth order Runge-Kutta method in time. The explicit time stepping is efficient as the model has a bounded eigenspectrum, and the stable time step sizes are governed by still water depth and the vertical resolution. The model is stabilised by using over-integration to effectively reduce aliasing errors and mild spectral model filtering to add some artificial viscosity to secure robustness for marginally resolved flows. 

The proposed model was shown to have a convergence rate of order $p$, although it is difficult to keep the sharp convergence rate for the most nonlinear waves due to high resolution requirements. Numerical experiments demonstrate that the stabilised model is both robust and accurate. It was shown how the spectral accuracy can be used to substantially reduce the number of degrees of freedom per wavelength and compare well to properties that can also be achieved with finite difference time domain schemes, but with the additional benefit of geometric flexibility. Also, we illustrate how the order of the basis function in vertical dimension can be used to control the range of validity of the model in terms of dispersive properties, i.e. a numerical truncation counterpart to the analytic truncation used in standard Boussinesq-type models. While the methodology is efficient in two space dimensions, particular attention is to be given to further improve numerical efficiency via efficient preconditioning methods that maintain high efficiency for general unstructured grids. 

In ongoing work, we aim at considering advanced nonlinear hydrodynamics problems by extending the current framework to also handle moving and floating objects. The present model needs to be further improved to handle run-up for calculations in the swash zone and a breaking wave model needs to be included for realistic applications. The proposed methodology in two space dimensions can be extended without conceptual modifications for three space dimensions, paving the road for marine hydrodynamics applications in 3D for offshore structures. Main challenges for enabling efficient computations in three space dimensions would be to (i) switch from a direct solver to an iterative solver strategy and identify an efficient and scalable preconditioning strategy, and (ii) via proper software design enable efficient mapping to modern and emerging many-core architectures for accelerated performance of the computational framework.

\section*{References}
\bibliographystyle{plain}
\bibliography{refs.bib}

\end{document}